\documentclass[aps,prd,reprint,twocolumn,superscriptaddress,preprintnumbers,nofootinbib]{revtex4-2}

\usepackage{amsthm}
\usepackage{slashed}
\usepackage{amssymb}
\usepackage{graphicx,color,xcolor}
\usepackage{amsmath,amssymb,amsfonts}
\usepackage{mathtools}
\usepackage[caption=false]{subfig}
\usepackage[colorlinks=True, citecolor=blue, urlcolor=blue, linkcolor=blue]{hyperref}
\usepackage{bbm}
\usepackage{scalerel,stackengine}
\usepackage{wasysym}
\usepackage{qcircuit}
\usepackage{physics}
\usepackage{bbold}
\usepackage{orcidlink}
\usepackage{comment}

\newcommand{\nn}{\nonumber}

\newcommand{\be}{\begin{eqnarray}}
\newcommand{\ee}{\end{eqnarray}}

\usepackage{soul}
\begin{document}

\title{Emergent Hydrodynamic Mode on SU(2) Plaquette Chains and Quantum Simulation}

\author{Francesco Turro \orcidlink{0000-0002-1107-2873}}
\email{francesco.turro@gmail.com}
\affiliation{InQubator for Quantum Simulation, University of Washington, Seattle, WA 98195, USA}

\author{Xiaojun Yao \orcidlink{0000-0002-8377-2203}}
\email{xjyao@uw.edu}
\affiliation{InQubator for Quantum Simulation, University of Washington, Seattle, WA 98195, USA}

\date{\today}
\preprint{IQuS@UW-21-095}
\begin{abstract}
We search for emergent hydrodynamic modes in real-time Hamiltonian dynamics of $2+1$-dimensional SU(2) lattice gauge theory on a quasi one dimensional plaquette chain, by numerically computing symmetric correlation functions of energy densities on lattice sizes of about $20$ with the local Hilbert space truncated at $j_{\rm max}=\frac{1}{2}$. Because of the Umklapp processes, we only find a mode for energy diffusion. The symmetric correlator exhibits transport peak near zero frequency with a width approximately proportional to momentum squared at small momentum, when the system is fully quantum ergodic, as indicated by the eigenenergy level statistics. This transport peak leads to a power-law $t^{-\frac{1}{2}}$ decay of the symmetric correlator at late time, also known as the long-time tail, as well as diffusion-like spreading in position space. We also introduce a quantum algorithm for computing the symmetric correlator on a quantum computer and find it gives results consistent with exact diagonalization when tested on the IBM emulator. Finally we discuss the future prospect of searching for the sound modes.
\end{abstract}

\maketitle

\section{Introduction}
Hydrodynamics is a universal effective description of late-time long-wavelength dynamics for systems obeying energy-momentum conservation. Its applications range from (sub)atomic systems of bosons and fermions~\cite{PhysRevA.86.033614,PhysRevA.89.053608,Brandstetter:2023jsy} to cosmology and astrophysics~\cite{Cen:1992zk,Vogelsberger:2014kha,Bea:2024bxu}.

In the field of relativistic heavy ion collisions, it has been used as a major tool to describe the evolution and expansion of the quark-gluon plasma (QGP)~\cite{Song:2010mg,Schenke:2010rr}, a deconfined phase of nuclear matter produced shortly after the initial collision. Analyses of experimental data based on relativistic viscous hydrodynamics~\cite{Bernhard:2019bmu,Nijs:2020ors} indicate that the shear viscosity of the QGP is small and the ratio of shear viscosity and entropy density is very close to that of certain consistent strongly coupled theories calculated via the holographic method~\cite{Policastro:2001yc,Brigante:2008gz}.

The values of shear viscosity and other transport coefficients of QCD at varying temperatures serve as important inputs for an array of phenomenological studies in heavy ion collisions. However, calculating them accurately in QCD is challenging due to nonperturbative effects and the sign problem in the Euclidean lattice QCD approach~\cite{Moore:2020pfu}, although progress has been made~\cite{Jeon:1994if,Arnold:2000dr,Arnold:2003zc,Ghiglieri:2018dib,Meyer:2007ic,Mages:2015rea,Itou:2020azb,Itou:2021hsj,Altenkort:2022yhb}. This challenge together with the rapid development of quantum computing technology motivated recent studies on using the Hamiltonian formulation of lattice gauge theory to calculate the shear viscosity~\cite{Cohen:2021imf,turro2024shear}. The Hamiltonian lattice method is limited to only small systems at the moment~\cite{turro2024shear}, due to the limitation of current classical computers and quantum devices, which can lead to large systematic uncertainty. In particular, the lattice size may be too small to support the full development of long-wavelength hydrodynamic behavior. In this paper, we aim at answering this question and demonstrating one case where hydrodynamic behavior emerges in the Hamiltonian dynamics on the lattice. Because of the real-time nature, such a demonstration is difficult in the Euclidean lattice setup. 

The main object that we will calculate in the Hamiltonian lattice approach is the real-time symmetric correlation function of stress-energy tensors, which is expected to evolve hydrodynamically at late time. Our results will show that a diffusive mode emerges in the SU(2) Hamiltonian dynamics on a lattice of size $\sim20$. Because of the Umklapp effect, sound modes are not observed at the couplings we study with the local Hilbert space truncation.

The paper is organized as follows: In Sec.~\ref{sec2}, we will briefly review relativistic hydrodynamics in $1+1$ dimensions in the continuum and on the lattice, in the latter of which Umklapp processes occur, and introduce the real-time symmetric correlator of stress-energy tensors. Then in Sec.~\ref{sec3}, the lattice Hamiltonian of SU(2) pure gauge theory in $2+1$ dimensions will be given for a plaquette chain, which renders the system quasi $1+1$ dimensional. This setup will be used to calculate the real-time symmetric correlator with the results presented and discussed in Sec.~\ref{sec4}. We will further introduce a quantum algorithm in Sec.~\ref{sec5} to calculate the symmetric correlator and show some test results of the algorithm. Finally, conclusions will be drawn in Sec.~\ref{sec6}, with a prospect of when the sound modes will emerge.

\section{Hydrodynamics and symmetric correlator}
\label{sec2}
In this section we will first briefly review the well-known relativistic hydrodynamics in the continuum and on the lattice with a focus on the case in $1+1$ dimensions. The motivation of the brief review is to highlight the characteristics of hydrodynamics that we want to look for in numerical lattice calculations of the real-time symmetric correlation functions, which will be introduced after the review of hydrodynamics.

\subsection{Hydrodynamics in the continuum}
Relativistic hydrodynamics describes conservation of energy and momentum (see e.g. Refs.~\cite{Romatschke:2009im,Teaney:2009qa,Jeon:2015dfa,Romatschke:2017ejr,rezzolla2013relativistic}, and Refs.~\cite{Crossley:2015evo,Glorioso:2017fpd,Vardhan:2024qdi} for effective field theory setups based on the action principles)
\begin{align}
\nabla_\mu T^{\mu\nu} = 0\,,
\end{align}
where $\nabla_\mu$ is the geometric covariant derivative and $T^{\mu\nu}$ is the stress-energy tensor. When the long-wavelength dynamics of a system close to local equilibrium is of major interest, one can express the stress-energy tensor in terms of local equilibrium properties such as energy density and fluid velocity and organize the expression according to a power counting given by the number of gradients. In $d$ dimensional spacetime, the stress-energy tensor expanded to linear order in gradients can be written as
\begin{align}
T^{\mu\nu} = \varepsilon u^\mu u^\nu - (P+\Pi) \Delta^{\mu\nu} + 2\eta \nabla^{<\mu} u^{\nu >}\,,
\end{align}
where $\varepsilon$ is the local energy density, $P$ denotes the pressure and $u^\mu$ stands for the fluid velocity. The shear and bulk viscous terms are given by
\begin{align}
\Pi =& -\zeta \nabla_\mu u^\mu \,,\nn\\
2\eta \nabla^{<\mu} u^{\nu >} =&\ \eta \Delta^{\mu\alpha} \Delta^{\nu\beta}(\nabla_\alpha u_\beta + \nabla_\beta u_\alpha) \nn\\
& - \frac{2\eta}{d-1}\Delta^{\mu\nu}\Delta^{\alpha\beta} \nabla_\alpha u_\beta \,,
\end{align}
where $\eta$ and $\zeta$ are the shear and bulk viscosities, respectively and $\Delta^{\mu\nu} = g^{\mu\nu} - u^\mu u^\nu$. We use the most negative convention for the spacetime metric $g^{\mu\nu}$.

In $1+1$ dimensions, if the system is at global equilibrium, the fluid is at rest $u_0^\mu = (1,0)$ and the energy density and pressure can be expressed in terms of their thermal expectation values, i.e., $\varepsilon = \varepsilon_0$ and $P=P_0$. Under a small perturbation, the fluid velocity, local energy density and pressure become $u^\mu = u_0^\mu + \delta u^\mu$, $\varepsilon = \varepsilon_0 + \delta\varepsilon$ and $P=P_0+\delta P$, respectively. Proper normalization of the fluid velocity $u_\mu u^\mu=1$ leads to $\delta u^t=0$. The pressure perturbation can be related to the energy density perturbation via the speed of sound $\delta P = c_s^2\delta \varepsilon$. Expanded to linear order in perturbation, the stress-energy tensor becomes
\begin{align}
T^{\mu\nu}  &= T_0^{\mu\nu}  + \delta T^{\mu\nu} \,,\nn\\
T_0^{\mu\nu}  &= \begin{pmatrix}
\varepsilon_0 & 0 \\
0 & P_0
\end{pmatrix} \,,\nn\\
\delta T^{\mu\nu} &= \begin{pmatrix}
\delta\varepsilon_0 & (\varepsilon_0+P_0)\delta u^x \\
(\varepsilon_0+P_0)\delta u^x & \delta P - \zeta\partial_x\delta u^x 
\end{pmatrix} \,.
\end{align}
For later convenience, we define the momentum density perturbation and the bulk viscous damping rate as
\begin{align}
g^x\equiv (\varepsilon_0+P_0) \delta u^x \,,\qquad \gamma_\zeta \equiv \frac{\zeta}{\varepsilon_0+P_0}\,,
\end{align}
respectively.
Then the linearized hydrodynamic equations $\nabla_\mu \delta T^{\mu\nu}=0$ can be expressed as
\begin{align}
\label{eqn:e_gx}
&\partial_t \delta\varepsilon + \partial_x g^x = 0 \,, \nn\\
& \partial_t g^x + c_s^2\partial_x \delta \varepsilon - \gamma_\zeta \partial_x^2 g^x = 0 \,,
\end{align}
where we have assumed the speed of sound and the bulk viscosity are constant. By using the first equation, the second equation can be equivalently written as 
\begin{align}
\label{eqn:gx}
\partial_t^2 g^x - c_s^2 \partial_x^2 g^x - \gamma_\zeta \partial_t \partial_x^2 g^x = 0\,.
\end{align}

Solutions to the hydrodynamic equation~\eqref{eqn:gx} can be found by studying the modes in frequency-momentum space, defined by
\begin{align}
g^x(t,x) = \int \frac{{\rm d}\omega {\rm d}k}{(2\pi)^2} \,e^{-i\omega t + ikx} g^x(\omega, k)\,.
\end{align}
Plugging into Eq.~\eqref{eqn:gx} leads to two sound modes specified by
\begin{align}
\omega_{s\pm} = \pm \sqrt{c_s^2k^2 - \frac{\gamma_\zeta^2k^4}{4}} - i\frac{\gamma_\zeta k^2}{2} \,.
\end{align}
For the validity of the gradient expansion in setting up the hydrodynamics, one expects $c_s k \gg \gamma_\zeta k^2$. Under this condition, the solution to $g^x(t,x)$ can be constructed by using Fourier transform (see e.g. Ref.~\cite{Arnold:1997gh})
\begin{align}
g^x(t,k) =&\ g^x(t=0,k) \cos(c_skt) e^{-\frac{\gamma_\zeta k^2t}{2}} \nn\\
&- ic_s \delta\varepsilon(t=0,k) \sin(c_skt) e^{-\frac{\gamma_\zeta k^2t}{2}} \,.
\end{align}
The solution to $\delta\varepsilon$ can be found by using the first line of Eq.~\eqref{eqn:e_gx}.

\subsection{Hydrodynamics on the lattice}
\label{sec:hydro_lattice}
Hydrodynamics on the lattice can be different from that in the continuum since the continuous translation symmetry is broken down to lattice translation symmetry. As a result, continuous momentum conservation breaks down to crystal momentum conservation. More crucially, crystal momentum is only conserved modulo $\frac{2\pi}{a}$~\cite{pitaevskii2017course} (we consider one spatial dimension here for simplicity) where $a$ denotes the lattice spacing. As an example, we consider a physical scattering process with two incoming particles with momenta $k_1$ and $k_2$ and two outgoing particles with momenta $k'_1$ and $k'_2$. Momentum conservation on the lattice is given by
\begin{align}
\label{eqn:umklapp}
k_1+k_2 = k_1' + k_2' +b\,,
\end{align}
where $b$ denotes any reciprocal lattice period and thus is an integer multiplying $\frac{2\pi}{a}$. Processes with $b\neq0$ are called Umklapp processes~\cite{pitaevskii2017course}, which break momentum conservation in the continuum sense. As a result, the momentum density is no longer an effective degree of freedom for hydrodynamics on the lattice~\cite{Arnold:1997gh}. So we only have the energy conservation equation that reads
\begin{align}
\partial_t \delta \varepsilon = -\partial_x j_\varepsilon \,,
\end{align}
where $j_\varepsilon$ is the energy flux density. The lowest-order gradient expansion gives
\begin{align}
j_\varepsilon = -D_\varepsilon \partial_x \varepsilon \,,
\end{align}
where $D_\varepsilon$ denotes the energy diffusion coefficient. The hydrodynamics on the lattice without momentum conservation is just energy diffusion equation
\begin{align}
\partial_t \delta \varepsilon -D_\varepsilon \partial_x^2 \delta \varepsilon = 0\,.
\end{align}
Mode analysis in frequency-momentum space leads to a diffusive mode given by
\begin{align}
\omega_d = -i D_\varepsilon k^2 \,.
\end{align}
The solution to the diffusion equation is 
\begin{align}
\delta \varepsilon(t,k) = e^{-D_\varepsilon k^2 t}\delta \varepsilon(t=0,k) \,.
\end{align}
If the initial condition is localized as a delta function $\delta\varepsilon(t=0,x) = \delta\varepsilon_0 \delta(x-x_0)$, the solution in spacetime is given by
\begin{align}
\label{eqn:diffu_sol}
\delta\varepsilon(t,x) = \frac{ \delta\varepsilon_0 }{\sqrt{4\pi D_\varepsilon t}} e^{-\frac{(x-x_0)^2}{4D_\varepsilon t}} \,.
\end{align}
At late time, it is expected that $\delta\varepsilon(t,x)\sim t^{-\frac{1}{2}}$ at some fixed $x$, which is known as the ``long-time tail''~\cite{Kovtun:2003vj,Caron-Huot:2009kyg,Kovtun:2012rj,Romatschke:2021imm,PhysRevA.89.053608,Akamatsu:2016llw,Martinez:2018wia,Shukla:2021ksb,Matthies:2024lqx}.

To recover the continuum hydrodynamics, one must take the continuum limit $a\to0$, which will suppress the Umklapp processes. More specifically, when the averaged energy density, or effectively the temperature, is much smaller than $\frac{2\pi}{a}$, Umklapp processes are unlikely to occur. If $b\neq0$ in Eq.~\eqref{eqn:umklapp}, its magnitude is at least $\frac{2\pi}{a}$ and then at least one momentum $k_i$ has a magnitude bigger than $\frac{\pi}{a}$. The existence of such a high momentum particle in the system at or close to thermal equilibrium is suppressed exponentially by $e^{-\beta E_{k_i}}$ with relativistic dispersion relation $E_{k_i}\sim \frac{\pi}{a}$. Therefore, normal processes with $b=0$ dominate and then momentum density is expected to be a good hydrodynamic degree of freedom.

\subsection{Real-time symmetric correlator}
Real-time symmetric correlation functions of stress-energy tensors are defined as
\begin{align}
\label{eqn:sym_Tmunu}
G_s^{\mu\nu}(t,x) = {\rm Tr}[ \{ T^{\mu\nu}(t,x), T^{\mu\nu}(t_0,x_0) \} \rho_T] \,,
\end{align}
where $\rho_T\equiv \frac{e^{-\beta H}}{Z}$ with $\beta=1/T$ denotes the thermal density matrix, and $t_0$ and $x_0$ are some arbitrary reference points that can be set to zero. The symmetric correlator characterizes the fluctuations in the system at thermal equilibrium.

The Onsager's postulate~\cite{Kovtun:2003vj} states that the symmetric correlators follow the same hydrodynamic equations for the classical stress-energy tensor. This can be understood by thinking of the symmetric correlator as the change of the expectation value of $T^{\mu\nu}(t,x)$ under the perturbation given by $\rho_T \to \rho_T + \{T^{\mu\nu}(t_0,x_0), \rho_T\}$, up to some normalization.

In the following, we will focus on the symmetric correlation function of the energy density by setting $\mu=\nu=0$ in Eq.~\eqref{eqn:sym_Tmunu} and investigate if its time evolution on the lattice exhibits the diffusive hydrodynamic behavior, as explained in Sec.~\ref{sec:hydro_lattice}.

\section{Hamiltonian of SU(2) Plaquette Chains}
\label{sec3}
\begin{figure}
\centering
\includegraphics[width=0.9\linewidth]{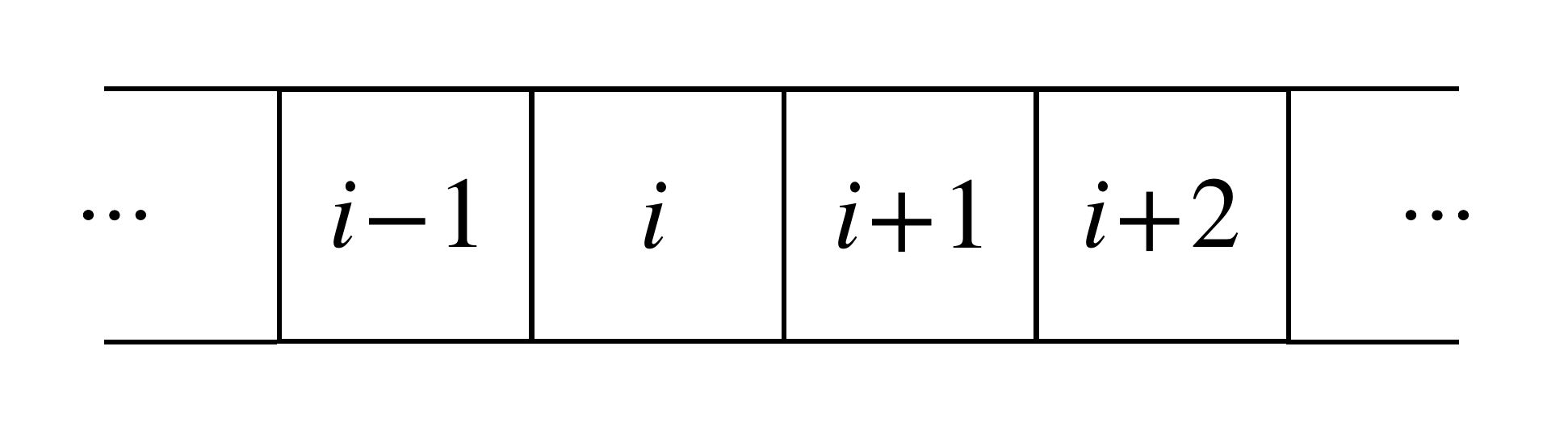}
\caption{A plaquette chain in $2+1$ dimensions where the plaquette location is labeled by $i$. In the spin map for the $j_{\rm max}=\frac{1}{2}$ case, each plaquette is mapped onto a spin. A spin-up state at site $i$ corresponds to the top and the bottom links of the $i$-th plaquette having $j=\frac{1}{2}$ electric fluxes while a spin-down state describes those links with no electric fluxes.}
\label{fig:chain}
\end{figure}

We will compute the symmetric correlator of the energy density for a specific lattice gauge theory, the SU(2) pure gauge theory in $2+1$ dimensions discretized on a chain of plaquettes, as shown in Fig.~\ref{fig:chain}. The Kogut-Susskind Hamiltonian of the SU(2) lattice gauge theory is well-known~\cite{PhysRevD.11.395}. In the electric basis~\cite{Byrnes:2005qx,Zohar:2014qma,Liu:2021tef}, one can manage to project onto physical states that obey the Gauss law at each vertex and explicitly write down matrix elements of the Hamiltonian in the physical Hilbert space~\cite{Klco:2019evd,ARahman:2021ktn,Hayata:2021kcp,ARahman:2022tkr}. One can then construct the physical Hamiltonian matrix on a lattice of a given size with the local Hilbert space truncated at some electric flux value $j_{\rm max}$.

When $j_{\rm max}=\frac{1}{2}$, the Hilbert space and the Hamiltonian can be greatly simplified and mapped onto a spin system~\cite{Hayata:2021kcp,ARahman:2022tkr,Yao:2023pht}. In this mapping, each square plaquette is identified with a spin. A spin-up state means the electric fluxes in the top and bottom links of the plaquette are $j=\frac{1}{2}$ in the original Hilbert space while a spin-down state means both of them are $j=0$. Details of this mapping and the convention we follow can be found in Ref.~\cite{Yao:2023pht}. After the mapping, the Hamiltonian can be written as
\begin{align}
\label{eqn:H_ising}
aH &= H^{\rm el}+H^{\rm mag} \nn\\
H^{\rm el} &= J \sum_{i=0}^{N-1}\sigma_i^z\sigma_{i+1}^z + h_z \sum_{i=0}^{N-1}\sigma_i^z \nn \\
H^{\rm mag} &= h_x \sum_{i=0}^{N-1} \frac{1-3\sigma_{i-1}^z}{4} \frac{1-3\sigma_{i+1}^z}{4} \sigma_i^x \,, 
\end{align}
where
\begin{align}
J = -\frac{3ag^2}{16}\,,\quad h_z=\frac{3ag^2}{8}\,,\quad h_x = -\frac{2}{ag^2}\,.
\end{align}
We have multiplied the Hamiltonian by the lattice spacing $a$. So we will express energy and other physical quantities in units of $a$.
For the periodic boundary condition we set $\sigma^z_i = \sigma^z_{i+N}$ while for the open boundary condition we choose $\sigma^z_{-1}=\sigma^z_N=-1$ (by this we mean the states at site $-1$ and $N$ are spin-down).
We define the Hamiltonian density (per plaquette) as
\begin{align}
H_i &= \frac{J}{2}(\sigma_i^z\sigma_{i+1}^z + \sigma_{i-1}^z\sigma_{i}^z) + h_z\sigma_i^z \nn\\
& \quad + h_x \frac{1-3\sigma_{i-1}^z}{4} \frac{1-3\sigma_{i+1}^z}{4} \sigma_i^x \,.
\end{align}

In the following, we will evaluate the real-time symmetric correlators of the energy density (Hamiltonian density) given by 
\begin{align}
\label{eqn:Gs}
G_s(t,j,i) &= {\rm Tr}[ \{H_j(t), H_i(0)\} \rho_T ] \,.
\end{align}
with $i$ and $j$ labeling lattice plaquette positions, and study their time evolution. In the continuum, $H_j(t)$ will correspond to $T^{00}(t,x)$.

\section{Classical Computing Results}
\label{sec4}
The results presented in this section are obtained by exactly diagonalizing the Hamiltonian. 

\subsection{Diffusive mode}
We first calculate the symmetric correlator in frequency-momentum space on a periodic plaquette chain. Eigenstates can be identified by their eigenenergies and crystal momenta, $|E_n(p)\rangle$, since lattice translation operators commute with the Hamiltonian. The crystal momentum on a lattice of size $N$ is a multiple of $\frac{2\pi}{Na}$, i.e., $p=\frac{2\pi n_p}{N}$ in lattice units with $n_p\in[-\lfloor N/2 \rfloor,-\lfloor N/2 \rfloor+1,\cdots,\lfloor (N-1)/2 \rfloor]$. If we treat lattice site labels $j$ and $i$ as continuous variables for the moment, inserting complete sets of eigenstates in Eq.~\eqref{eqn:Gs} and Fourier transforming gives
\begin{align}
G_s(\omega, k) =& \int {\rm d}t {\rm d}x \, e^{i\omega t - ikx} G_s(t,x) \nn\\
=& \sum_p \sum_q \sum_{E_n(p)} \sum_{E_m(q)} |\langle E_n(p) | H_i | E_m(q) \rangle |^2 \nn\\
&\times (2\pi)^2 \delta[\omega + E_n(p)-E_m(q)] \delta(k+p-q) \nn\\
&\times \frac{1}{Z}\Big[ e^{-\beta E_n(p)} + e^{-\beta E_m(q)} \Big] \,,
\end{align}
where $H_i$ is the Hamiltonian density in the Schr\"odinger picture and it suffices to use any lattice site $i$ in evaluating its matrix elements, due to the translation invariance. 

On the lattice, we need to modify this expression in two aspects. First, lattice positions are discrete so the spatial integration should be replaced with a summation over the lattice points. As a result, the Dirac delta function for momentum conservation turns to a Kronecker delta function
\begin{align}
2\pi(k+p-q) \to N\delta_{k, -p+q} \,.
\end{align}
Second, we have a finite number of eigenstates on the lattice and thus the energy spectrum is not continuous. We regularize the Dirac delta function for energy conservation with a rectangular function
\begin{align}
2\pi \delta[\omega + E_n(p)-E_m(q)] \to \frac{2\pi}{\Delta \omega}\bigg|_{ |\omega + E_n(p)- E_m(q)| < \frac{\Delta\omega}{2}} \,,
\end{align}
where $\Delta\omega$ is the size of the frequency window. It should be much smaller than the range of the eigenenergy spectrum but bigger than the averaged energy gap.

\begin{figure*}[t]
\centering
\subfloat[$ag^2=1.2,T=5$.\label{fig:Gs_wk1.2_T5}]{%
  \includegraphics[width=0.33\linewidth]{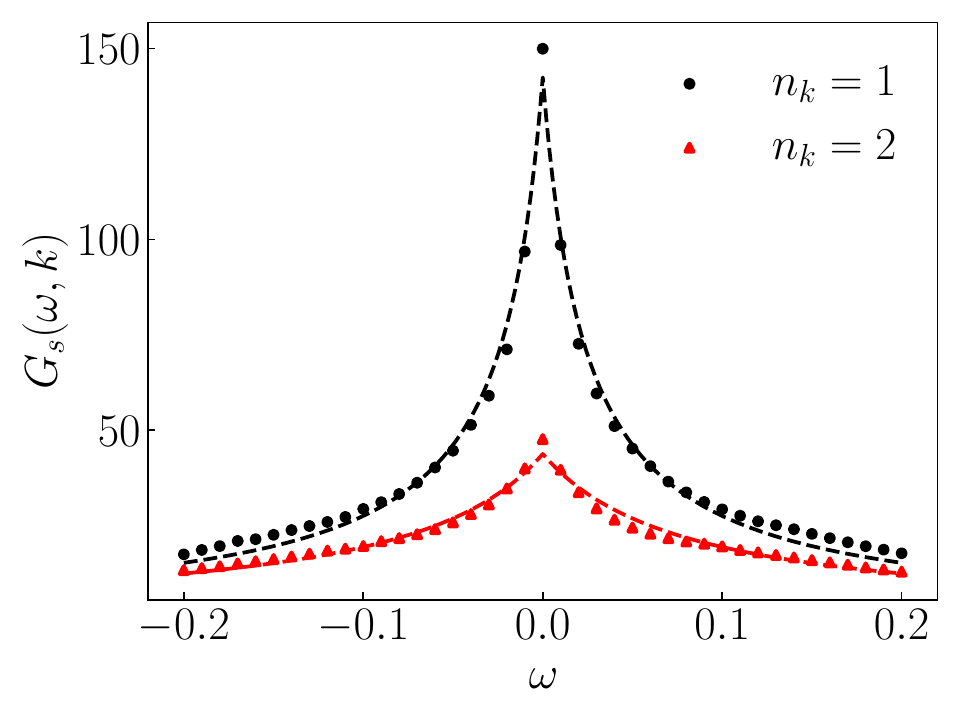}%
}\hfill
\subfloat[$ag^2=1.2,T=10$.\label{fig:Gs_wk1.2_T10}]{%
  \includegraphics[width=0.33\linewidth]{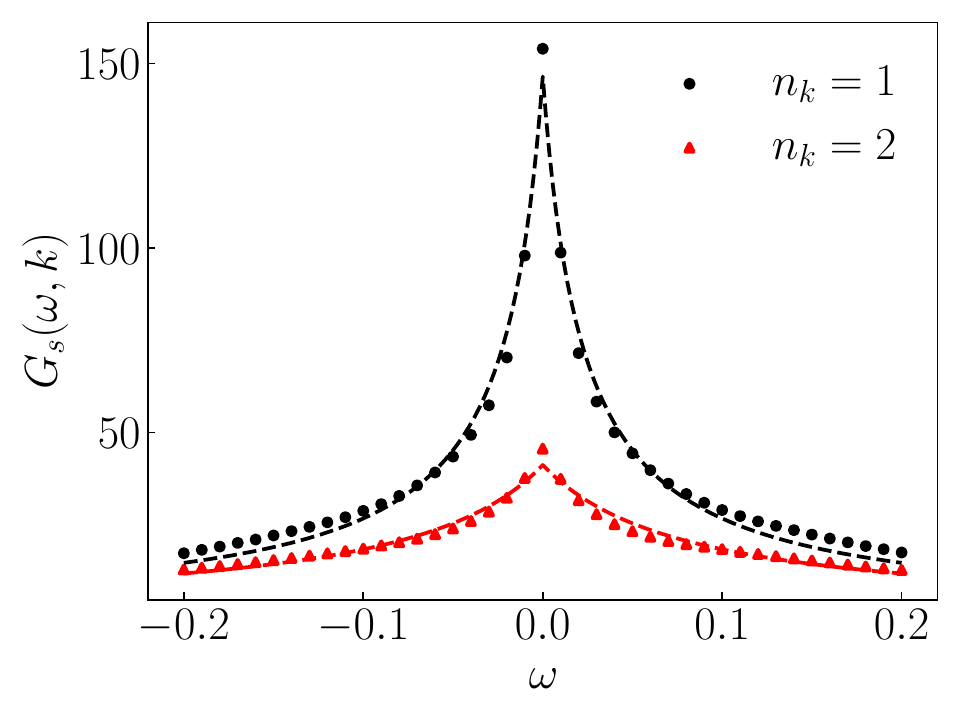}%
}\hfill
\subfloat[$ag^2=1.2,T=20$.\label{fig:Gs_wk1.2_T20}]{%
  \includegraphics[width=0.33\linewidth]{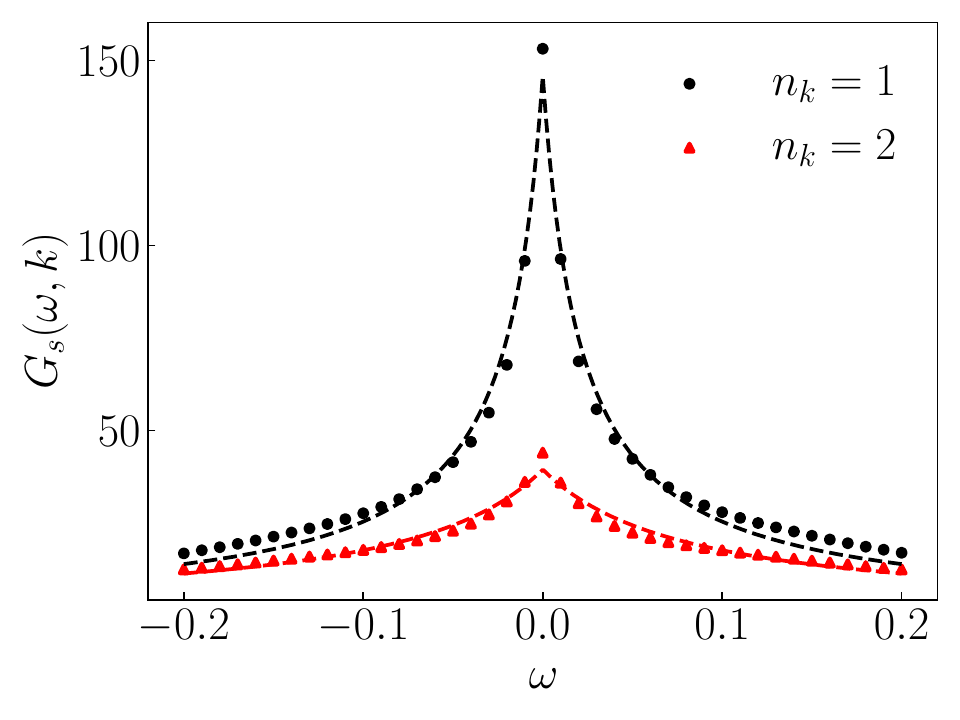}%
}

\subfloat[$ag^2=0.8,T=5$.\label{fig:Gs_wk0.8_T5}]{%
  \includegraphics[width=0.33\linewidth]{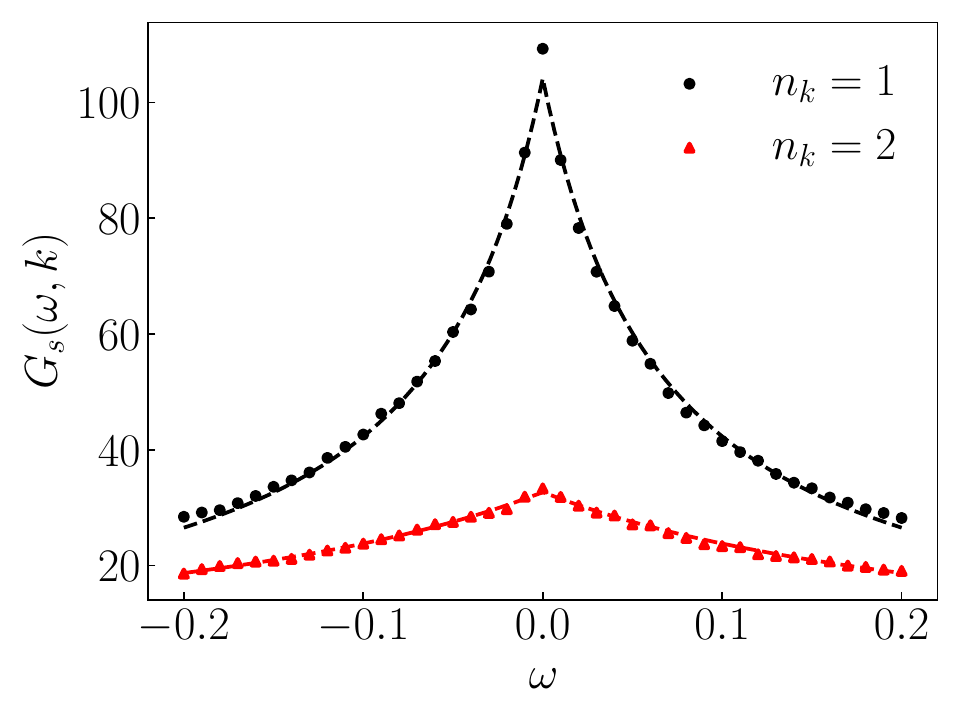}%
}\hfill
\subfloat[$ag^2=0.8,T=10$.\label{fig:Gs_wk0.8_T10}]{%
  \includegraphics[width=0.33\linewidth]{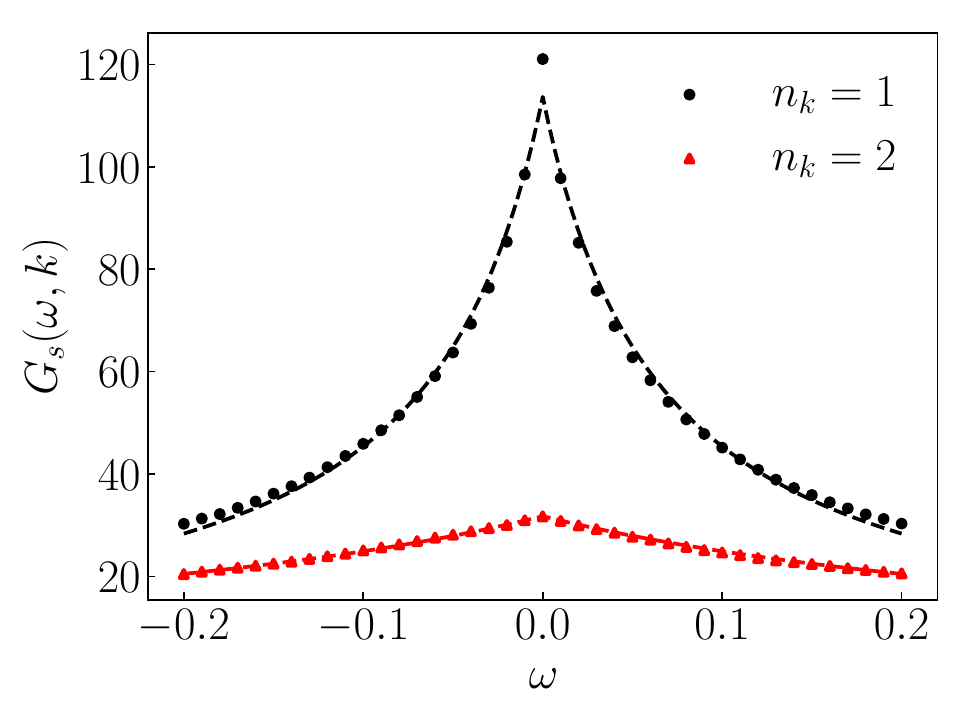}%
}\hfill
\subfloat[$ag^2=0.8,T=20$.\label{fig:Gs_wk0.8_T20}]{%
  \includegraphics[width=0.33\linewidth]{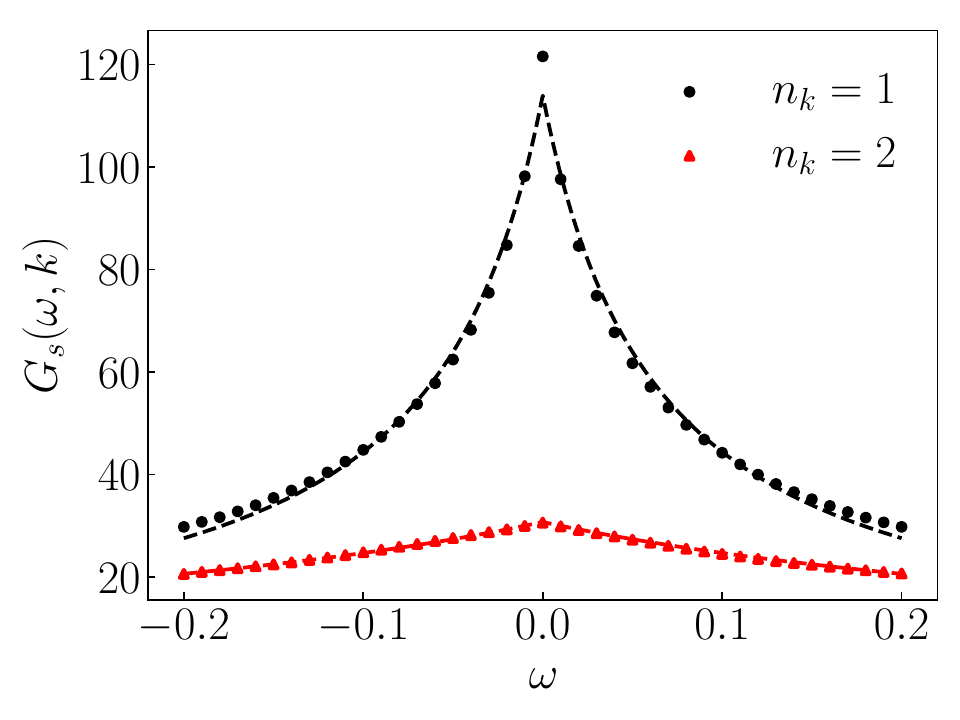}%
}
\caption{Real-time symmetric correlation function of energy densities in frequency space with two different crystal momenta $k=\frac{2\pi n_k}{N}$ on a periodic $N=20$ plaquette chain at two different couplings and three different temperatures. Dashed lines are fits of the function $\frac{a}{|\omega|+b}$. The fitting results are listed in Table~\ref{tab:1}, together with the momentum dependence of $a$ and $b$.}
\label{fig:Gs_wk}
\end{figure*}

\begin{table*}[t]
\centering
\begin{tabular}{ |c|c|c|c|c|c|c|c|c|c| } 
 \hline
 ~$N$~ & ~$ag^2$~ & ~$T$~ & $a(n_k=1)$ & $a(n_k=2)$ & ~$b(n_k=1)$~ & ~$b(n_k=2)$~ & ~$\frac{a(n_k=2)}{a(n_k=1)}$~ & ~$\frac{b(n_k=2)}{b(n_k=1)}$~ & ~$D_\varepsilon=\frac{N^2}{4\pi^2}b(n_k=1)$~ \\ 
 \hline
 20 & 1.2 & 5 & $3.41$ & $3.48$ & 0.0239 & 0.0794 & 1.02 & 3.32 & 0.243 \\
 \hline
 20 & 1.2 & 10 & $3.27$ & $3.31$ & 0.0223 & 0.0803 & 1.01 & 3.60 & 0.226\\
 \hline
 20 & 1.2 & 20 & ~$3.07$~ & ~$3.19$~ & 0.0210 & 0.0807 & 1.04 & 3.83 & 0.213\\
 \hline
 20 & 0.8 & 5 & $7.13$ & $8.75$ & 0.0684 & 0.267 & 1.23 & 3.91 & 0.693\\
 \hline
 20 & 0.8 & 10 & $7.55$ & $11.5$ &  0.0664 & 0.363 & 1.53 & 5.47 & 0.673\\
 \hline
 20 & 0.8 & 20 & $7.28$ & $12.5$ & 0.0639 & 0.408 & 1.72 & 6.39 & 0.647 \\
 \hline
\end{tabular}
\caption{Parameter values of Eq.~\eqref{eqn:fit_mode} fitted from the symmetric correlator in frequency-momentum space in Fig.~\ref{fig:Gs_wk} and the extracted energy diffusion coefficients for two couplings and three temperatures. Numbers are rounded to three significant figures.}
\label{tab:1}
\end{table*}

Combining everything together gives
\begin{align}
\label{eqn:final_Gs}
&G_s(\omega, k) \nn\\
=& \sum_p \sum_{E_n(p)} \sum_{E_m(p+k)} \frac{2\pi N} {\Delta\omega Z} |\langle E_n(p) | H_i | E_m(p+k) \rangle |^2 \nn\\
&\times  \Big[ e^{-\beta E_n(p)} + e^{-\beta E_m(p+k)} \Big]
\bigg|_{ |\omega + E_n(p)- E_m(p+k)| < \frac{\Delta\omega}{2}} \,\nn\\
\approx&\ \sum_p \sum_{E_n(p)} \frac{2\pi N}{Z} \overline{f^2_{H_i}(E_n,p,\omega,k)} \, e^{-\beta E_n(p)} \big( 1 + e^{-\beta \omega} \big)\,,
\end{align}
where we have assumed $\Delta\omega$ is small and defined
\begin{align}
\label{eqn:overline_f}
\overline{f^2_{H_i}(E_n,p,\omega,k)} \equiv&\ \frac{1}{\Delta\omega}  \sum_{E_m(p+k)} \bigg|_{ |\omega + E_n(p)- E_m(p+k)| < \frac{\Delta\omega}{2}} \nn\\
&\qquad |\langle E_n(p) | H_i | E_m(p+k) \rangle |^2 \,.
\end{align}

We then evaluate $G_s(\omega, k)$ on a $N=20$ periodic plaquette chain with $j_{\rm max}=\frac{1}{2}$ for two couplings $ag^2=1.2$ and $ag^2=0.8$. The frequency window size is chosen to be $\Delta\omega=0.01$. The energy gaps between the first excited state and the ground state are $E_1-E_0=4.1$ and $5.1$, and the averaged energy gaps are $9.3\times10^{-4}$ and $1.3\times10^{-3}$ in the $k=0$ momentum sector for $ag^2=1.2$ and $ag^2=0.8$, respectively. The results are shown in Fig.~\ref{fig:Gs_wk} for two values of momenta $\frac{2\pi n_k}{N}$ with $n_k=1$ and $n_k=2$ and three temperatures in lattice units. When summing over the $p$ momentum values in Eq.~\eqref{eqn:final_Gs}, we use $n_p\in[0,1,2,\cdots,9]$ for the $n_k=1$ case while $n_p\in[0,1,2\cdots,8]$ for the $n_k=2$ case. This truncation in $p$ is justified by the parity symmetry connecting the $p$ and $-p$ momentum sectors and the fact that $G_s$ is an even function in $k$. 

We observe transport peaks near zero frequency for both momenta, with the width in the $n_k=2$ case bigger than that in the $n_k=1$ case. Diffusive hydrodynamics predict the width to be $D_\varepsilon k^2=\frac{4\pi^2 n_k^2}{N^2} D_\varepsilon$, i.e., scale quadratically with momentum. To extract the width of the transport peak, we fit the numerical results in the frequency window $\omega\in[-0.2,0.2]$ with the function 
\begin{align}
\label{eqn:fit_mode}
\frac{a}{|\omega|+b} \,.
\end{align}
The fitted parameter values are listed in Table~\ref{tab:1}. One may wonder why the shape is of the form~\eqref{eqn:fit_mode} rather than the standard Lorentzian shape
\begin{align}
\label{eqn:lorentzian}
\frac{D_\varepsilon k^2}{\omega^2 + (D_\varepsilon k^2)^2} \,.
\end{align}
This is likely a finite size effect. In Appendix~\ref{app:1}, we estimate the time it takes for a perturbation introduced at $t=0$ at the center of the plaquette chain to reach the boundary. If we take this time to be $12.5$, it corresponds to the frequency $\frac{2\pi}{12.5}\approx 0.5$. It means that physics at frequency below $0.5$ is affected by the boundary condition. The transport peak observed here is located at frequency below $0.1$. On the other hand, by calculating the symmetric correlator in the microcanonical ensemble, we find the transport peak in a certain energy shell can be described by a Lorentzian shape, which is shown in Appendix~\ref{app:2}. To further understand this, one needs to perform calculations on a lattice big enough so that the transport peak appears before the perturbation induced at the center of the lattice reaches the boundary.

The physical meaning of the parameter $b$ is $D_\varepsilon k^2$. If the transport peak is diffusive, the ratio $\frac{b(n_k=2)}{b(n_k=1)}$ should be $4$. The best of our numerical results are $3.83$ for the $ag^2=1.2$ case at $T=20$ and $3.91$ for the $ag^2=0.8$ case at $T=5$. In general, the results in the $ag^2=1.2$ case are closer to $4$ than those in the $ag^2=0.8$ case. This can be understood from the fact that the $ag^2=1.2$ case is more quantum ergodic than the $ag^2=0.8$ case\footnote{The averaged restricted gap ratios are $0.5291$ and $0.5228$ in the $n_k=1$ momentum sector for $ag^2=1.2$ and $ag^2=0.8$, respectively~\cite{Ebner:2023ixq}, which are indicators of quantum ergodicity. For Gaussian Orthogonal Ensemble (GOE) of random matrices, the averaged restricted gap ratio is approximately $0.5307$. So the $ag^2=1.2$ case behaves more like the continuum SU(2) gauge theory in the quantum ergodicity perspective. In order to reach the complete quantum ergodicity in the $ag^2=0.8$ case, one has to use a bigger lattice and/or increase the $j_{\rm max}$ cutoff~\cite{Ebner:2023ixq}.}. An estimate shows that the minimum $j_{\rm max}$ required for physical results at energy $E$ scales as $\frac{E}{g^2\epsilon}$ for an accuracy $\epsilon$~\cite{turro2024shear}. This means that one has to increase $j_{\rm max}$ at smaller coupling and at higher temperature in order to achieve physical results, which explains the large deviation seen in the $ag^2=0.8$ case at $T=10$ and $T=20$. The result at $T=5$ in the $ag^2=1.2$ case is worse than the two results at higher temperatures. This is because the temperature $T=5$ is on the order of the first excitation energy ($E_1-E_0=4.1$) and thus not many states are active in the dynamics. In other words, the systems at such a low temperature are not quantum ergodic enough to exhibit complete thermalization behavior.

If the transport peak is diffusive, we would also have $\frac{a(n_k=2)}{a(n_k=1)}=1$. The ratios of the fitted $a$ parameters are listed in Table~\ref{tab:1} too. Deviations from unity are observed, especially in the $ag^2=0.8$ case at high temperatures, and are believed to have the same origin as explained above for $\frac{b(n_k=2)}{b(n_k=1)}$. 
It is expected that on bigger lattices with larger $j_{\rm max}$ these finite-size effects will disappear.

The energy diffusion transport coefficient can be extracted from the fitting via $D_\varepsilon = \frac{N^2}{4\pi^2}b(n_k=1)$. The results in lattice units are listed in Table~\ref{tab:1}. A mild increase with temperature can be seen, which is probably a finite-size effect, since one expects to increase $j_{\rm max}$ as the temperature increases. We emphasize that if the ratio $\frac{b(n_k=2)}{b(n_k=1)}$ obtained from the fitting results deviates significantly from 4, as in the $ag^2=0.8$ case, the diffusion coefficient extracted this way will not be accurate. 

\subsection{Long-time tail}

\begin{figure*}[t]
\centering
\subfloat[$ag^2=1.2,T=5$.\label{fig:1.2_T5}]{%
  \includegraphics[width=0.48\linewidth]{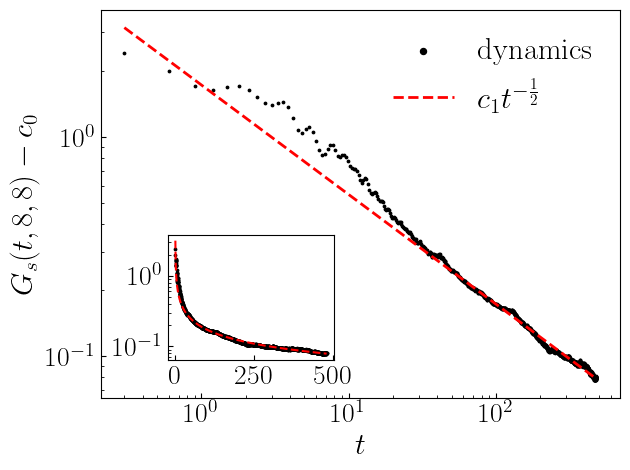}%
}\hfill
\subfloat[$ag^2=1.2,T=20$.\label{fig:1.2_T20}]{%
  \includegraphics[width=0.48\linewidth]{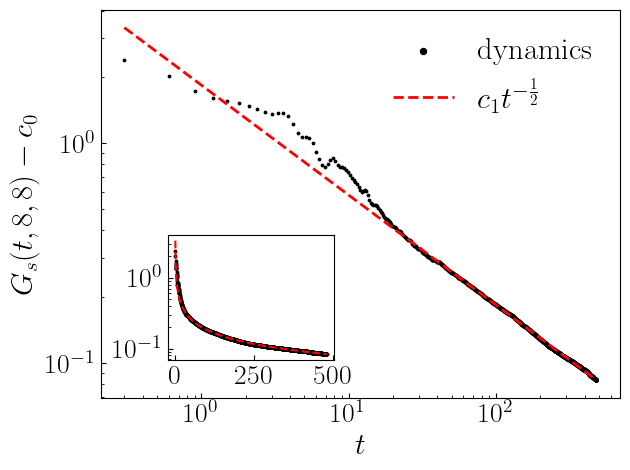}%
}

\subfloat[$ag^2=0.8,T=5$.\label{fig:0.8_T5}]{%
  \includegraphics[width=0.48\linewidth]{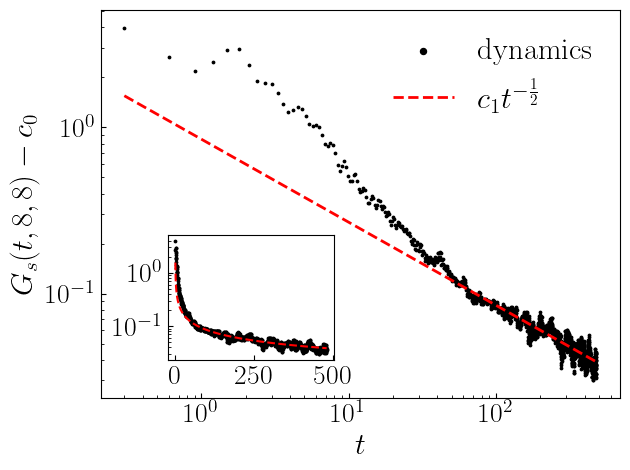}%
}\hfill
\subfloat[$ag^2=0.8,T=20$.\label{fig:0.8_T20}]{%
  \includegraphics[width=0.48\linewidth]{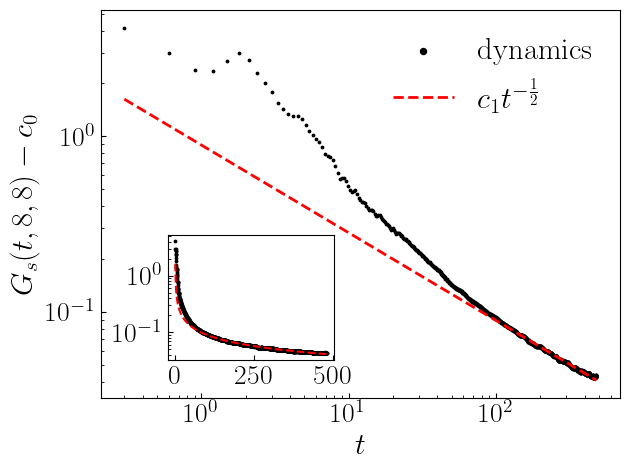}%
}
\caption{Real-time symmetric correlation function $G_s(t,j,i)$ at the same site $i=j=8$ on an open $N=17$ plaquette chain for two different couplings and two different temperatures. Red dashed lines are fits of the functional form $c_0+c_1t^{-\frac{1}{2}}$ with the fitted parameters shown in Table~\ref{tab:2}.}
\label{fig:Gs_t88}
\end{figure*}

Next we study the long-time tail predicted by the diffusive hydrodynamics as indicated in Eq.~\eqref{eqn:diffu_sol}. To this end, we calculate the symmetric correlator at the same position on an open plaquette chain with length $N=17$. The position of the Hamiltonian density operator is chosen to be the center of the lattice, i.e., $i=j=8$ in $G_s(t,j,i)$. The results are shown as black points in Fig.~\ref{fig:Gs_t88}, together with a fit of the form
\begin{align}
\label{eqn:power_tail}
c_0+c_1t^{-\frac{1}{2}} \,,
\end{align}
which is labeled by the red dashed line. The fitting is performed by using the results at $t>100$. The fitted parameter values are listed in Table~\ref{tab:2}. At the two couplings and two temperatures studied, $G_s(t,8,8)$ exhibits the power-law decay at late time, which is consistent with the long-time tail of diffusive hydrodynamics. At the lower temperature $T=5$, some oscillating behavior is seen in $G_s(t,8,8)$, particularly in the $ag^2=0.8$ case. This is caused by the fact that at such a low temperature, not many states are active and thus the system is less quantum ergodic, as explained in the previous subsection. As a result, the time evolution phases of different active states are not canceled completely and the oscillation is still manifest.

In Fig.~\ref{fig:Gs_t88}, we note that the hydrodynamic behavior $t^{-\frac{1}{2}}$ occurs earlier at higher temperature and larger coupling, qualitatively consistent with the scaling $\frac{1}{g^4T}$ estimated in Ref.~\cite{Arnold:1997gh}. More quantitative comparison requires a precise definition of the onset of hydrodynamics, which we will not pursue here.

One may wonder if the functional form of the transport peak~\eqref{eqn:fit_mode} we found in the numerical results can give the same long-time tail predicted by the standard diffusive peak of the form~\eqref{eqn:lorentzian}. The answer is yes and we give a proof here. First, we note
\begin{align}
I(t,k)&\equiv \int_{-\infty}^{+\infty} \frac{{\rm d} \omega}{2\pi} \frac{e^{-i \omega t}}{|\omega| + D_\varepsilon k^2} \nn\\
&= 2{\rm Re}\int_0^{+\infty} \frac{{\rm d} \omega}{2\pi} \frac{e^{i \omega t}}{\omega + D_\varepsilon k^2} \,.
\end{align}
The integrand $\frac{e^{i \omega t}}{\omega + D_\varepsilon k^2}$ has no pole in the first quadrant. Closing a contour along the positive $x$ and $y$ axes in the first quadrant, we find that the Cauchy's integral theorem under the change of variable $\omega=iD_\varepsilon k^2 y$ leads to
\begin{align}
\label{eqn:Itk}
I(t,k) = \int_0^{+\infty}\frac{{\rm d}y}{2\pi} \frac{2y \,e^{-D_\varepsilon k^2 ty}}{y^2+1} \,.
\end{align}
Finally performing the momentum integral with the phase $e^{ikx}=1$ for $x=0$ we find
\begin{align}
\int_{-\infty}^{+\infty}\frac{{\rm d}k}{2\pi} I(t,k) = \frac{1}{\sqrt{8\pi D_\varepsilon t}} \,.
\end{align}
In a nutshell, the functional form of the transport peak we found numerically gives the same power-law long-time tail as the standard diffusive mode.  

\begin{table}[h]
\centering
\begin{tabular}{ |c|c|c|c|c| } 
 \hline
 ~$N$~ & ~$ag^2$~ & ~$T$~ & ~$c_0$~ & ~$c_1$~ \\ 
 \hline
 17 & 1.2 & 5 & ~0.390~ & ~1.73~ \\
 \hline
 17 & 1.2 & 20 & 0.149 & 1.83 \\
 \hline
 17 & 0.8 & 5 & 1.02 & ~0.850~ \\
 \hline
 17 & 0.8 & 20 & 0.337 & ~0.891~ \\
 \hline
\end{tabular}
\caption{Parameter values of Eq.~\eqref{eqn:power_tail} fitted from the long-time tails at $t>100$ in Fig.~\ref{fig:Gs_t88} for two couplings and two temperatures. Numbers are rounded to three significant figures.}
\label{tab:2}
\end{table}

\subsection{Spatial diffusion}

\begin{figure*}
\centering
\subfloat[Fit using Eq.~\eqref{eqn:fresnel_fit}.\label{fig:Gs_x_continuum}]{%
  \includegraphics[width=0.48\linewidth]{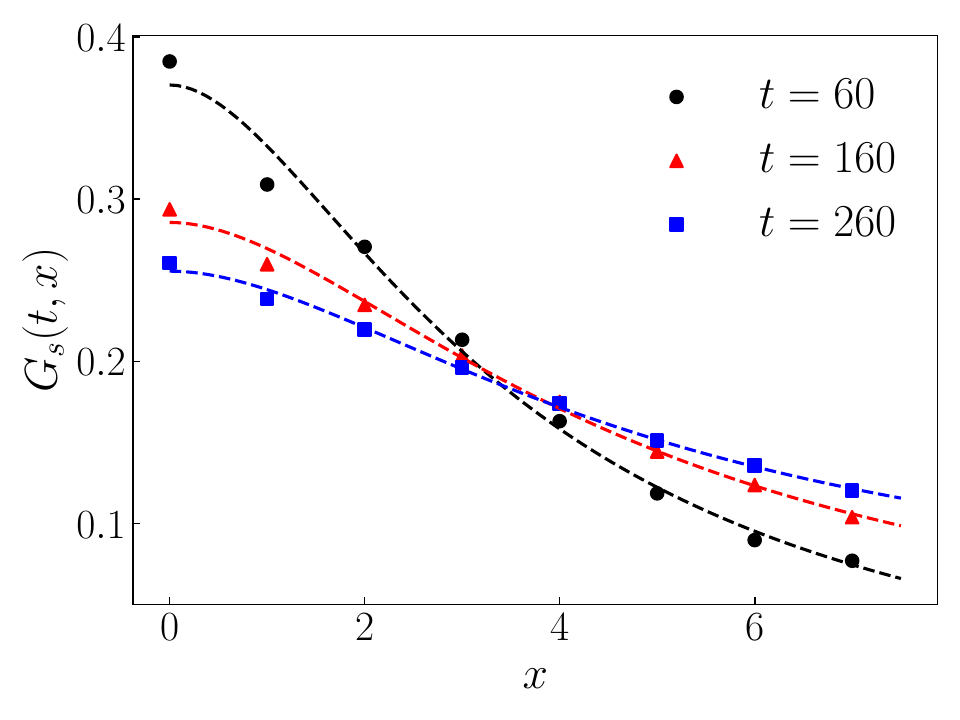}%
}\hfill
\subfloat[Fit using Eq.~\eqref{eqn:finite_sum_k}.\label{fig:Gs_x_discrete}]{%
  \includegraphics[width=0.48\linewidth]{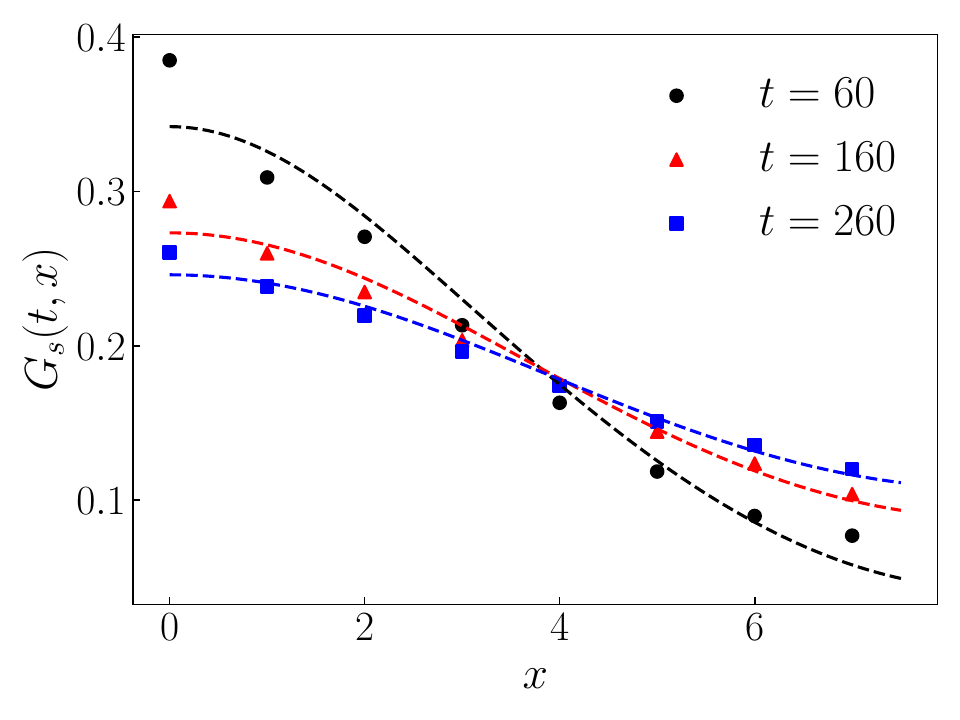}%
}
\caption{Symmetric correlators $G_s(t,x)$ as functions of $x$ on an open $N=17$ plaquette chain with $ag^2=1.2$ and $j_{\rm max}=\frac{1}{2}$ at three different times. Dashed lines are fits of the forms of (a) Eq.~\eqref{eqn:fresnel_fit} and (b) Eq.~\eqref{eqn:finite_sum_k}. The fitted parameter values are listed in Tables~\ref{tab:3} and~\ref{tab:4}, respectively. }
\label{fig:Gs_x}
\end{figure*}
Finally we investigate the real-time evolution of symmetric correlator in position space. Fourier transforming $I(t,k)$ back to position space we find
\begin{align}
\label{eqn:fresnel}
\int_{-\infty}^{+\infty}\frac{{\rm d}k}{2\pi} e^{ikx} I(t,k) =&\ \frac{1}{\sqrt{8\pi D_\varepsilon t}} \bigg\{ \cos(z) \Big[1-2S\Big(\sqrt{\frac{2z}{\pi}}\Big) \Big] \nn\\
& - \sin(z)\Big[ 1-2C\Big(\sqrt{\frac{2z}{\pi}}\Big) \Big] \bigg\}\,,
\end{align}
where $z=\frac{x^2}{4D_\varepsilon t}$, and $S$ and $C$ denote the Fresnel integrals. Qualitatively, Eq.~\eqref{eqn:fresnel} behaves similarly to the diffusion equation~\eqref{eqn:diffu_sol}.

We calculate the symmetric correlator $G_s(t,j,i)$ at different locations $j$ with $i=8$ fixed on an open $N=17$ plaquette chain at $ag^2=1.2$ and $j_{\rm max}=\frac{1}{2}$ and plot the results of $G_s(t,x=i-j)$ at $T=20$ in Fig.~\ref{fig:Gs_x_continuum}. The numerical results are marked as points at three different times. The dashed lines are fits of the functional form
\begin{align}
\label{eqn:fresnel_fit}
c_0' + c_1'\bigg\{ \cos(c_2'x^2)\Big[1-2S\Big(\sqrt{\frac{2c_2'x^2}{\pi}}\Big) \Big] \nn\\
- \sin(c_2'x^2)\Big[1-2C\Big(\sqrt{\frac{2c_2'x^2}{\pi}}\Big) \Big]
\bigg\} \,,
\end{align}
which is motivated from Eq.~\eqref{eqn:fresnel}. The fitted parameter values are listed in Table~\ref{tab:3}. 

\begin{table}[t]
\centering
\begin{tabular}{ |c|c|c|c|c|c|c|c|c| } 
 \hline
 ~$N$~ & $ag^2$ & ~$T$~ & ~$t$~ & ~$c_0'$~ & ~$c_1'$~ & ~$c_2'$~ & ~$\sqrt{t}c_1'$~ & ~$\frac{1}{4tc_2'}$~ \\ [0.11cm]
 \hline
 17 & 1.2 & 20 & 60 & $-0.0597$ & 0.430 & 0.129 & 3.33 & ~0.0323~ \\
 \hline
 17 & 1.2 & 20 & ~160~ & $-0.0196$ & 0.305 & ~0.0702~ & 3.86 & ~0.0223~ \\
 \hline
 17 & 1.2 & 20 & ~260~ & ~0.0204~ & ~0.235~ & ~0.0634~ & 3.79 & ~0.0152~ \\
 \hline
\end{tabular}
\caption{Parameter values of Eq.~\eqref{eqn:fresnel_fit} fitted from the $G_s(t,x)$ results in Fig.~\ref{fig:Gs_x_continuum}. Numbers are rounded to three significant figures.}
\label{tab:3}
\end{table}

\begin{table}[t]
\centering
\begin{tabular}{ |c|c|c|c|c|c| } 
 \hline
 ~$N$~ & $ag^2$ & ~$T$~ & ~$t$~ & ~$c_0''$~ & ~$c_1''$~ \\
 \hline
 17 & 1.2 & 20 & 60 & ~$-2.76$~ & ~22.7~  \\
 \hline
 17 & 1.2 & 20 & ~160~ & ~$-8.30$~ & ~65.6~ \\
 \hline
 17 & 1.2 & 20 & ~260~ & ~$-15.6$~ & ~121~ \\
 \hline
\end{tabular}
\caption{Parameter values of Eq.~\eqref{eqn:finite_sum_k} fitted from the $G_s(t,x)$ results in Fig.~\ref{fig:Gs_x_discrete}. Numbers are rounded to three significant figures.}
\label{tab:4}
\end{table}

The physical meaning of $c_1'$ is $c_1'\propto t^{-\frac{1}{2}}$ so we expect $\sqrt{t}c_1'$ to be approximately independent of $t$ in the fit. This is approximately true as indicated in Table~\ref{tab:3}. The physical meaning of $c_2'$ is $c_2' = \frac{1}{4D_\varepsilon t}$ so we expect to be able to extract $D_\varepsilon = \frac{1}{4tc_2'}$ from the fitted $c_2'$. However, the values of $\frac{1}{4tc_2'}$ as listed in Table~\ref{tab:3} clearly show $t$ dependence. Furthermore, their magnitudes are one-order-of-magnitude smaller than the $D_\varepsilon$ values extracted from the transport peak in $G_s(\omega, k)$ as shown in Table~\ref{tab:1}. These two points indicate that using Eq.~\eqref{eqn:fresnel_fit} to fit $G_x(t,x)$ is not a suitable approach of extracting the energy diffusion coefficient in practice. The reason for the failure is that Eq.~\eqref{eqn:fresnel_fit} is obtained from the integral over $k$ as in Eq.~\eqref{eqn:fresnel}, which can differ significantly from the summation over $k=\frac{2\pi n_k}{N}$ on a small lattice. To demonstrate this claim, we fit $G_s(t,x)$ using the finitely summed version
\begin{align}
\label{eqn:finite_sum_k}
c_0'' + c_1''\frac{1}{N} \sum_{n_k=-\lfloor N/2 \rfloor}^{\lfloor (N-1)/2\rfloor} e^{ikx}I(t,k) \,,
\end{align}
where $k=\frac{2\pi n_k}{N}$ and $I(t,k)$ is given in Eq.~\eqref{eqn:Itk} with the diffusion coefficient $D_\varepsilon=0.213$ as shown in Table~\ref{tab:1} for $ag^2=1.2$ and $T=20$. The fitted results are shown in Fig.~\ref{fig:Gs_x_discrete} and Table~\ref{tab:4}. We find the fitting is of similar quality except for the two end points at $t=60$. This may indicate that at such an early time, the dynamics cannot be fully described by the transport peak we found. It also shows that the diffusion coefficient extracted from Fig.~\ref{fig:Gs_wk} can indeed describe $G_s(t,x)$ at different locations at late time.

\section{Quantum Computing}
\label{sec5}
\subsection{Quantum algorithm}
\label{QA}
Motivated by the recent rapid developments of using the Hamiltonian lattice formulation, classical and quantum computing to study dynamics of gauge theory~\cite{banuls2020simulating,klco2022standard,Bauer_2023,beck2023quantum,bauer2023quantum,Lamm:2019bik,Raychowdhury:2019iki,Ciavarella:2021nmj,DeJong:2020riy,Rajput:2021trn,deJong:2021wsd,Kadam:2022ipf,Farrell:2022wyt,Farrell:2022vyh,Honda:2022edn,Cataldi:2023xki,Halimeh:2023lid,Liu:2023lsr,Hayata:2023pkw,Farrell:2023fgd,Farrell:2024fit,Watson:2023oov,Kavaki:2024ijd,Ciavarella:2024fzw,Illa:2024kmf,Farrell:2024mgu,Lamm:2024jnl,Li:2024lrl,Florio:2024aix,Kadam:2024zkj,Gustafson:2024bww,Fontana:2024rux,Lee:2024jnt,Burbano:2024uvn,Ciavarella:2024lsp,Araz:2024kkg,Zhang:2024fgv,Dhaulakhandi:2024tox,Zemlevskiy:2024vxt,Mueller:2024mmk,Davoudi:2024osg,Jeyaretnam:2024tkj,Araz:2024bgg,Itou:2024psm,Lin:2024eiz,Ballini:2024qmr,Ciavarella:2025zqf,Janik:2025bbz,Chandrasekharan:2025smw}, we propose a quantum algorithm to compute the real-time symmetric correlator. 
The algorithm is based on implementing the following four steps:
\begin{enumerate}
\item Thermal state preparation. The result of this step is to have the thermal density matrix, $\rho_T=\frac{1}{Z}e^{-\beta H}$ on the system qubits. We use the Quantum Imaginary Time Propagation (QITP) method~\cite{turro_QITP_2022,franceschino_thermal2023,turro2024shear,Motta:2019yya}. Details in the implementation for the SU(2) lattice gauge theory can be found in Ref.~\cite{turro2024shear}.

\item Applying the real-exponential of the first Hamiltonian density operator $H_i$ with a small parameter $\Gamma$, $e^{-\Gamma H_i}$ on the system qubits. This can be implemented with the QITP algorithm or the block-encoding~\cite{Low2019hamiltonian}. When using the QITP method, we add an extra ancilla qubit and apply the following unitary operator that couples the ancilla qubit and the system qubits:
\begin{align}
\qquad 
\begin{pmatrix}
    e^{-\Gamma (H_i-E_T)} & -\sqrt{1-e^{-2\Gamma (H_i-E_T)}}\\
    \sqrt{1-e^{-2\Gamma (H_i-E_T)}} & e^{-\Gamma (H_i-E_T)}
\end{pmatrix} \,,
\end{align}
where $E_T$ is a constant and can be chosen to be the ground state energy of $H_i$, which is easy to obtain since $H_i$ is local and increases the success probability of the QITP algorithm. The non-unitary operator $e^{-\Gamma(H_i-E_T)}$ is successfully applied on the system qubits if the ancilla state is measured in the $\ket{0}$ state. We have applied this method in similar scenarios in our previous work~\cite{Lee:2024jnt}, where more details can be found.
\item Implementing the real-time evolution. Generally, this is done by the Trotter decomposition of the Hamiltonian. Trotterization is well known and details for the case of SU(2) lattice gauge theory can be found in Ref.~\cite{turro2024shear}.
\item Measuring the expectation value of the second Hamiltonian density operator $H_j$. The expectation value can be obtained in two different ways. One can decompose the Hamiltonian density operator $H_j$ in terms of Pauli strings and then measure the expectation value of each Pauli string and then sum over the individual results. Alternatively, one can implement a unitary operator that diagonalizes the operator $H_j$, which is efficient in our case since the Hamiltonian density operator $H_j$ is local and only acts on a small number of qubits. By doing so, we change the basis from the spin computational basis to the eigenbasis of $H_j$. Measuring probabilities in this eigenbasis, we obtain the expectation value of the Hamiltonian density operator by summing over the products of the measured probabilities and the corresponding eigenvalues of $H_j$. In this work, $H_j$ only acts on three qubits, so we use the second method.
\end{enumerate}

The symmetric correlator ${\rm Tr} [\{H_j(t),H_i(0)\} \rho_T ]$ can be computed by taking the difference between the measurement results of two quantum circuits. One quantum circuit implements the second step with a small value of $\Gamma$, while the other skips the second step (or implements the $\Gamma=0$ case). Let $p_n(\Gamma)$ denote the probability of measuring the eigenbasis state $\ket{n(j)}$ of $H_j$ in the circuit with $\Gamma$. The eigenstates of $H_j$ are specified by $H_j\ket{n(j)} = E_n(j)\ket{n(j)}$. We can show the symmetric correlator is given by
\begin{align}   
\label{eq:hydro_qc}
G_s(t,j,i) =& \sum_n E_n(j) \left[\frac{ p_n(0) - p_n(\Gamma)}{\Gamma}+2 E_T p_n(0)\right] \nn\\
& + O(\Gamma) \,,
\end{align}
where $p_n(\Gamma)$ depends on $j$ and $i$ implicitly.
We give a proof of this formula in Appendix~\ref{app:3}.

\subsection{Efficiency of the algorithm}
The total qubit cost of the algorithm is given by $2N+5$ with $N$ the number of qubits that encode the physical system of the plaquette chain. Additional $N+4$ qubits are needed for the thermal state preparation, where we use $N$ qubits for preparing the thermal state at infinite temperature, $\rho_{\infty}=\frac{1}{2^{N}}$. Then, we cool the system down to the desired temperature with the Trotterized thermal propagator $e^{-\frac{\beta}{2} H }$, where we decompose the full propagator into products of three magnetic propagators and one electric one, 
\begin{equation}
e^{-\frac{\beta}{2} H }= e^{-\frac{\beta}{2} H^{\rm el} } e^{-\frac{\beta}{2} H^{\rm mag}_{{\rm com}0} } e^{-\frac{\beta}{2} H^{\rm mag}_{{\rm com}1} } e^{-\frac{\beta}{2} H^{\rm mag}_{{\rm com}2} }\,,
\end{equation}
where the magnetic Hamiltonian $H^{\rm mag}_{{\rm com}\,n}$ is obtained by summing a maximal set of commuting magnetic Hamiltonians. Since the magnetic Hamiltonian density operator is a three-body term, we have three such sets, denoted by the index $n=1,2,3$. Mathematically, we have
\begin{align}
H^{\rm mag}_{{\rm com}\,n} = \sum_{i=0}^{3i+n<N} H^{\rm mag}_{3i+n} \,,
\end{align}
where the summation terminates if $3i+n\geq N$. Each Trotterized thermal propagator $e^{-\frac{\beta}{2}H_{\alpha}}$ where $H_\alpha=\{ H^{\rm el}, H^{\rm mag}_{\rm com0}, H^{\rm mag}_{\rm com1}, H^{\rm mag}_{\rm com2} \}$ requires one ancilla qubit to implement in the QITP, resulting in four ancilla qubits. Finally, one more qubit is used to implement the $H_i$ operator as discussed in the second step.

The success probability of the QITP algorithm for thermal state preparation is known to decrease exponentially with the system volume. However, the prefactor of the volume in the exponential is a very small number at high temperature, indicating that the QITP algorithm can be used for high-temperature and small-volume calculations such as our test of the algorithm discussed later. More efficient quantum algorithms exist such as those based on the Lindblad equation~\cite{Chen:2023cuc,Chen:2023zpu,Ding:2024mxo,Chen:2024btm,Brunner:2024ejl}. For lattice gauge theory applications, Ref.~\cite{Lee:2023urk} showed that the time required for an initial state to reach approximate thermal state at high temperature in the Lindbladian evolution scales polynomially with the system volume in the Schwinger model case, with a power between 1 and 2.

The resources required in the QITP algorithm for implementing $e^{-\Gamma H_i}$ do not scale with the system volume since $H_i$ is the Hamiltonian density operator and thus local. Furthermore, we want $\Gamma$ to be small for the validity of Eq.~\eqref{eq:hydro_qc}, which corresponds to high temperature, justifying the QITP algorithm. Because of the same reason of locality, finding the unitary transformation that diagonalizes $H_j$ in the final measurement does not scale with the volume either. On the other hand, we expect resources for both steps to scale up with the cutoff $j_{\rm max}$. Since the size of the local Hilbert space grows polynomially with $j_{\rm max}$ roughly as $2j^2_{\rm max}$, and the minimum of $j_{\rm max}$ is inversely proportional to the accuracy of the calculation and the coupling~\cite{turro2024shear}, we expect that the resources for the implementation of $e^{-\Gamma H_i}$ and the diagonalization of $H_j$ will only scale polynomially with the final accuracy.

The resources (mainly the number of shots for a given accuracy) required in linearly combining the measurement results of the two quantum circuits in the fourth step only induce a overall constant factor of two due to the standard propagation of measurement uncertainties.

\subsection{Emulator results}
\begin{figure*}[t]
\centering
\subfloat[$i=j=4$.\label{fig:9_44}]{%
  \includegraphics[width=0.47\linewidth]{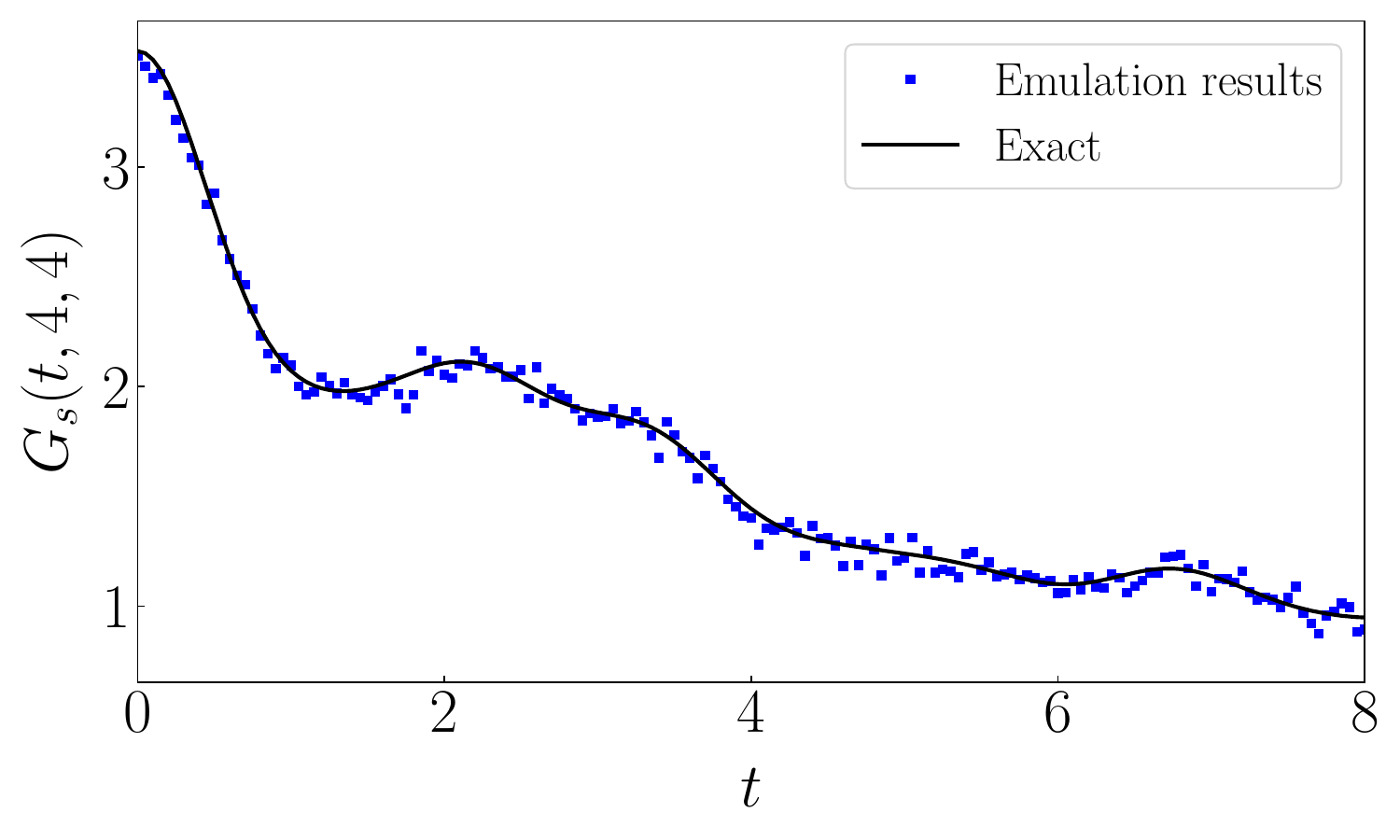}%
}\hfill
\subfloat[$i=4,j=1$.\label{fig:eqn:9_14}]{%
  \includegraphics[width=0.47\linewidth]{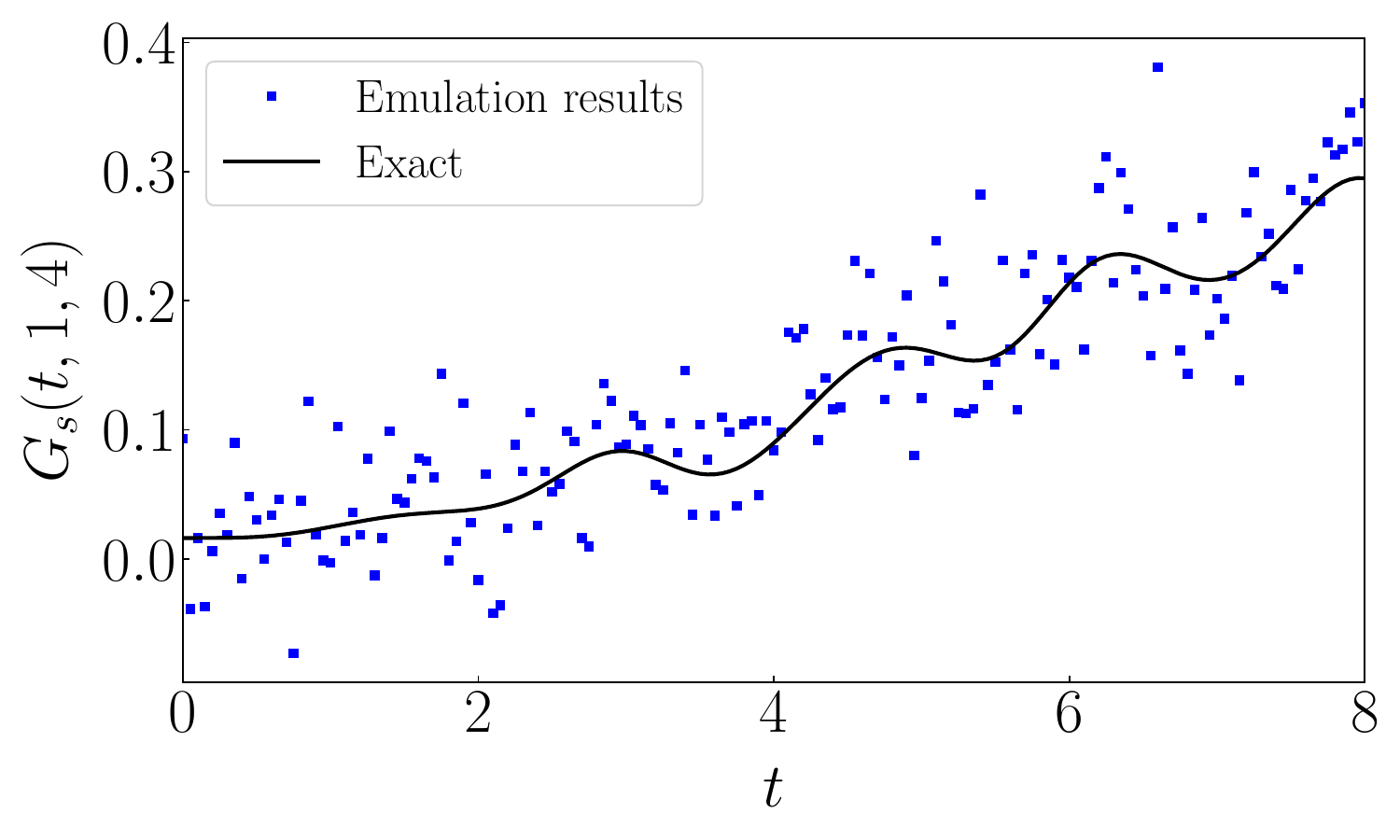}
}
\caption{IBM emulator results (blue squares) of $G_s(t,j,i)$ for (a) $i=j=4$ and (b) $i=4,j=1$ on an open $N=9$ plaquette chain with $ag^2=1$ and $j_{\rm max}=\frac{1}{2}$ at the inverse temperature $\beta=0.05$. We use $2\times10^6$ shots in the emulation. Black solid line represents the exact evolution.}
\label{fig:qc_results_9}
\end{figure*}

We test the described quantum algorithm for an open plaquette chain of size $N=9$ with $ag^2=1$ and $j_{\rm max}=\frac{1}{2}$.  Figure~\ref{fig:qc_results_9} shows the obtained results of the symmetric correlators $G_s(t,j,i)$ for $i=j=4$ and $i=4,j=1$ at $\beta=0.05$, using a Trotterization time step of $\Delta t=0.05$ and $\Gamma=0.005$. The number of shots used in each case is $2\times10^6$. Black solid line represents the exact evolution while the blue points represent the quantum circuit results obtained from the IBM emulator. We observe that the emulator results follow the exact evolution, up to statistical uncertainties. With the same number of shots, the uncertainty in the $i=4,j=1$ case looks relatively larger than the $i=j=4$ case, since the value of $G_s(t,1,4)$ is much smaller at early time. The uncertainty can be further reduced by using more shots. If we decrease the temperature $T$ in our calculations, more shots are also needed for the same accuracy because of the decreased success probability of thermal state preparation, as explained in the previous subsection.

\section{Conclusions}
\label{sec6}
In this paper, we scrutinize emergent hydrodynamic behavior in real-time Hamiltonian dynamics of $2+1$-dimensional SU(2) lattice gauge theory on a plaquette chain, by computing symmetric correlation functions of stress-energy tensors. Because of the Umklapp processes on the lattice, momentum density is no longer an effective degree of freedom at high temperature and thus the hydrodynamics becomes energy diffusion. We found a transport peak near zero frequency in the symmetric correlator of energy densities. The width of the transport peak is approximately quadratically proportional to momentum at small momentum when the lattice system is fully quantum ergodic. The transport peak leads to a power-law decay of the symmetric correlator at late time and can also explain the position dependence.

We also introduced a quantum algorithm to compute the symmetric correlator and tested the algorithm on a small lattice classically. The results obtained from the quantum algorithm agree well with those obtained by classical exact diagonalization.

In future work, we plan to investigate when the momentum density will become an effective degree of freedom again and thus the sound modes will become manifest. A necessary condition is the suppression of the Umklapp processes. As explained in Sec.~\ref{sec:hydro_lattice}, the suppression happens when $T_{\rm phy} \ll \frac{\pi}{a}$. In physical units, we would like $T_{\rm phy}$ to be fixed. So in the continuum limit $a\to0$, this condition will be satisfied. In lattice units, we want $T=aT_{\rm phy}\ll \pi$. However, the temperatures we used in this work are all above $5$, violating this condition. This is why we did not observe the sound modes. We cannot further lower the temperature in our current studies with $j_{\rm max}=\frac{1}{2}$, since the energy gap between the first excited state and the ground state is $E_1-E_0=4.1$ and $5.1$ for $ag^2=1.2$ and $ag^2=0.8$, respectively. Further lowering the temperature will only probe dynamics near the ground state rather than highly excited states in the deconfined region. As $a\to0$, it is expected that $E_1-E_0\to0$ in lattice units such that the physical value $\frac{E_1-E_0}{a}$ is finite and fixed. Then one is able to study the deconfined dynamics at a temperature satisfying $E_1-E_0 \ll T \ll \pi$ in lattice units, with the Umklapp processes suppressed. For SU(2) pure gauge theory in $2+1$ dimensions, it is known that $ag^2\to0$ as $a\to0$~\cite{Romatschke:2019nmo}. The fact that as we decrease $ag^2$ from 1.2 to 0.8, the energy gap $E_1-E_0$ increases, indicates that the finite-size artifacts in our studies with $j_{\rm max}=\frac{1}{2}$ are large and we have to increase the $j_{\rm max}$ cutoff. Currently, exact diagonalization can be done for $j_{\rm max}=1$ on an open $N=7$ chain~\cite{Ebner:2024mee} or a periodic $N=9$ chain~\cite{Ebner:2024qtu}. However, these lattice sizes that allow exact diagonalization might be too small for the system to exhibit long-wavelength hydrodynamics. Other classical methods such as tensor networks or neutral networks should be explored, as well as quantum computing. It is unclear at the moment if these classical methods can lead to the expected physics. If not, demonstrating the sounds modes for SU($N_c$) non-Abelian lattice gauge theory could be ``quantum supremacy'' for high-energy nuclear physics. 

\begin{acknowledgements}
We thank Anton Andreev, Paul Romatschke and Larry Yaffe for discussions that inspired this study. This work 
is supported by the U.S. Department of Energy, Office of Science, Office of Nuclear Physics, InQubator for Quantum Simulation (IQuS) (https://iqus.uw.edu) under Award Number DOE (NP) Award DE-SC0020970 via the program on Quantum Horizons: QIS Research and Innovation for Nuclear Science. This research used resources of the National Energy Research Scientific Computing Center (NERSC), a Department of Energy Office of Science User Facility using NERSC award NP-ERCAP0032083. This work was enabled, in part, by the use of advanced computational, storage and networking infrastructure provided by the Hyak supercomputer system at the University of Washington.
\end{acknowledgements}

\appendix

\section{Time to reach boundary}
\label{app:1}
\begin{figure}
\centering
\includegraphics[width=0.9\linewidth]{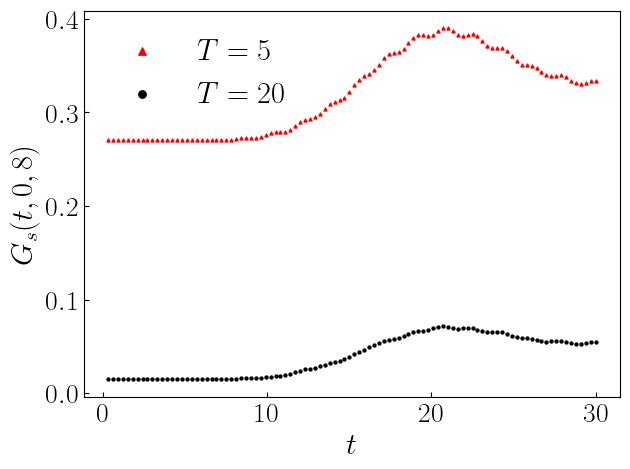}
\caption{Real-time symmetric correlation functions of energy densities at two different sites $j=0$ and $i=8$ on an open $N=17$ plaquette chain with $ag^2=1.2$ and $j_{\rm max}=\frac{1}{2}$ for two different temperatures: $T=5$ (red) and $T=20$ (black).}
\label{fig:Gs_t08}
\end{figure}

To study the boundary effect, we estimate the time it takes for a physical signal to reach the boundary. In particular, we calculate the symmetric correlator $G_s(t,j,i)$ at two different positions $j=0$ and $i=8$ on an open $N=17$ plaquette chain with $ag^2=1.2$ and $j_{\rm max}=\frac{1}{2}$. Results at two different temperatures $T=5$ and $T=20$ are shown in Fig.~\ref{fig:Gs_t08}. We observe that the perturbation introduced at the center of the lattice $i=8$ at $t=0$ takes about $t=12.5$ to reach the boundary, at which point it bounces back on the open lattice. It is estimated that the perturbation bounced back from the boundary reaches the center of the lattice at roughly $t=25$.

\section{Symmetric correlator in microcanonical ensemble}
\label{app:2}
\begin{figure}
\centering
\includegraphics[width=0.9\linewidth]{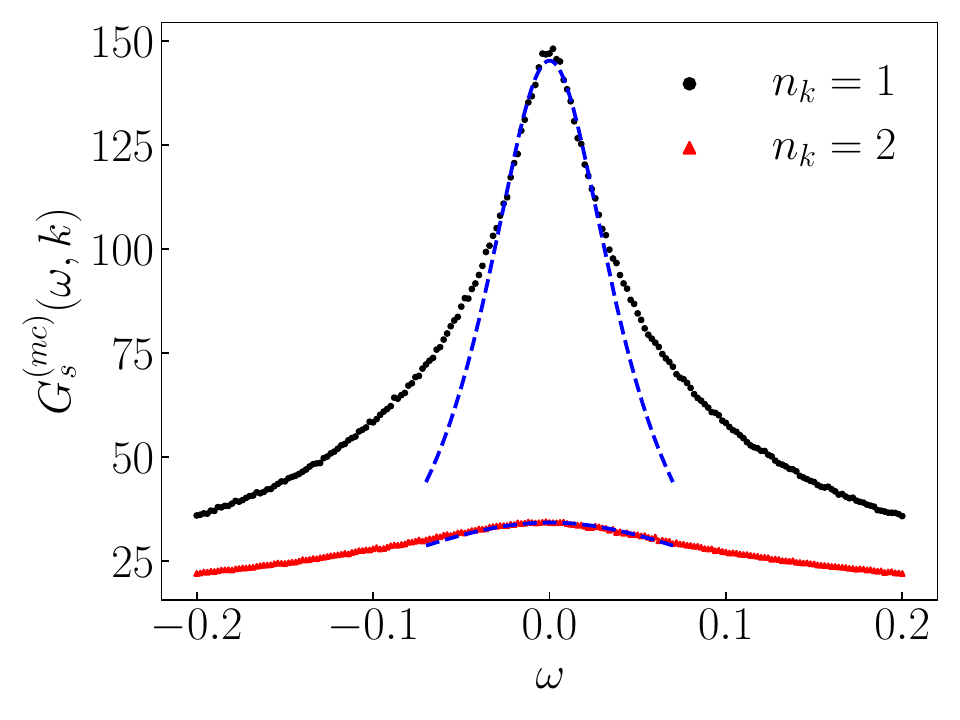}
\caption{Symmetric correlator of Hamiltonian densities in frequency-momentum space in the microcanonical ensemble for a $N=20$ periodic plaquette chain with $ag^2=0.8$ and $j_{\rm max}=\frac{1}{2}$ in the energy shell $[-5,-4]$. Blue dashed lines are fits of the Lorentzian shape with the fitted parameter values shown in Table~\ref{tab:5}.}
\label{fig:f_omega_mc}
\end{figure}

\begin{table}
\centering
\begin{tabular}{ |c|c|c|c|c|c| } 
 \hline
 ~$N$~ & ~$ag^2$~ & $a'(n_k=1)$ & $a'(n_k=2)$ & $b'(n_k=1)$ & $b'(n_k=2)$\\
 \hline
 20 & 0.8 & ~0.309~ & ~0.874~ & 0.0461 & 0.160 \\
 \hline
\end{tabular}
\caption{Parameter values of Eq.~\eqref{eqn:lorentz_fit} fitted from the $G_s^{(mc)}(\omega,k)$ results in Fig.~\ref{fig:f_omega_mc}. Numbers are rounded to three significant figures.}
\label{tab:5}
\end{table}

Here we explain the evaluation of the symmetric correlator in frequency-momentum space in the microcanonical ensemble. We specify a thin energy shell as $[E_*-\frac{\Delta E}{2}, E_*+\frac{\Delta E}{2}]$. Inserting complete sets of energy and momentum eigenstates in Eq.~\eqref{eqn:Gs}, we can show
\begin{align}
& G_s^{(mc)}(\omega, k) = 2\pi N\sum_p  \nn\\
&\quad \bigg[ \frac{1}{N_{E_n(p)}}\sum_{E_n(p)}\bigg|_{|E_*-E_n(p)|<\frac{\Delta E}{2}} \overline{f^2_{H_i}(E_n,p,\omega,k)} \nn\\
&+ \frac{1}{N_{E_n(p+k)}}\sum_{E_n(p+k)}\bigg|_{|E_*-E_n(p+k)|<\frac{\Delta E}{2}} \overline{g^2_{H_i}(E_n,p,\omega,k)} \bigg]\,,
\end{align}
where $N_{E_n(p)}$ and $N_{E_n(p+k)}$ are the total numbers of energy eigenstates with momentum $p$ and $p+k$, respectively, in the thin energy shell.
$\overline{f^2_{H_i}(E_n,p,\omega,k)}$ is defined in Eq.~\eqref{eqn:overline_f} and $\overline{g^2_{H_i}(E_n,p,\omega,k)}$ is defined similarly as
\begin{align}
\overline{g^2_{H_i}(E_n,p,\omega,k)} \equiv&\ \frac{1}{\Delta\omega}  \sum_{E_m(p)} \bigg|_{ |\omega + E_m(p)- E_n(p+k)| < \frac{\Delta\omega}{2}} \nn\\
&\qquad |\langle E_m(p) | H_i | E_n(p+k) \rangle |^2 \,.
\end{align}

Results of the symmetric correlator in the microcanonical ensemble are shown in Fig.~\ref{fig:f_omega_mc} for a $N=20$ periodic plaquette chain with $ag^2=0.8$ and $j_{\rm max}=\frac{1}{2}$ in the energy shell $[-5,-4]$, i.e., $E_*=-4.5$ and $\Delta E = 1$. We have chosen $\Delta \omega = 0.002$. The blue dashed lines are fits of the Lorentzian shape
\begin{align}
\label{eqn:lorentz_fit}
\frac{a'}{\omega^2 + b'^2} \,.
\end{align}
The physical meaning of $a'$ and $b'$ can be found by comparing with Eq.~\eqref{eqn:lorentzian}, which gives $a'\propto D_\varepsilon k^2$ and $b'=D_\varepsilon k^2$.
The fitted parameter values are listed in Table~\ref{tab:5}. The symmetric correlator in the microcanonical ensemble can be well described by the Lorentzian shape at small frequency. Diffusive hydrodynamics predicts that $\frac{b'(n_k=2)}{b'(n_k=1)}=4$ and our results give $\frac{b'(n_k=2)}{b'(n_k=1)}\approx 3.46$.

In another energy shell $[-3,-2]$ at the same coupling $ag^2=0.8$, or the same energy shell $[-5,-4]$ at a different coupling $ag^2=1.2$, we find the symmetric correlator in the microcanonical ensemble is better described by Eq.~\eqref{eqn:fit_mode} rather than the Lorentzian shape. We leave the investigation of when the Lorentzian shape becomes manifest to future work.

\section{Proof of Eq.~\eqref{eq:hydro_qc}}
\label{app:3}
Here we provide a proof of Eq.~\eqref{eq:hydro_qc}. After applying all the four steps of the quantum algorithm introduced in Sec.~\ref{QA}, the final density matrix is given by
\begin{align}
  \rho_{f}(\Gamma)=U_j e^{-iHt} e^{-\Gamma (H_i-E_T)} \rho_T e^{-\Gamma (H_i-E_T)}  e^{iHt} U_j^\dagger\,,
\end{align}
where $U_j$ denotes the unitary transformation that diagonalizes the Hamiltonian density operator $H_j$, i.e., $U_j H_j U_j^\dagger = \sum_n E_n(j)|n(j)\rangle \langle n(j)|$. The probability of measuring the final state to be $\ket{n(j)}$ is 
\begin{align}
p_n(\Gamma) = \langle n(j) | \rho_f(\Gamma) | n(j) \rangle \,.
\end{align}
Summation over all the measurement results leads to
\begin{align}
& \sum_n E_n(j) p_n(\Gamma) = {\rm Tr} \Big[ \sum_n E_n(j)  \langle n(j)| \rho_f(\Gamma) | n(j) \rangle\Big] \nn\\
& = {\rm Tr} \Big[ \sum_n E_n(j) | n(j) \rangle \langle n(j)| \rho_f(\Gamma) \Big] \nn\\
&= {\rm Tr} \big[ H_j  e^{-iHt} e^{-\Gamma (H_i-E_T)} \rho_T e^{-\Gamma (H_i-E_T)}  e^{iHt} \big] \nn\\
& = {\rm Tr} \big[ H_j(t) e^{-\Gamma (H_i-E_T)} \rho_T e^{-\Gamma (H_i-E_T)} \big] \,.
\end{align}
Some algebra finally leads to 
\begin{align}
& \sum_n E_n(j) [p_n(0) - p_n(\Gamma)] \nn\\
& = \Gamma {\rm Tr}[ \{H_j(t), H_i \} \rho_T ] - 2\Gamma E_T{\rm Tr}[H_j(t)\rho_T] + O(\Gamma^2) \,.
\end{align}
Dividing by $\Gamma$ gives Eq.~\eqref{eq:hydro_qc} in the main text.

\bibliography{main.bib}

\begin{thebibliography}{114}%
\makeatletter
\providecommand \@ifxundefined [1]{%
 \@ifx{#1\undefined}
}%
\providecommand \@ifnum [1]{%
 \ifnum #1\expandafter \@firstoftwo
 \else \expandafter \@secondoftwo
 \fi
}%
\providecommand \@ifx [1]{%
 \ifx #1\expandafter \@firstoftwo
 \else \expandafter \@secondoftwo
 \fi
}%
\providecommand \natexlab [1]{#1}%
\providecommand \enquote  [1]{``#1''}%
\providecommand \bibnamefont  [1]{#1}%
\providecommand \bibfnamefont [1]{#1}%
\providecommand \citenamefont [1]{#1}%
\providecommand \href@noop [0]{\@secondoftwo}%
\providecommand \href [0]{\begingroup \@sanitize@url \@href}%
\providecommand \@href[1]{\@@startlink{#1}\@@href}%
\providecommand \@@href[1]{\endgroup#1\@@endlink}%
\providecommand \@sanitize@url [0]{\catcode `\\12\catcode `\$12\catcode
  `\&12\catcode `\#12\catcode `\^12\catcode `\_12\catcode `\%12\relax}%
\providecommand \@@startlink[1]{}%
\providecommand \@@endlink[0]{}%
\providecommand \url  [0]{\begingroup\@sanitize@url \@url }%
\providecommand \@url [1]{\endgroup\@href {#1}{\urlprefix }}%
\providecommand \urlprefix  [0]{URL }%
\providecommand \Eprint [0]{\href }%
\providecommand \doibase [0]{http://dx.doi.org/}%
\providecommand \selectlanguage [0]{\@gobble}%
\providecommand \bibinfo  [0]{\@secondoftwo}%
\providecommand \bibfield  [0]{\@secondoftwo}%
\providecommand \translation [1]{[#1]}%
\providecommand \BibitemOpen [0]{}%
\providecommand \bibitemStop [0]{}%
\providecommand \bibitemNoStop [0]{.\EOS\space}%
\providecommand \EOS [0]{\spacefactor3000\relax}%
\providecommand \BibitemShut  [1]{\csname bibitem#1\endcsname}%
\let\auto@bib@innerbib\@empty
\bibitem [{\citenamefont {Kulkarni}\ and\ \citenamefont
  {Abanov}(2012)}]{PhysRevA.86.033614}%
  \BibitemOpen
  \bibfield  {author} {\bibinfo {author} {\bibfnamefont {M.}~\bibnamefont
  {Kulkarni}}\ and\ \bibinfo {author} {\bibfnamefont {A.~G.}\ \bibnamefont
  {Abanov}},\ }\href {\doibase 10.1103/PhysRevA.86.033614} {\bibfield
  {journal} {\bibinfo  {journal} {Phys. Rev. A}\ }\textbf {\bibinfo {volume}
  {86}},\ \bibinfo {pages} {033614} (\bibinfo {year} {2012})}\BibitemShut
  {NoStop}%
\bibitem [{\citenamefont {Lux}\ \emph {et~al.}(2014)\citenamefont {Lux},
  \citenamefont {M\"uller}, \citenamefont {Mitra},\ and\ \citenamefont
  {Rosch}}]{PhysRevA.89.053608}%
  \BibitemOpen
  \bibfield  {author} {\bibinfo {author} {\bibfnamefont {J.}~\bibnamefont
  {Lux}}, \bibinfo {author} {\bibfnamefont {J.}~\bibnamefont {M\"uller}},
  \bibinfo {author} {\bibfnamefont {A.}~\bibnamefont {Mitra}}, \ and\ \bibinfo
  {author} {\bibfnamefont {A.}~\bibnamefont {Rosch}},\ }\href {\doibase
  10.1103/PhysRevA.89.053608} {\bibfield  {journal} {\bibinfo  {journal} {Phys.
  Rev. A}\ }\textbf {\bibinfo {volume} {89}},\ \bibinfo {pages} {053608}
  (\bibinfo {year} {2014})}\BibitemShut {NoStop}%
\bibitem [{\citenamefont {Brandstetter}\ \emph {et~al.}(2025)\citenamefont
  {Brandstetter} \emph {et~al.}}]{Brandstetter:2023jsy}%
  \BibitemOpen
  \bibfield  {author} {\bibinfo {author} {\bibfnamefont {S.}~\bibnamefont
  {Brandstetter}} \emph {et~al.},\ }\href {\doibase 10.1038/s41567-024-02705-8}
  {\bibfield  {journal} {\bibinfo  {journal} {Nature Phys.}\ }\textbf {\bibinfo
  {volume} {21}},\ \bibinfo {pages} {52} (\bibinfo {year} {2025})},\ \Eprint
  {http://arxiv.org/abs/2308.09699} {arXiv:2308.09699 [cond-mat.quant-gas]}
  \BibitemShut {NoStop}%
\bibitem [{\citenamefont {Cen}(1992)}]{Cen:1992zk}%
  \BibitemOpen
  \bibfield  {author} {\bibinfo {author} {\bibfnamefont {R.}~\bibnamefont
  {Cen}},\ }\href {\doibase 10.1086/191630} {\bibfield  {journal} {\bibinfo
  {journal} {Astrophys. J. Suppl.}\ }\textbf {\bibinfo {volume} {78}},\
  \bibinfo {pages} {341} (\bibinfo {year} {1992})}\BibitemShut {NoStop}%
\bibitem [{\citenamefont {Vogelsberger}\ \emph {et~al.}(2014)\citenamefont
  {Vogelsberger}, \citenamefont {Genel}, \citenamefont {Springel},
  \citenamefont {Torrey}, \citenamefont {Sijacki}, \citenamefont {Xu},
  \citenamefont {Snyder}, \citenamefont {Bird}, \citenamefont {Nelson},\ and\
  \citenamefont {Hernquist}}]{Vogelsberger:2014kha}%
  \BibitemOpen
  \bibfield  {author} {\bibinfo {author} {\bibfnamefont {M.}~\bibnamefont
  {Vogelsberger}}, \bibinfo {author} {\bibfnamefont {S.}~\bibnamefont {Genel}},
  \bibinfo {author} {\bibfnamefont {V.}~\bibnamefont {Springel}}, \bibinfo
  {author} {\bibfnamefont {P.}~\bibnamefont {Torrey}}, \bibinfo {author}
  {\bibfnamefont {D.}~\bibnamefont {Sijacki}}, \bibinfo {author} {\bibfnamefont
  {D.}~\bibnamefont {Xu}}, \bibinfo {author} {\bibfnamefont {G.~F.}\
  \bibnamefont {Snyder}}, \bibinfo {author} {\bibfnamefont {S.}~\bibnamefont
  {Bird}}, \bibinfo {author} {\bibfnamefont {D.}~\bibnamefont {Nelson}}, \ and\
  \bibinfo {author} {\bibfnamefont {L.}~\bibnamefont {Hernquist}},\ }\href
  {\doibase 10.1038/nature13316} {\bibfield  {journal} {\bibinfo  {journal}
  {Nature}\ }\textbf {\bibinfo {volume} {509}},\ \bibinfo {pages} {177}
  (\bibinfo {year} {2014})},\ \Eprint {http://arxiv.org/abs/1405.1418}
  {arXiv:1405.1418 [astro-ph.CO]} \BibitemShut {NoStop}%
\bibitem [{\citenamefont {Bea}\ \emph {et~al.}(2024)\citenamefont {Bea},
  \citenamefont {Casalderrey-Solana}, \citenamefont {Mateos},\ and\
  \citenamefont {Sanchez-Garitaonandia}}]{Bea:2024bxu}%
  \BibitemOpen
  \bibfield  {author} {\bibinfo {author} {\bibfnamefont {Y.}~\bibnamefont
  {Bea}}, \bibinfo {author} {\bibfnamefont {J.}~\bibnamefont
  {Casalderrey-Solana}}, \bibinfo {author} {\bibfnamefont {D.}~\bibnamefont
  {Mateos}}, \ and\ \bibinfo {author} {\bibfnamefont {M.}~\bibnamefont
  {Sanchez-Garitaonandia}},\ }\href@noop {} {\  (\bibinfo {year} {2024})},\
  \Eprint {http://arxiv.org/abs/2406.14450} {arXiv:2406.14450 [hep-th]}
  \BibitemShut {NoStop}%
\bibitem [{\citenamefont {Song}\ \emph {et~al.}(2011)\citenamefont {Song},
  \citenamefont {Bass}, \citenamefont {Heinz}, \citenamefont {Hirano},\ and\
  \citenamefont {Shen}}]{Song:2010mg}%
  \BibitemOpen
  \bibfield  {author} {\bibinfo {author} {\bibfnamefont {H.}~\bibnamefont
  {Song}}, \bibinfo {author} {\bibfnamefont {S.~A.}\ \bibnamefont {Bass}},
  \bibinfo {author} {\bibfnamefont {U.}~\bibnamefont {Heinz}}, \bibinfo
  {author} {\bibfnamefont {T.}~\bibnamefont {Hirano}}, \ and\ \bibinfo {author}
  {\bibfnamefont {C.}~\bibnamefont {Shen}},\ }\href {\doibase
  10.1103/PhysRevLett.106.192301} {\bibfield  {journal} {\bibinfo  {journal}
  {Phys. Rev. Lett.}\ }\textbf {\bibinfo {volume} {106}},\ \bibinfo {pages}
  {192301} (\bibinfo {year} {2011})},\ \bibinfo {note} {[Erratum:
  Phys.Rev.Lett. 109, 139904 (2012)]},\ \Eprint
  {http://arxiv.org/abs/1011.2783} {arXiv:1011.2783 [nucl-th]} \BibitemShut
  {NoStop}%
\bibitem [{\citenamefont {Schenke}\ \emph {et~al.}(2011)\citenamefont
  {Schenke}, \citenamefont {Jeon},\ and\ \citenamefont
  {Gale}}]{Schenke:2010rr}%
  \BibitemOpen
  \bibfield  {author} {\bibinfo {author} {\bibfnamefont {B.}~\bibnamefont
  {Schenke}}, \bibinfo {author} {\bibfnamefont {S.}~\bibnamefont {Jeon}}, \
  and\ \bibinfo {author} {\bibfnamefont {C.}~\bibnamefont {Gale}},\ }\href
  {\doibase 10.1103/PhysRevLett.106.042301} {\bibfield  {journal} {\bibinfo
  {journal} {Phys. Rev. Lett.}\ }\textbf {\bibinfo {volume} {106}},\ \bibinfo
  {pages} {042301} (\bibinfo {year} {2011})},\ \Eprint
  {http://arxiv.org/abs/1009.3244} {arXiv:1009.3244 [hep-ph]} \BibitemShut
  {NoStop}%
\bibitem [{\citenamefont {Bernhard}\ \emph {et~al.}(2019)\citenamefont
  {Bernhard}, \citenamefont {Moreland},\ and\ \citenamefont
  {Bass}}]{Bernhard:2019bmu}%
  \BibitemOpen
  \bibfield  {author} {\bibinfo {author} {\bibfnamefont {J.~E.}\ \bibnamefont
  {Bernhard}}, \bibinfo {author} {\bibfnamefont {J.~S.}\ \bibnamefont
  {Moreland}}, \ and\ \bibinfo {author} {\bibfnamefont {S.~A.}\ \bibnamefont
  {Bass}},\ }\href {\doibase 10.1038/s41567-019-0611-8} {\bibfield  {journal}
  {\bibinfo  {journal} {Nature Phys.}\ }\textbf {\bibinfo {volume} {15}},\
  \bibinfo {pages} {1113} (\bibinfo {year} {2019})}\BibitemShut {NoStop}%
\bibitem [{\citenamefont {Nijs}\ \emph {et~al.}(2021)\citenamefont {Nijs},
  \citenamefont {van~der Schee}, \citenamefont {G\"ursoy},\ and\ \citenamefont
  {Snellings}}]{Nijs:2020ors}%
  \BibitemOpen
  \bibfield  {author} {\bibinfo {author} {\bibfnamefont {G.}~\bibnamefont
  {Nijs}}, \bibinfo {author} {\bibfnamefont {W.}~\bibnamefont {van~der Schee}},
  \bibinfo {author} {\bibfnamefont {U.}~\bibnamefont {G\"ursoy}}, \ and\
  \bibinfo {author} {\bibfnamefont {R.}~\bibnamefont {Snellings}},\ }\href
  {\doibase 10.1103/PhysRevLett.126.202301} {\bibfield  {journal} {\bibinfo
  {journal} {Phys. Rev. Lett.}\ }\textbf {\bibinfo {volume} {126}},\ \bibinfo
  {pages} {202301} (\bibinfo {year} {2021})},\ \Eprint
  {http://arxiv.org/abs/2010.15130} {arXiv:2010.15130 [nucl-th]} \BibitemShut
  {NoStop}%
\bibitem [{\citenamefont {Policastro}\ \emph {et~al.}(2001)\citenamefont
  {Policastro}, \citenamefont {Son},\ and\ \citenamefont
  {Starinets}}]{Policastro:2001yc}%
  \BibitemOpen
  \bibfield  {author} {\bibinfo {author} {\bibfnamefont {G.}~\bibnamefont
  {Policastro}}, \bibinfo {author} {\bibfnamefont {D.~T.}\ \bibnamefont {Son}},
  \ and\ \bibinfo {author} {\bibfnamefont {A.~O.}\ \bibnamefont {Starinets}},\
  }\href {\doibase 10.1103/PhysRevLett.87.081601} {\bibfield  {journal}
  {\bibinfo  {journal} {Phys. Rev. Lett.}\ }\textbf {\bibinfo {volume} {87}},\
  \bibinfo {pages} {081601} (\bibinfo {year} {2001})},\ \Eprint
  {http://arxiv.org/abs/hep-th/0104066} {arXiv:hep-th/0104066} \BibitemShut
  {NoStop}%
\bibitem [{\citenamefont {Brigante}\ \emph {et~al.}(2008)\citenamefont
  {Brigante}, \citenamefont {Liu}, \citenamefont {Myers}, \citenamefont
  {Shenker},\ and\ \citenamefont {Yaida}}]{Brigante:2008gz}%
  \BibitemOpen
  \bibfield  {author} {\bibinfo {author} {\bibfnamefont {M.}~\bibnamefont
  {Brigante}}, \bibinfo {author} {\bibfnamefont {H.}~\bibnamefont {Liu}},
  \bibinfo {author} {\bibfnamefont {R.~C.}\ \bibnamefont {Myers}}, \bibinfo
  {author} {\bibfnamefont {S.}~\bibnamefont {Shenker}}, \ and\ \bibinfo
  {author} {\bibfnamefont {S.}~\bibnamefont {Yaida}},\ }\href {\doibase
  10.1103/PhysRevLett.100.191601} {\bibfield  {journal} {\bibinfo  {journal}
  {Phys. Rev. Lett.}\ }\textbf {\bibinfo {volume} {100}},\ \bibinfo {pages}
  {191601} (\bibinfo {year} {2008})},\ \Eprint {http://arxiv.org/abs/0802.3318}
  {arXiv:0802.3318 [hep-th]} \BibitemShut {NoStop}%
\bibitem [{\citenamefont {Moore}(2020)}]{Moore:2020pfu}%
  \BibitemOpen
  \bibfield  {author} {\bibinfo {author} {\bibfnamefont {G.~D.}\ \bibnamefont
  {Moore}},\ }in\ \href@noop {} {\emph {\bibinfo {booktitle} {{Criticality in
  QCD and the Hadron Resonance Gas}}}}\ (\bibinfo {year} {2020})\ \Eprint
  {http://arxiv.org/abs/2010.15704} {arXiv:2010.15704 [hep-ph]} \BibitemShut
  {NoStop}%
\bibitem [{\citenamefont {Jeon}(1995)}]{Jeon:1994if}%
  \BibitemOpen
  \bibfield  {author} {\bibinfo {author} {\bibfnamefont {S.}~\bibnamefont
  {Jeon}},\ }\href {\doibase 10.1103/PhysRevD.52.3591} {\bibfield  {journal}
  {\bibinfo  {journal} {Phys. Rev. D}\ }\textbf {\bibinfo {volume} {52}},\
  \bibinfo {pages} {3591} (\bibinfo {year} {1995})},\ \Eprint
  {http://arxiv.org/abs/hep-ph/9409250} {arXiv:hep-ph/9409250} \BibitemShut
  {NoStop}%
\bibitem [{\citenamefont {Arnold}\ \emph {et~al.}(2000)\citenamefont {Arnold},
  \citenamefont {Moore},\ and\ \citenamefont {Yaffe}}]{Arnold:2000dr}%
  \BibitemOpen
  \bibfield  {author} {\bibinfo {author} {\bibfnamefont {P.}~\bibnamefont
  {Arnold}}, \bibinfo {author} {\bibfnamefont {G.~D.}\ \bibnamefont {Moore}}, \
  and\ \bibinfo {author} {\bibfnamefont {L.~G.}\ \bibnamefont {Yaffe}},\ }\href
  {\doibase 10.1088/1126-6708/2000/11/001} {\bibfield  {journal} {\bibinfo
  {journal} {JHEP}\ }\textbf {\bibinfo {volume} {2000}},\ \bibinfo {pages}
  {001} (\bibinfo {year} {2000})},\ \Eprint
  {http://arxiv.org/abs/hep-ph/0010177} {arXiv:hep-ph/0010177} \BibitemShut
  {NoStop}%
\bibitem [{\citenamefont {Arnold}\ \emph {et~al.}(2003)\citenamefont {Arnold},
  \citenamefont {Moore},\ and\ \citenamefont {Yaffe}}]{Arnold:2003zc}%
  \BibitemOpen
  \bibfield  {author} {\bibinfo {author} {\bibfnamefont {P.}~\bibnamefont
  {Arnold}}, \bibinfo {author} {\bibfnamefont {G.~D.}\ \bibnamefont {Moore}}, \
  and\ \bibinfo {author} {\bibfnamefont {L.~G.}\ \bibnamefont {Yaffe}},\ }\href
  {\doibase 10.1088/1126-6708/2003/05/051} {\bibfield  {journal} {\bibinfo
  {journal} {JHEP}\ }\textbf {\bibinfo {volume} {2003}},\ \bibinfo {pages}
  {051} (\bibinfo {year} {2003})},\ \Eprint
  {http://arxiv.org/abs/hep-ph/0302165} {arXiv:hep-ph/0302165} \BibitemShut
  {NoStop}%
\bibitem [{\citenamefont {Ghiglieri}\ \emph {et~al.}(2018)\citenamefont
  {Ghiglieri}, \citenamefont {Moore},\ and\ \citenamefont
  {Teaney}}]{Ghiglieri:2018dib}%
  \BibitemOpen
  \bibfield  {author} {\bibinfo {author} {\bibfnamefont {J.}~\bibnamefont
  {Ghiglieri}}, \bibinfo {author} {\bibfnamefont {G.~D.}\ \bibnamefont
  {Moore}}, \ and\ \bibinfo {author} {\bibfnamefont {D.}~\bibnamefont
  {Teaney}},\ }\href {\doibase 10.1007/JHEP03(2018)179} {\bibfield  {journal}
  {\bibinfo  {journal} {JHEP}\ ,\ \bibinfo {pages} {1}} (\bibinfo {year}
  {2018})},\ \Eprint {http://arxiv.org/abs/1802.09535} {arXiv:1802.09535
  [hep-ph]} \BibitemShut {NoStop}%
\bibitem [{\citenamefont {Meyer}(2007)}]{Meyer:2007ic}%
  \BibitemOpen
  \bibfield  {author} {\bibinfo {author} {\bibfnamefont {H.~B.}\ \bibnamefont
  {Meyer}},\ }\href {\doibase 10.1103/PhysRevD.76.101701} {\bibfield  {journal}
  {\bibinfo  {journal} {Phys. Rev. D}\ }\textbf {\bibinfo {volume} {76}},\
  \bibinfo {pages} {101701} (\bibinfo {year} {2007})},\ \Eprint
  {http://arxiv.org/abs/0704.1801} {arXiv:0704.1801 [hep-lat]} \BibitemShut
  {NoStop}%
\bibitem [{\citenamefont {Mages}\ \emph {et~al.}(2015)\citenamefont {Mages},
  \citenamefont {Bors\'anyi}, \citenamefont {Fodor}, \citenamefont
  {Sch\"afer},\ and\ \citenamefont {Szab\'o}}]{Mages:2015rea}%
  \BibitemOpen
  \bibfield  {author} {\bibinfo {author} {\bibfnamefont {S.~W.}\ \bibnamefont
  {Mages}}, \bibinfo {author} {\bibfnamefont {S.}~\bibnamefont {Bors\'anyi}},
  \bibinfo {author} {\bibfnamefont {Z.}~\bibnamefont {Fodor}}, \bibinfo
  {author} {\bibfnamefont {A.}~\bibnamefont {Sch\"afer}}, \ and\ \bibinfo
  {author} {\bibfnamefont {K.}~\bibnamefont {Szab\'o}},\ }\href {\doibase
  10.22323/1.214.0232} {\bibfield  {journal} {\bibinfo  {journal} {PoS}\
  }\textbf {\bibinfo {volume} {LATTICE2014}},\ \bibinfo {pages} {232} (\bibinfo
  {year} {2015})}\BibitemShut {NoStop}%
\bibitem [{\citenamefont {Itou}\ and\ \citenamefont
  {Nagai}(2020)}]{Itou:2020azb}%
  \BibitemOpen
  \bibfield  {author} {\bibinfo {author} {\bibfnamefont {E.}~\bibnamefont
  {Itou}}\ and\ \bibinfo {author} {\bibfnamefont {Y.}~\bibnamefont {Nagai}},\
  }\href {\doibase 10.1007/JHEP07(2020)007} {\bibfield  {journal} {\bibinfo
  {journal} {JHEP}\ }\textbf {\bibinfo {volume} {07}},\ \bibinfo {pages} {007}
  (\bibinfo {year} {2020})},\ \Eprint {http://arxiv.org/abs/2004.02426}
  {arXiv:2004.02426 [hep-lat]} \BibitemShut {NoStop}%
\bibitem [{\citenamefont {Itou}\ and\ \citenamefont
  {Nagai}(2022)}]{Itou:2021hsj}%
  \BibitemOpen
  \bibfield  {author} {\bibinfo {author} {\bibfnamefont {E.}~\bibnamefont
  {Itou}}\ and\ \bibinfo {author} {\bibfnamefont {Y.}~\bibnamefont {Nagai}},\
  }\href {\doibase 10.22323/1.396.0214} {\bibfield  {journal} {\bibinfo
  {journal} {PoS}\ }\textbf {\bibinfo {volume} {LATTICE2021}},\ \bibinfo
  {pages} {214} (\bibinfo {year} {2022})},\ \Eprint
  {http://arxiv.org/abs/2110.13417} {arXiv:2110.13417 [hep-lat]} \BibitemShut
  {NoStop}%
\bibitem [{\citenamefont {Altenkort}\ \emph {et~al.}(2023)\citenamefont
  {Altenkort}, \citenamefont {Eller}, \citenamefont {Francis}, \citenamefont
  {Kaczmarek}, \citenamefont {Mazur}, \citenamefont {Moore},\ and\
  \citenamefont {Shu}}]{Altenkort:2022yhb}%
  \BibitemOpen
  \bibfield  {author} {\bibinfo {author} {\bibfnamefont {L.}~\bibnamefont
  {Altenkort}}, \bibinfo {author} {\bibfnamefont {A.~M.}\ \bibnamefont
  {Eller}}, \bibinfo {author} {\bibfnamefont {A.}~\bibnamefont {Francis}},
  \bibinfo {author} {\bibfnamefont {O.}~\bibnamefont {Kaczmarek}}, \bibinfo
  {author} {\bibfnamefont {L.}~\bibnamefont {Mazur}}, \bibinfo {author}
  {\bibfnamefont {G.~D.}\ \bibnamefont {Moore}}, \ and\ \bibinfo {author}
  {\bibfnamefont {H.-T.}\ \bibnamefont {Shu}},\ }\href {\doibase
  10.1103/PhysRevD.108.014503} {\bibfield  {journal} {\bibinfo  {journal}
  {Phys. Rev. D}\ }\textbf {\bibinfo {volume} {108}},\ \bibinfo {pages}
  {014503} (\bibinfo {year} {2023})}\BibitemShut {NoStop}%
\bibitem [{\citenamefont {Cohen}\ \emph {et~al.}(2021)\citenamefont {Cohen},
  \citenamefont {Lamm}, \citenamefont {Lawrence},\ and\ \citenamefont
  {Yamauchi}}]{Cohen:2021imf}%
  \BibitemOpen
  \bibfield  {author} {\bibinfo {author} {\bibfnamefont {T.~D.}\ \bibnamefont
  {Cohen}}, \bibinfo {author} {\bibfnamefont {H.}~\bibnamefont {Lamm}},
  \bibinfo {author} {\bibfnamefont {S.}~\bibnamefont {Lawrence}}, \ and\
  \bibinfo {author} {\bibfnamefont {Y.}~\bibnamefont {Yamauchi}} (\bibinfo
  {collaboration} {NuQS}),\ }\href {\doibase 10.1103/PhysRevD.104.094514}
  {\bibfield  {journal} {\bibinfo  {journal} {Phys. Rev. D}\ }\textbf {\bibinfo
  {volume} {104}},\ \bibinfo {pages} {094514} (\bibinfo {year} {2021})},\
  \Eprint {http://arxiv.org/abs/2104.02024} {arXiv:2104.02024 [hep-lat]}
  \BibitemShut {NoStop}%
\bibitem [{\citenamefont {Turro}\ \emph {et~al.}(2024)\citenamefont {Turro},
  \citenamefont {Ciavarella},\ and\ \citenamefont {Yao}}]{turro2024shear}%
  \BibitemOpen
  \bibfield  {author} {\bibinfo {author} {\bibfnamefont {F.}~\bibnamefont
  {Turro}}, \bibinfo {author} {\bibfnamefont {A.}~\bibnamefont {Ciavarella}}, \
  and\ \bibinfo {author} {\bibfnamefont {X.}~\bibnamefont {Yao}},\ }\href
  {\doibase 10.1103/PhysRevD.109.114511} {\bibfield  {journal} {\bibinfo
  {journal} {Phys. Rev. D}\ }\textbf {\bibinfo {volume} {109}},\ \bibinfo
  {pages} {114511} (\bibinfo {year} {2024})},\ \Eprint
  {http://arxiv.org/abs/2402.04221} {arXiv:2402.04221 [hep-lat]} \BibitemShut
  {NoStop}%
\bibitem [{\citenamefont {Romatschke}(2010)}]{Romatschke:2009im}%
  \BibitemOpen
  \bibfield  {author} {\bibinfo {author} {\bibfnamefont {P.}~\bibnamefont
  {Romatschke}},\ }\href {\doibase 10.1142/S0218301310014613} {\bibfield
  {journal} {\bibinfo  {journal} {Int. J. Mod. Phys. E}\ }\textbf {\bibinfo
  {volume} {19}},\ \bibinfo {pages} {1} (\bibinfo {year} {2010})},\ \Eprint
  {http://arxiv.org/abs/0902.3663} {arXiv:0902.3663 [hep-ph]} \BibitemShut
  {NoStop}%
\bibitem [{\citenamefont {Teaney}(2010)}]{Teaney:2009qa}%
  \BibitemOpen
  \bibfield  {author} {\bibinfo {author} {\bibfnamefont {D.~A.}\ \bibnamefont
  {Teaney}},\ }\enquote {\bibinfo {title} {{Viscous Hydrodynamics and the Quark
  Gluon Plasma}},}\ in\ \href {\doibase 10.1142/9789814293297_0004} {\emph
  {\bibinfo {booktitle} {{Quark-gluon plasma 4}}}},\ \bibinfo {editor} {edited
  by\ \bibinfo {editor} {\bibfnamefont {R.~C.}\ \bibnamefont {Hwa}}\ and\
  \bibinfo {editor} {\bibfnamefont {X.-N.}\ \bibnamefont {Wang}}}\ (\bibinfo
  {year} {2010})\ pp.\ \bibinfo {pages} {207--266},\ \Eprint
  {http://arxiv.org/abs/0905.2433} {arXiv:0905.2433 [nucl-th]} \BibitemShut
  {NoStop}%
\bibitem [{\citenamefont {Jeon}\ and\ \citenamefont
  {Heinz}(2015)}]{Jeon:2015dfa}%
  \BibitemOpen
  \bibfield  {author} {\bibinfo {author} {\bibfnamefont {S.}~\bibnamefont
  {Jeon}}\ and\ \bibinfo {author} {\bibfnamefont {U.}~\bibnamefont {Heinz}},\
  }\href {\doibase 10.1142/S0218301315300106} {\bibfield  {journal} {\bibinfo
  {journal} {Int. J. Mod. Phys. E}\ }\textbf {\bibinfo {volume} {24}},\
  \bibinfo {pages} {1530010} (\bibinfo {year} {2015})},\ \Eprint
  {http://arxiv.org/abs/1503.03931} {arXiv:1503.03931 [hep-ph]} \BibitemShut
  {NoStop}%
\bibitem [{\citenamefont {Romatschke}\ and\ \citenamefont
  {Romatschke}(2019)}]{Romatschke:2017ejr}%
  \BibitemOpen
  \bibfield  {author} {\bibinfo {author} {\bibfnamefont {P.}~\bibnamefont
  {Romatschke}}\ and\ \bibinfo {author} {\bibfnamefont {U.}~\bibnamefont
  {Romatschke}},\ }\href {\doibase 10.1017/9781108651998} {\emph {\bibinfo
  {title} {{Relativistic Fluid Dynamics In and Out of Equilibrium}}}},\
  Cambridge Monographs on Mathematical Physics\ (\bibinfo  {publisher}
  {Cambridge University Press},\ \bibinfo {year} {2019})\ \Eprint
  {http://arxiv.org/abs/1712.05815} {arXiv:1712.05815 [nucl-th]} \BibitemShut
  {NoStop}%
\bibitem [{\citenamefont {Rezzolla}\ and\ \citenamefont
  {Zanotti}(2013)}]{rezzolla2013relativistic}%
  \BibitemOpen
  \bibfield  {author} {\bibinfo {author} {\bibfnamefont {L.}~\bibnamefont
  {Rezzolla}}\ and\ \bibinfo {author} {\bibfnamefont {O.}~\bibnamefont
  {Zanotti}},\ }\href@noop {} {\emph {\bibinfo {title} {Relativistic
  hydrodynamics}}}\ (\bibinfo  {publisher} {Oxford University Press, USA},\
  \bibinfo {year} {2013})\BibitemShut {NoStop}%
\bibitem [{\citenamefont {Crossley}\ \emph {et~al.}(2017)\citenamefont
  {Crossley}, \citenamefont {Glorioso},\ and\ \citenamefont
  {Liu}}]{Crossley:2015evo}%
  \BibitemOpen
  \bibfield  {author} {\bibinfo {author} {\bibfnamefont {M.}~\bibnamefont
  {Crossley}}, \bibinfo {author} {\bibfnamefont {P.}~\bibnamefont {Glorioso}},
  \ and\ \bibinfo {author} {\bibfnamefont {H.}~\bibnamefont {Liu}},\ }\href
  {\doibase 10.1007/JHEP09(2017)095} {\bibfield  {journal} {\bibinfo  {journal}
  {JHEP}\ }\textbf {\bibinfo {volume} {09}},\ \bibinfo {pages} {095} (\bibinfo
  {year} {2017})},\ \Eprint {http://arxiv.org/abs/1511.03646} {arXiv:1511.03646
  [hep-th]} \BibitemShut {NoStop}%
\bibitem [{\citenamefont {Glorioso}\ \emph {et~al.}(2017)\citenamefont
  {Glorioso}, \citenamefont {Crossley},\ and\ \citenamefont
  {Liu}}]{Glorioso:2017fpd}%
  \BibitemOpen
  \bibfield  {author} {\bibinfo {author} {\bibfnamefont {P.}~\bibnamefont
  {Glorioso}}, \bibinfo {author} {\bibfnamefont {M.}~\bibnamefont {Crossley}},
  \ and\ \bibinfo {author} {\bibfnamefont {H.}~\bibnamefont {Liu}},\ }\href
  {\doibase 10.1007/JHEP09(2017)096} {\bibfield  {journal} {\bibinfo  {journal}
  {JHEP}\ }\textbf {\bibinfo {volume} {09}},\ \bibinfo {pages} {096} (\bibinfo
  {year} {2017})},\ \Eprint {http://arxiv.org/abs/1701.07817} {arXiv:1701.07817
  [hep-th]} \BibitemShut {NoStop}%
\bibitem [{\citenamefont {Vardhan}\ \emph {et~al.}(2024)\citenamefont
  {Vardhan}, \citenamefont {Grozdanov}, \citenamefont {Leutheusser},\ and\
  \citenamefont {Liu}}]{Vardhan:2024qdi}%
  \BibitemOpen
  \bibfield  {author} {\bibinfo {author} {\bibfnamefont {S.}~\bibnamefont
  {Vardhan}}, \bibinfo {author} {\bibfnamefont {S.}~\bibnamefont {Grozdanov}},
  \bibinfo {author} {\bibfnamefont {S.}~\bibnamefont {Leutheusser}}, \ and\
  \bibinfo {author} {\bibfnamefont {H.}~\bibnamefont {Liu}},\ }\href@noop {} {\
   (\bibinfo {year} {2024})},\ \Eprint {http://arxiv.org/abs/2408.12868}
  {arXiv:2408.12868 [hep-th]} \BibitemShut {NoStop}%
\bibitem [{\citenamefont {Arnold}\ and\ \citenamefont
  {Yaffe}(1998)}]{Arnold:1997gh}%
  \BibitemOpen
  \bibfield  {author} {\bibinfo {author} {\bibfnamefont {P.~B.}\ \bibnamefont
  {Arnold}}\ and\ \bibinfo {author} {\bibfnamefont {L.~G.}\ \bibnamefont
  {Yaffe}},\ }\href {\doibase 10.1103/PhysRevD.57.1178} {\bibfield  {journal}
  {\bibinfo  {journal} {Phys. Rev. D}\ }\textbf {\bibinfo {volume} {57}},\
  \bibinfo {pages} {1178} (\bibinfo {year} {1998})},\ \Eprint
  {http://arxiv.org/abs/hep-ph/9709449} {arXiv:hep-ph/9709449} \BibitemShut
  {NoStop}%
\bibitem [{\citenamefont {Pitaevskii}\ \emph {et~al.}(2017)\citenamefont
  {Pitaevskii}, \citenamefont {Lifshitz},\ and\ \citenamefont
  {Sykes}}]{pitaevskii2017course}%
  \BibitemOpen
  \bibfield  {author} {\bibinfo {author} {\bibfnamefont {L.}~\bibnamefont
  {Pitaevskii}}, \bibinfo {author} {\bibfnamefont {E.}~\bibnamefont
  {Lifshitz}}, \ and\ \bibinfo {author} {\bibfnamefont {J.~B.}\ \bibnamefont
  {Sykes}},\ }\href@noop {} {\emph {\bibinfo {title} {Course of theoretical
  physics: physical kinetics}}},\ Vol.~\bibinfo {volume} {10}\ (\bibinfo
  {publisher} {Elsevier},\ \bibinfo {year} {2017})\BibitemShut {NoStop}%
\bibitem [{\citenamefont {Kovtun}\ and\ \citenamefont
  {Yaffe}(2003)}]{Kovtun:2003vj}%
  \BibitemOpen
  \bibfield  {author} {\bibinfo {author} {\bibfnamefont {P.}~\bibnamefont
  {Kovtun}}\ and\ \bibinfo {author} {\bibfnamefont {L.~G.}\ \bibnamefont
  {Yaffe}},\ }\href {\doibase 10.1103/PhysRevD.68.025007} {\bibfield  {journal}
  {\bibinfo  {journal} {Phys. Rev. D}\ }\textbf {\bibinfo {volume} {68}},\
  \bibinfo {pages} {025007} (\bibinfo {year} {2003})},\ \Eprint
  {http://arxiv.org/abs/hep-th/0303010} {arXiv:hep-th/0303010} \BibitemShut
  {NoStop}%
\bibitem [{\citenamefont {Caron-Huot}\ and\ \citenamefont
  {Saremi}(2010)}]{Caron-Huot:2009kyg}%
  \BibitemOpen
  \bibfield  {author} {\bibinfo {author} {\bibfnamefont {S.}~\bibnamefont
  {Caron-Huot}}\ and\ \bibinfo {author} {\bibfnamefont {O.}~\bibnamefont
  {Saremi}},\ }\href {\doibase 10.1007/JHEP11(2010)013} {\bibfield  {journal}
  {\bibinfo  {journal} {JHEP}\ }\textbf {\bibinfo {volume} {11}},\ \bibinfo
  {pages} {013} (\bibinfo {year} {2010})},\ \Eprint
  {http://arxiv.org/abs/0909.4525} {arXiv:0909.4525 [hep-th]} \BibitemShut
  {NoStop}%
\bibitem [{\citenamefont {Kovtun}(2012)}]{Kovtun:2012rj}%
  \BibitemOpen
  \bibfield  {author} {\bibinfo {author} {\bibfnamefont {P.}~\bibnamefont
  {Kovtun}},\ }\href {\doibase 10.1088/1751-8113/45/47/473001} {\bibfield
  {journal} {\bibinfo  {journal} {J. Phys. A}\ }\textbf {\bibinfo {volume}
  {45}},\ \bibinfo {pages} {473001} (\bibinfo {year} {2012})},\ \Eprint
  {http://arxiv.org/abs/1205.5040} {arXiv:1205.5040 [hep-th]} \BibitemShut
  {NoStop}%
\bibitem [{\citenamefont {Romatschke}(2021)}]{Romatschke:2021imm}%
  \BibitemOpen
  \bibfield  {author} {\bibinfo {author} {\bibfnamefont {P.}~\bibnamefont
  {Romatschke}},\ }\href {\doibase 10.1103/PhysRevLett.127.111603} {\bibfield
  {journal} {\bibinfo  {journal} {Phys. Rev. Lett.}\ }\textbf {\bibinfo
  {volume} {127}},\ \bibinfo {pages} {111603} (\bibinfo {year} {2021})},\
  \Eprint {http://arxiv.org/abs/2104.06435} {arXiv:2104.06435 [hep-th]}
  \BibitemShut {NoStop}%
\bibitem [{\citenamefont {Akamatsu}\ \emph {et~al.}(2017)\citenamefont
  {Akamatsu}, \citenamefont {Mazeliauskas},\ and\ \citenamefont
  {Teaney}}]{Akamatsu:2016llw}%
  \BibitemOpen
  \bibfield  {author} {\bibinfo {author} {\bibfnamefont {Y.}~\bibnamefont
  {Akamatsu}}, \bibinfo {author} {\bibfnamefont {A.}~\bibnamefont
  {Mazeliauskas}}, \ and\ \bibinfo {author} {\bibfnamefont {D.}~\bibnamefont
  {Teaney}},\ }\href {\doibase 10.1103/PhysRevC.95.014909} {\bibfield
  {journal} {\bibinfo  {journal} {Phys. Rev. C}\ }\textbf {\bibinfo {volume}
  {95}},\ \bibinfo {pages} {014909} (\bibinfo {year} {2017})},\ \Eprint
  {http://arxiv.org/abs/1606.07742} {arXiv:1606.07742 [nucl-th]} \BibitemShut
  {NoStop}%
\bibitem [{\citenamefont {Martinez}\ and\ \citenamefont
  {Sch\"afer}(2019)}]{Martinez:2018wia}%
  \BibitemOpen
  \bibfield  {author} {\bibinfo {author} {\bibfnamefont {M.}~\bibnamefont
  {Martinez}}\ and\ \bibinfo {author} {\bibfnamefont {T.}~\bibnamefont
  {Sch\"afer}},\ }\href {\doibase 10.1103/PhysRevC.99.054902} {\bibfield
  {journal} {\bibinfo  {journal} {Phys. Rev. C}\ }\textbf {\bibinfo {volume}
  {99}},\ \bibinfo {pages} {054902} (\bibinfo {year} {2019})},\ \Eprint
  {http://arxiv.org/abs/1812.05279} {arXiv:1812.05279 [hep-th]} \BibitemShut
  {NoStop}%
\bibitem [{\citenamefont {Shukla}(2021)}]{Shukla:2021ksb}%
  \BibitemOpen
  \bibfield  {author} {\bibinfo {author} {\bibfnamefont {A.}~\bibnamefont
  {Shukla}},\ }\href {\doibase 10.1016/j.nuclphysb.2021.115442} {\bibfield
  {journal} {\bibinfo  {journal} {Nucl. Phys. B}\ }\textbf {\bibinfo {volume}
  {968}},\ \bibinfo {pages} {115442} (\bibinfo {year} {2021})},\ \Eprint
  {http://arxiv.org/abs/2101.10000} {arXiv:2101.10000 [hep-th]} \BibitemShut
  {NoStop}%
\bibitem [{\citenamefont {Matthies}\ \emph {et~al.}(2024)\citenamefont
  {Matthies}, \citenamefont {Dannenfeld}, \citenamefont {Pappalardi},\ and\
  \citenamefont {Rosch}}]{Matthies:2024lqx}%
  \BibitemOpen
  \bibfield  {author} {\bibinfo {author} {\bibfnamefont {A.}~\bibnamefont
  {Matthies}}, \bibinfo {author} {\bibfnamefont {N.}~\bibnamefont
  {Dannenfeld}}, \bibinfo {author} {\bibfnamefont {S.}~\bibnamefont
  {Pappalardi}}, \ and\ \bibinfo {author} {\bibfnamefont {A.}~\bibnamefont
  {Rosch}},\ }\href@noop {} {\  (\bibinfo {year} {2024})},\ \Eprint
  {http://arxiv.org/abs/2410.16182} {arXiv:2410.16182 [quant-ph]} \BibitemShut
  {NoStop}%
\bibitem [{\citenamefont {Kogut}\ and\ \citenamefont
  {Susskind}(1975)}]{PhysRevD.11.395}%
  \BibitemOpen
  \bibfield  {author} {\bibinfo {author} {\bibfnamefont {J.}~\bibnamefont
  {Kogut}}\ and\ \bibinfo {author} {\bibfnamefont {L.}~\bibnamefont
  {Susskind}},\ }\href {\doibase 10.1103/PhysRevD.11.395} {\bibfield  {journal}
  {\bibinfo  {journal} {Phys. Rev. D}\ }\textbf {\bibinfo {volume} {11}},\
  \bibinfo {pages} {395} (\bibinfo {year} {1975})}\BibitemShut {NoStop}%
\bibitem [{\citenamefont {Byrnes}\ and\ \citenamefont
  {Yamamoto}(2006)}]{Byrnes:2005qx}%
  \BibitemOpen
  \bibfield  {author} {\bibinfo {author} {\bibfnamefont {T.}~\bibnamefont
  {Byrnes}}\ and\ \bibinfo {author} {\bibfnamefont {Y.}~\bibnamefont
  {Yamamoto}},\ }\href {\doibase 10.1103/PhysRevA.73.022328} {\bibfield
  {journal} {\bibinfo  {journal} {Phys. Rev. A}\ }\textbf {\bibinfo {volume}
  {73}},\ \bibinfo {pages} {022328} (\bibinfo {year} {2006})}\BibitemShut
  {NoStop}%
\bibitem [{\citenamefont {Zohar}\ and\ \citenamefont
  {Burrello}(2015)}]{Zohar:2014qma}%
  \BibitemOpen
  \bibfield  {author} {\bibinfo {author} {\bibfnamefont {E.}~\bibnamefont
  {Zohar}}\ and\ \bibinfo {author} {\bibfnamefont {M.}~\bibnamefont
  {Burrello}},\ }\href {\doibase 10.1103/PhysRevD.91.054506} {\bibfield
  {journal} {\bibinfo  {journal} {Phys. Rev. D}\ }\textbf {\bibinfo {volume}
  {91}},\ \bibinfo {pages} {054506} (\bibinfo {year} {2015})},\ \Eprint
  {http://arxiv.org/abs/1409.3085} {arXiv:1409.3085 [quant-ph]} \BibitemShut
  {NoStop}%
\bibitem [{\citenamefont {Liu}\ and\ \citenamefont
  {Chandrasekharan}(2022)}]{Liu:2021tef}%
  \BibitemOpen
  \bibfield  {author} {\bibinfo {author} {\bibfnamefont {H.}~\bibnamefont
  {Liu}}\ and\ \bibinfo {author} {\bibfnamefont {S.}~\bibnamefont
  {Chandrasekharan}},\ }\href {\doibase 10.3390/sym14020305} {\bibfield
  {journal} {\bibinfo  {journal} {Symmetry}\ }\textbf {\bibinfo {volume}
  {14}},\ \bibinfo {pages} {305} (\bibinfo {year} {2022})},\ \Eprint
  {http://arxiv.org/abs/2112.02090} {arXiv:2112.02090 [hep-lat]} \BibitemShut
  {NoStop}%
\bibitem [{\citenamefont {Klco}\ \emph {et~al.}(2020)\citenamefont {Klco},
  \citenamefont {Stryker},\ and\ \citenamefont {Savage}}]{Klco:2019evd}%
  \BibitemOpen
  \bibfield  {author} {\bibinfo {author} {\bibfnamefont {N.}~\bibnamefont
  {Klco}}, \bibinfo {author} {\bibfnamefont {J.~R.}\ \bibnamefont {Stryker}}, \
  and\ \bibinfo {author} {\bibfnamefont {M.~J.}\ \bibnamefont {Savage}},\
  }\href {\doibase 10.1103/PhysRevD.101.074512} {\bibfield  {journal} {\bibinfo
   {journal} {Phys. Rev. D}\ }\textbf {\bibinfo {volume} {101}},\ \bibinfo
  {pages} {074512} (\bibinfo {year} {2020})},\ \Eprint
  {http://arxiv.org/abs/1908.06935} {arXiv:1908.06935 [quant-ph]} \BibitemShut
  {NoStop}%
\bibitem [{\citenamefont {A~Rahman}\ \emph {et~al.}(2021)\citenamefont
  {A~Rahman}, \citenamefont {Lewis}, \citenamefont {Mendicelli},\ and\
  \citenamefont {Powell}}]{ARahman:2021ktn}%
  \BibitemOpen
  \bibfield  {author} {\bibinfo {author} {\bibfnamefont {S.}~\bibnamefont
  {A~Rahman}}, \bibinfo {author} {\bibfnamefont {R.}~\bibnamefont {Lewis}},
  \bibinfo {author} {\bibfnamefont {E.}~\bibnamefont {Mendicelli}}, \ and\
  \bibinfo {author} {\bibfnamefont {S.}~\bibnamefont {Powell}},\ }\href
  {\doibase 10.1103/PhysRevD.104.034501} {\bibfield  {journal} {\bibinfo
  {journal} {Phys. Rev. D}\ }\textbf {\bibinfo {volume} {104}},\ \bibinfo
  {pages} {034501} (\bibinfo {year} {2021})},\ \Eprint
  {http://arxiv.org/abs/2103.08661} {arXiv:2103.08661 [hep-lat]} \BibitemShut
  {NoStop}%
\bibitem [{\citenamefont {Hayata}\ \emph {et~al.}(2021)\citenamefont {Hayata},
  \citenamefont {Hidaka},\ and\ \citenamefont {Kikuchi}}]{Hayata:2021kcp}%
  \BibitemOpen
  \bibfield  {author} {\bibinfo {author} {\bibfnamefont {T.}~\bibnamefont
  {Hayata}}, \bibinfo {author} {\bibfnamefont {Y.}~\bibnamefont {Hidaka}}, \
  and\ \bibinfo {author} {\bibfnamefont {Y.}~\bibnamefont {Kikuchi}},\ }\href
  {\doibase 10.1103/PhysRevD.104.074518} {\bibfield  {journal} {\bibinfo
  {journal} {Phys. Rev. D}\ }\textbf {\bibinfo {volume} {104}},\ \bibinfo
  {pages} {074518} (\bibinfo {year} {2021})},\ \Eprint
  {http://arxiv.org/abs/2103.05179} {arXiv:2103.05179 [quant-ph]} \BibitemShut
  {NoStop}%
\bibitem [{\citenamefont {A~Rahman}\ \emph {et~al.}(2022)\citenamefont
  {A~Rahman}, \citenamefont {Lewis}, \citenamefont {Mendicelli},\ and\
  \citenamefont {Powell}}]{ARahman:2022tkr}%
  \BibitemOpen
  \bibfield  {author} {\bibinfo {author} {\bibfnamefont {S.}~\bibnamefont
  {A~Rahman}}, \bibinfo {author} {\bibfnamefont {R.}~\bibnamefont {Lewis}},
  \bibinfo {author} {\bibfnamefont {E.}~\bibnamefont {Mendicelli}}, \ and\
  \bibinfo {author} {\bibfnamefont {S.}~\bibnamefont {Powell}},\ }\href
  {\doibase 10.1103/PhysRevD.106.074502} {\bibfield  {journal} {\bibinfo
  {journal} {Phys. Rev. D}\ }\textbf {\bibinfo {volume} {106}},\ \bibinfo
  {pages} {074502} (\bibinfo {year} {2022})},\ \Eprint
  {http://arxiv.org/abs/2205.09247} {arXiv:2205.09247 [hep-lat]} \BibitemShut
  {NoStop}%
\bibitem [{\citenamefont {Yao}(2023)}]{Yao:2023pht}%
  \BibitemOpen
  \bibfield  {author} {\bibinfo {author} {\bibfnamefont {X.}~\bibnamefont
  {Yao}},\ }\href {\doibase 10.1103/PhysRevD.108.L031504} {\bibfield  {journal}
  {\bibinfo  {journal} {Phys. Rev. D}\ }\textbf {\bibinfo {volume} {108}},\
  \bibinfo {pages} {L031504} (\bibinfo {year} {2023})},\ \Eprint
  {http://arxiv.org/abs/2303.14264} {arXiv:2303.14264 [hep-lat]} \BibitemShut
  {NoStop}%
\bibitem [{\citenamefont {Ebner}\ \emph
  {et~al.}(2024{\natexlab{a}})\citenamefont {Ebner}, \citenamefont {M\"uller},
  \citenamefont {Sch\"afer}, \citenamefont {Seidl},\ and\ \citenamefont
  {Yao}}]{Ebner:2023ixq}%
  \BibitemOpen
  \bibfield  {author} {\bibinfo {author} {\bibfnamefont {L.}~\bibnamefont
  {Ebner}}, \bibinfo {author} {\bibfnamefont {B.}~\bibnamefont {M\"uller}},
  \bibinfo {author} {\bibfnamefont {A.}~\bibnamefont {Sch\"afer}}, \bibinfo
  {author} {\bibfnamefont {C.}~\bibnamefont {Seidl}}, \ and\ \bibinfo {author}
  {\bibfnamefont {X.}~\bibnamefont {Yao}},\ }\href {\doibase
  10.1103/PhysRevD.109.014504} {\bibfield  {journal} {\bibinfo  {journal}
  {Phys. Rev. D}\ }\textbf {\bibinfo {volume} {109}},\ \bibinfo {pages}
  {014504} (\bibinfo {year} {2024}{\natexlab{a}})},\ \Eprint
  {http://arxiv.org/abs/2308.16202} {arXiv:2308.16202 [hep-lat]} \BibitemShut
  {NoStop}%
\bibitem [{\citenamefont {Banuls}\ \emph {et~al.}(2020)\citenamefont {Banuls},
  \citenamefont {Blatt}, \citenamefont {Catani}, \citenamefont {Celi},
  \citenamefont {Cirac}, \citenamefont {Dalmonte}, \citenamefont {Fallani},
  \citenamefont {Jansen}, \citenamefont {Lewenstein}, \citenamefont
  {Montangero} \emph {et~al.}}]{banuls2020simulating}%
  \BibitemOpen
  \bibfield  {author} {\bibinfo {author} {\bibfnamefont {M.~C.}\ \bibnamefont
  {Banuls}}, \bibinfo {author} {\bibfnamefont {R.}~\bibnamefont {Blatt}},
  \bibinfo {author} {\bibfnamefont {J.}~\bibnamefont {Catani}}, \bibinfo
  {author} {\bibfnamefont {A.}~\bibnamefont {Celi}}, \bibinfo {author}
  {\bibfnamefont {J.~I.}\ \bibnamefont {Cirac}}, \bibinfo {author}
  {\bibfnamefont {M.}~\bibnamefont {Dalmonte}}, \bibinfo {author}
  {\bibfnamefont {L.}~\bibnamefont {Fallani}}, \bibinfo {author} {\bibfnamefont
  {K.}~\bibnamefont {Jansen}}, \bibinfo {author} {\bibfnamefont
  {M.}~\bibnamefont {Lewenstein}}, \bibinfo {author} {\bibfnamefont
  {S.}~\bibnamefont {Montangero}},  \emph {et~al.},\ }\href {\doibase
  10.1140/epjd/e2020-100571-8} {\bibfield  {journal} {\bibinfo  {journal} {Eur.
  Phys. J.}\ }\textbf {\bibinfo {volume} {74}},\ \bibinfo {pages} {1} (\bibinfo
  {year} {2020})}\BibitemShut {NoStop}%
\bibitem [{\citenamefont {Klco}\ \emph {et~al.}(2022)\citenamefont {Klco},
  \citenamefont {Roggero},\ and\ \citenamefont {Savage}}]{klco2022standard}%
  \BibitemOpen
  \bibfield  {author} {\bibinfo {author} {\bibfnamefont {N.}~\bibnamefont
  {Klco}}, \bibinfo {author} {\bibfnamefont {A.}~\bibnamefont {Roggero}}, \
  and\ \bibinfo {author} {\bibfnamefont {M.~J.}\ \bibnamefont {Savage}},\
  }\href {\doibase 10.1088/1361-6633/ac58a4} {\bibfield  {journal} {\bibinfo
  {journal} {Rep. Prog. Phys.}\ }\textbf {\bibinfo {volume} {85}},\ \bibinfo
  {pages} {064301} (\bibinfo {year} {2022})}\BibitemShut {NoStop}%
\bibitem [{\citenamefont {Bauer}\ \emph
  {et~al.}(2023{\natexlab{a}})\citenamefont {Bauer}, \citenamefont {Davoudi},
  \citenamefont {Balantekin}, \citenamefont {Bhattacharya}, \citenamefont
  {Carena}, \citenamefont {de~Jong}, \citenamefont {Draper}, \citenamefont
  {El-Khadra}, \citenamefont {Gemelke}, \citenamefont {Hanada}, \citenamefont
  {Kharzeev}, \citenamefont {Lamm}, \citenamefont {Li}, \citenamefont {Liu},
  \citenamefont {Lukin}, \citenamefont {Meurice}, \citenamefont {Monroe},
  \citenamefont {Nachman}, \citenamefont {Pagano}, \citenamefont {Preskill},
  \citenamefont {Rinaldi}, \citenamefont {Roggero}, \citenamefont {Santiago},
  \citenamefont {Savage}, \citenamefont {Siddiqi}, \citenamefont {Siopsis},
  \citenamefont {Van~Zanten}, \citenamefont {Wiebe}, \citenamefont {Yamauchi},
  \citenamefont {Yeter-Aydeniz},\ and\ \citenamefont {Zorzetti}}]{Bauer_2023}%
  \BibitemOpen
  \bibfield  {author} {\bibinfo {author} {\bibfnamefont {C.~W.}\ \bibnamefont
  {Bauer}}, \bibinfo {author} {\bibfnamefont {Z.}~\bibnamefont {Davoudi}},
  \bibinfo {author} {\bibfnamefont {A.~B.}\ \bibnamefont {Balantekin}},
  \bibinfo {author} {\bibfnamefont {T.}~\bibnamefont {Bhattacharya}}, \bibinfo
  {author} {\bibfnamefont {M.}~\bibnamefont {Carena}}, \bibinfo {author}
  {\bibfnamefont {W.~A.}\ \bibnamefont {de~Jong}}, \bibinfo {author}
  {\bibfnamefont {P.}~\bibnamefont {Draper}}, \bibinfo {author} {\bibfnamefont
  {A.}~\bibnamefont {El-Khadra}}, \bibinfo {author} {\bibfnamefont
  {N.}~\bibnamefont {Gemelke}}, \bibinfo {author} {\bibfnamefont
  {M.}~\bibnamefont {Hanada}}, \bibinfo {author} {\bibfnamefont
  {D.}~\bibnamefont {Kharzeev}}, \bibinfo {author} {\bibfnamefont
  {H.}~\bibnamefont {Lamm}}, \bibinfo {author} {\bibfnamefont {Y.-Y.}\
  \bibnamefont {Li}}, \bibinfo {author} {\bibfnamefont {J.}~\bibnamefont
  {Liu}}, \bibinfo {author} {\bibfnamefont {M.}~\bibnamefont {Lukin}}, \bibinfo
  {author} {\bibfnamefont {Y.}~\bibnamefont {Meurice}}, \bibinfo {author}
  {\bibfnamefont {C.}~\bibnamefont {Monroe}}, \bibinfo {author} {\bibfnamefont
  {B.}~\bibnamefont {Nachman}}, \bibinfo {author} {\bibfnamefont
  {G.}~\bibnamefont {Pagano}}, \bibinfo {author} {\bibfnamefont
  {J.}~\bibnamefont {Preskill}}, \bibinfo {author} {\bibfnamefont
  {E.}~\bibnamefont {Rinaldi}}, \bibinfo {author} {\bibfnamefont
  {A.}~\bibnamefont {Roggero}}, \bibinfo {author} {\bibfnamefont {D.~I.}\
  \bibnamefont {Santiago}}, \bibinfo {author} {\bibfnamefont {M.~J.}\
  \bibnamefont {Savage}}, \bibinfo {author} {\bibfnamefont {I.}~\bibnamefont
  {Siddiqi}}, \bibinfo {author} {\bibfnamefont {G.}~\bibnamefont {Siopsis}},
  \bibinfo {author} {\bibfnamefont {D.}~\bibnamefont {Van~Zanten}}, \bibinfo
  {author} {\bibfnamefont {N.}~\bibnamefont {Wiebe}}, \bibinfo {author}
  {\bibfnamefont {Y.}~\bibnamefont {Yamauchi}}, \bibinfo {author}
  {\bibfnamefont {K.}~\bibnamefont {Yeter-Aydeniz}}, \ and\ \bibinfo {author}
  {\bibfnamefont {S.}~\bibnamefont {Zorzetti}},\ }\href {\doibase
  10.1103/prxquantum.4.027001} {\bibfield  {journal} {\bibinfo  {journal} {PRX
  Quantum}\ }\textbf {\bibinfo {volume} {4}} (\bibinfo {year}
  {2023}{\natexlab{a}}),\ 10.1103/prxquantum.4.027001}\BibitemShut {NoStop}%
\bibitem [{\citenamefont {Beck}\ and\ \citenamefont
  {et~al}(2023)}]{beck2023quantum}%
  \BibitemOpen
  \bibfield  {author} {\bibinfo {author} {\bibfnamefont {D.}~\bibnamefont
  {Beck}}\ and\ \bibinfo {author} {\bibnamefont {et~al}},\ }\href {\doibase
  https://arxiv.org/abs/2303.00113} {\enquote {\bibinfo {title} {Quantum
  information science and technology for nuclear physics. input into u.s.
  long-range planning, 2023},}\ } (\bibinfo {year} {2023}),\ \Eprint
  {http://arxiv.org/abs/2303.00113} {arXiv:2303.00113 [nucl-ex]} \BibitemShut
  {NoStop}%
\bibitem [{\citenamefont {Bauer}\ \emph
  {et~al.}(2023{\natexlab{b}})\citenamefont {Bauer}, \citenamefont {Davoudi},
  \citenamefont {Klco},\ and\ \citenamefont {Savage}}]{bauer2023quantum}%
  \BibitemOpen
  \bibfield  {author} {\bibinfo {author} {\bibfnamefont {C.~W.}\ \bibnamefont
  {Bauer}}, \bibinfo {author} {\bibfnamefont {Z.}~\bibnamefont {Davoudi}},
  \bibinfo {author} {\bibfnamefont {N.}~\bibnamefont {Klco}}, \ and\ \bibinfo
  {author} {\bibfnamefont {M.~J.}\ \bibnamefont {Savage}},\ }\href {\doibase
  https://doi.org/10.1038/s42254-023-00599-8} {\bibfield  {journal} {\bibinfo
  {journal} {Nat. Rev. Phys.}\ ,\ \bibinfo {pages} {1}} (\bibinfo {year}
  {2023}{\natexlab{b}})}\BibitemShut {NoStop}%
\bibitem [{\citenamefont {Lamm}\ \emph {et~al.}(2019)\citenamefont {Lamm},
  \citenamefont {Lawrence},\ and\ \citenamefont {Yamauchi}}]{Lamm:2019bik}%
  \BibitemOpen
  \bibfield  {author} {\bibinfo {author} {\bibfnamefont {H.}~\bibnamefont
  {Lamm}}, \bibinfo {author} {\bibfnamefont {S.}~\bibnamefont {Lawrence}}, \
  and\ \bibinfo {author} {\bibfnamefont {Y.}~\bibnamefont {Yamauchi}} (\bibinfo
  {collaboration} {NuQS}),\ }\href {\doibase 10.1103/PhysRevD.100.034518}
  {\bibfield  {journal} {\bibinfo  {journal} {Phys. Rev. D}\ }\textbf {\bibinfo
  {volume} {100}},\ \bibinfo {pages} {034518} (\bibinfo {year} {2019})},\
  \Eprint {http://arxiv.org/abs/1903.08807} {arXiv:1903.08807 [hep-lat]}
  \BibitemShut {NoStop}%
\bibitem [{\citenamefont {Raychowdhury}\ and\ \citenamefont
  {Stryker}(2020)}]{Raychowdhury:2019iki}%
  \BibitemOpen
  \bibfield  {author} {\bibinfo {author} {\bibfnamefont {I.}~\bibnamefont
  {Raychowdhury}}\ and\ \bibinfo {author} {\bibfnamefont {J.~R.}\ \bibnamefont
  {Stryker}},\ }\href {\doibase 10.1103/PhysRevD.101.114502} {\bibfield
  {journal} {\bibinfo  {journal} {Phys. Rev. D}\ }\textbf {\bibinfo {volume}
  {101}},\ \bibinfo {pages} {114502} (\bibinfo {year} {2020})},\ \Eprint
  {http://arxiv.org/abs/1912.06133} {arXiv:1912.06133 [hep-lat]} \BibitemShut
  {NoStop}%
\bibitem [{\citenamefont {Ciavarella}\ \emph {et~al.}(2021)\citenamefont
  {Ciavarella}, \citenamefont {Klco},\ and\ \citenamefont
  {Savage}}]{Ciavarella:2021nmj}%
  \BibitemOpen
  \bibfield  {author} {\bibinfo {author} {\bibfnamefont {A.}~\bibnamefont
  {Ciavarella}}, \bibinfo {author} {\bibfnamefont {N.}~\bibnamefont {Klco}}, \
  and\ \bibinfo {author} {\bibfnamefont {M.~J.}\ \bibnamefont {Savage}},\
  }\href {\doibase 10.1103/PhysRevD.103.094501} {\bibfield  {journal} {\bibinfo
   {journal} {Phys. Rev. D}\ }\textbf {\bibinfo {volume} {103}},\ \bibinfo
  {pages} {094501} (\bibinfo {year} {2021})},\ \Eprint
  {http://arxiv.org/abs/2101.10227} {arXiv:2101.10227 [quant-ph]} \BibitemShut
  {NoStop}%
\bibitem [{\citenamefont {De~Jong}\ \emph {et~al.}(2021)\citenamefont
  {De~Jong}, \citenamefont {Metcalf}, \citenamefont {Mulligan}, \citenamefont
  {P\l{}osko\'n}, \citenamefont {Ringer},\ and\ \citenamefont
  {Yao}}]{DeJong:2020riy}%
  \BibitemOpen
  \bibfield  {author} {\bibinfo {author} {\bibfnamefont {W.~A.}\ \bibnamefont
  {De~Jong}}, \bibinfo {author} {\bibfnamefont {M.}~\bibnamefont {Metcalf}},
  \bibinfo {author} {\bibfnamefont {J.}~\bibnamefont {Mulligan}}, \bibinfo
  {author} {\bibfnamefont {M.}~\bibnamefont {P\l{}osko\'n}}, \bibinfo {author}
  {\bibfnamefont {F.}~\bibnamefont {Ringer}}, \ and\ \bibinfo {author}
  {\bibfnamefont {X.}~\bibnamefont {Yao}},\ }\href {\doibase
  10.1103/PhysRevD.104.L051501} {\bibfield  {journal} {\bibinfo  {journal}
  {Phys. Rev. D}\ }\textbf {\bibinfo {volume} {104}},\ \bibinfo {pages}
  {051501} (\bibinfo {year} {2021})},\ \Eprint
  {http://arxiv.org/abs/2010.03571} {arXiv:2010.03571 [hep-ph]} \BibitemShut
  {NoStop}%
\bibitem [{\citenamefont {Rajput}\ \emph {et~al.}(2023)\citenamefont {Rajput},
  \citenamefont {Roggero},\ and\ \citenamefont {Wiebe}}]{Rajput:2021trn}%
  \BibitemOpen
  \bibfield  {author} {\bibinfo {author} {\bibfnamefont {A.}~\bibnamefont
  {Rajput}}, \bibinfo {author} {\bibfnamefont {A.}~\bibnamefont {Roggero}}, \
  and\ \bibinfo {author} {\bibfnamefont {N.}~\bibnamefont {Wiebe}},\ }\href
  {\doibase 10.1038/s41534-023-00706-8} {\bibfield  {journal} {\bibinfo
  {journal} {npj Quantum Inf.}\ }\textbf {\bibinfo {volume} {9}},\ \bibinfo
  {pages} {41} (\bibinfo {year} {2023})},\ \Eprint
  {http://arxiv.org/abs/2112.05186} {arXiv:2112.05186 [quant-ph]} \BibitemShut
  {NoStop}%
\bibitem [{\citenamefont {de~Jong}\ \emph {et~al.}(2022)\citenamefont
  {de~Jong}, \citenamefont {Lee}, \citenamefont {Mulligan}, \citenamefont
  {P\l{}osko\'n}, \citenamefont {Ringer},\ and\ \citenamefont
  {Yao}}]{deJong:2021wsd}%
  \BibitemOpen
  \bibfield  {author} {\bibinfo {author} {\bibfnamefont {W.~A.}\ \bibnamefont
  {de~Jong}}, \bibinfo {author} {\bibfnamefont {K.}~\bibnamefont {Lee}},
  \bibinfo {author} {\bibfnamefont {J.}~\bibnamefont {Mulligan}}, \bibinfo
  {author} {\bibfnamefont {M.}~\bibnamefont {P\l{}osko\'n}}, \bibinfo {author}
  {\bibfnamefont {F.}~\bibnamefont {Ringer}}, \ and\ \bibinfo {author}
  {\bibfnamefont {X.}~\bibnamefont {Yao}},\ }\href {\doibase
  10.1103/PhysRevD.106.054508} {\bibfield  {journal} {\bibinfo  {journal}
  {Phys. Rev. D}\ }\textbf {\bibinfo {volume} {106}},\ \bibinfo {pages}
  {054508} (\bibinfo {year} {2022})},\ \Eprint
  {http://arxiv.org/abs/2106.08394} {arXiv:2106.08394 [quant-ph]} \BibitemShut
  {NoStop}%
\bibitem [{\citenamefont {Kadam}\ \emph {et~al.}(2023)\citenamefont {Kadam},
  \citenamefont {Raychowdhury},\ and\ \citenamefont {Stryker}}]{Kadam:2022ipf}%
  \BibitemOpen
  \bibfield  {author} {\bibinfo {author} {\bibfnamefont {S.~V.}\ \bibnamefont
  {Kadam}}, \bibinfo {author} {\bibfnamefont {I.}~\bibnamefont {Raychowdhury}},
  \ and\ \bibinfo {author} {\bibfnamefont {J.~R.}\ \bibnamefont {Stryker}},\
  }\href {\doibase 10.1103/PhysRevD.107.094513} {\bibfield  {journal} {\bibinfo
   {journal} {Phys. Rev. D}\ }\textbf {\bibinfo {volume} {107}},\ \bibinfo
  {pages} {094513} (\bibinfo {year} {2023})},\ \Eprint
  {http://arxiv.org/abs/2212.04490} {arXiv:2212.04490 [hep-lat]} \BibitemShut
  {NoStop}%
\bibitem [{\citenamefont {Farrell}\ \emph
  {et~al.}(2023{\natexlab{a}})\citenamefont {Farrell}, \citenamefont
  {Chernyshev}, \citenamefont {Powell}, \citenamefont {Zemlevskiy},
  \citenamefont {Illa},\ and\ \citenamefont {Savage}}]{Farrell:2022wyt}%
  \BibitemOpen
  \bibfield  {author} {\bibinfo {author} {\bibfnamefont {R.~C.}\ \bibnamefont
  {Farrell}}, \bibinfo {author} {\bibfnamefont {I.~A.}\ \bibnamefont
  {Chernyshev}}, \bibinfo {author} {\bibfnamefont {S.~J.~M.}\ \bibnamefont
  {Powell}}, \bibinfo {author} {\bibfnamefont {N.~A.}\ \bibnamefont
  {Zemlevskiy}}, \bibinfo {author} {\bibfnamefont {M.}~\bibnamefont {Illa}}, \
  and\ \bibinfo {author} {\bibfnamefont {M.~J.}\ \bibnamefont {Savage}},\
  }\href {\doibase 10.1103/PhysRevD.107.054512} {\bibfield  {journal} {\bibinfo
   {journal} {Phys. Rev. D}\ }\textbf {\bibinfo {volume} {107}},\ \bibinfo
  {pages} {054512} (\bibinfo {year} {2023}{\natexlab{a}})},\ \Eprint
  {http://arxiv.org/abs/2207.01731} {arXiv:2207.01731 [quant-ph]} \BibitemShut
  {NoStop}%
\bibitem [{\citenamefont {Farrell}\ \emph
  {et~al.}(2023{\natexlab{b}})\citenamefont {Farrell}, \citenamefont
  {Chernyshev}, \citenamefont {Powell}, \citenamefont {Zemlevskiy},
  \citenamefont {Illa},\ and\ \citenamefont {Savage}}]{Farrell:2022vyh}%
  \BibitemOpen
  \bibfield  {author} {\bibinfo {author} {\bibfnamefont {R.~C.}\ \bibnamefont
  {Farrell}}, \bibinfo {author} {\bibfnamefont {I.~A.}\ \bibnamefont
  {Chernyshev}}, \bibinfo {author} {\bibfnamefont {S.~J.~M.}\ \bibnamefont
  {Powell}}, \bibinfo {author} {\bibfnamefont {N.~A.}\ \bibnamefont
  {Zemlevskiy}}, \bibinfo {author} {\bibfnamefont {M.}~\bibnamefont {Illa}}, \
  and\ \bibinfo {author} {\bibfnamefont {M.~J.}\ \bibnamefont {Savage}},\
  }\href {\doibase 10.1103/PhysRevD.107.054513} {\bibfield  {journal} {\bibinfo
   {journal} {Phys. Rev. D}\ }\textbf {\bibinfo {volume} {107}},\ \bibinfo
  {pages} {054513} (\bibinfo {year} {2023}{\natexlab{b}})},\ \Eprint
  {http://arxiv.org/abs/2209.10781} {arXiv:2209.10781 [quant-ph]} \BibitemShut
  {NoStop}%
\bibitem [{\citenamefont {Honda}\ \emph {et~al.}(2022)\citenamefont {Honda},
  \citenamefont {Itou},\ and\ \citenamefont {Tanizaki}}]{Honda:2022edn}%
  \BibitemOpen
  \bibfield  {author} {\bibinfo {author} {\bibfnamefont {M.}~\bibnamefont
  {Honda}}, \bibinfo {author} {\bibfnamefont {E.}~\bibnamefont {Itou}}, \ and\
  \bibinfo {author} {\bibfnamefont {Y.}~\bibnamefont {Tanizaki}},\ }\href
  {\doibase 10.1007/JHEP11(2022)141} {\bibfield  {journal} {\bibinfo  {journal}
  {JHEP}\ }\textbf {\bibinfo {volume} {11}},\ \bibinfo {pages} {141} (\bibinfo
  {year} {2022})},\ \Eprint {http://arxiv.org/abs/2210.04237} {arXiv:2210.04237
  [hep-lat]} \BibitemShut {NoStop}%
\bibitem [{\citenamefont {Cataldi}\ \emph {et~al.}(2023)\citenamefont
  {Cataldi}, \citenamefont {Magnifico}, \citenamefont {Silvi},\ and\
  \citenamefont {Montangero}}]{Cataldi:2023xki}%
  \BibitemOpen
  \bibfield  {author} {\bibinfo {author} {\bibfnamefont {G.}~\bibnamefont
  {Cataldi}}, \bibinfo {author} {\bibfnamefont {G.}~\bibnamefont {Magnifico}},
  \bibinfo {author} {\bibfnamefont {P.}~\bibnamefont {Silvi}}, \ and\ \bibinfo
  {author} {\bibfnamefont {S.}~\bibnamefont {Montangero}},\ }\href@noop {} {\
  (\bibinfo {year} {2023})},\ \Eprint {http://arxiv.org/abs/2307.09396}
  {arXiv:2307.09396 [hep-lat]} \BibitemShut {NoStop}%
\bibitem [{\citenamefont {Halimeh}\ \emph {et~al.}(2025)\citenamefont
  {Halimeh}, \citenamefont {Aidelsburger}, \citenamefont {Grusdt},
  \citenamefont {Hauke},\ and\ \citenamefont {Yang}}]{Halimeh:2023lid}%
  \BibitemOpen
  \bibfield  {author} {\bibinfo {author} {\bibfnamefont {J.~C.}\ \bibnamefont
  {Halimeh}}, \bibinfo {author} {\bibfnamefont {M.}~\bibnamefont
  {Aidelsburger}}, \bibinfo {author} {\bibfnamefont {F.}~\bibnamefont
  {Grusdt}}, \bibinfo {author} {\bibfnamefont {P.}~\bibnamefont {Hauke}}, \
  and\ \bibinfo {author} {\bibfnamefont {B.}~\bibnamefont {Yang}},\ }\href
  {\doibase 10.1038/s41567-024-02721-8} {\bibfield  {journal} {\bibinfo
  {journal} {Nature Phys.}\ }\textbf {\bibinfo {volume} {21}},\ \bibinfo
  {pages} {25} (\bibinfo {year} {2025})},\ \Eprint
  {http://arxiv.org/abs/2310.12201} {arXiv:2310.12201 [cond-mat.quant-gas]}
  \BibitemShut {NoStop}%
\bibitem [{\citenamefont {Liu}\ \emph {et~al.}(2023)\citenamefont {Liu},
  \citenamefont {Bhattacharya}, \citenamefont {Chandrasekharan},\ and\
  \citenamefont {Gupta}}]{Liu:2023lsr}%
  \BibitemOpen
  \bibfield  {author} {\bibinfo {author} {\bibfnamefont {H.}~\bibnamefont
  {Liu}}, \bibinfo {author} {\bibfnamefont {T.}~\bibnamefont {Bhattacharya}},
  \bibinfo {author} {\bibfnamefont {S.}~\bibnamefont {Chandrasekharan}}, \ and\
  \bibinfo {author} {\bibfnamefont {R.}~\bibnamefont {Gupta}},\ }\href@noop {}
  {\  (\bibinfo {year} {2023})},\ \Eprint {http://arxiv.org/abs/2312.17734}
  {arXiv:2312.17734 [hep-lat]} \BibitemShut {NoStop}%
\bibitem [{\citenamefont {Hayata}\ \emph {et~al.}(2024)\citenamefont {Hayata},
  \citenamefont {Hidaka},\ and\ \citenamefont {Nishimura}}]{Hayata:2023pkw}%
  \BibitemOpen
  \bibfield  {author} {\bibinfo {author} {\bibfnamefont {T.}~\bibnamefont
  {Hayata}}, \bibinfo {author} {\bibfnamefont {Y.}~\bibnamefont {Hidaka}}, \
  and\ \bibinfo {author} {\bibfnamefont {K.}~\bibnamefont {Nishimura}},\ }\href
  {\doibase 10.1007/JHEP07(2024)106} {\bibfield  {journal} {\bibinfo  {journal}
  {JHEP}\ }\textbf {\bibinfo {volume} {07}},\ \bibinfo {pages} {106} (\bibinfo
  {year} {2024})},\ \Eprint {http://arxiv.org/abs/2311.11643} {arXiv:2311.11643
  [hep-lat]} \BibitemShut {NoStop}%
\bibitem [{\citenamefont {Farrell}\ \emph
  {et~al.}(2024{\natexlab{a}})\citenamefont {Farrell}, \citenamefont {Illa},
  \citenamefont {Ciavarella},\ and\ \citenamefont {Savage}}]{Farrell:2023fgd}%
  \BibitemOpen
  \bibfield  {author} {\bibinfo {author} {\bibfnamefont {R.~C.}\ \bibnamefont
  {Farrell}}, \bibinfo {author} {\bibfnamefont {M.}~\bibnamefont {Illa}},
  \bibinfo {author} {\bibfnamefont {A.~N.}\ \bibnamefont {Ciavarella}}, \ and\
  \bibinfo {author} {\bibfnamefont {M.~J.}\ \bibnamefont {Savage}},\ }\href
  {\doibase 10.1103/PRXQuantum.5.020315} {\bibfield  {journal} {\bibinfo
  {journal} {PRX Quantum}\ }\textbf {\bibinfo {volume} {5}},\ \bibinfo {pages}
  {020315} (\bibinfo {year} {2024}{\natexlab{a}})},\ \Eprint
  {http://arxiv.org/abs/2308.04481} {arXiv:2308.04481 [quant-ph]} \BibitemShut
  {NoStop}%
\bibitem [{\citenamefont {Farrell}\ \emph
  {et~al.}(2024{\natexlab{b}})\citenamefont {Farrell}, \citenamefont {Illa},
  \citenamefont {Ciavarella},\ and\ \citenamefont {Savage}}]{Farrell:2024fit}%
  \BibitemOpen
  \bibfield  {author} {\bibinfo {author} {\bibfnamefont {R.~C.}\ \bibnamefont
  {Farrell}}, \bibinfo {author} {\bibfnamefont {M.}~\bibnamefont {Illa}},
  \bibinfo {author} {\bibfnamefont {A.~N.}\ \bibnamefont {Ciavarella}}, \ and\
  \bibinfo {author} {\bibfnamefont {M.~J.}\ \bibnamefont {Savage}},\ }\href
  {\doibase 10.1103/PhysRevD.109.114510} {\bibfield  {journal} {\bibinfo
  {journal} {Phys. Rev. D}\ }\textbf {\bibinfo {volume} {109}},\ \bibinfo
  {pages} {114510} (\bibinfo {year} {2024}{\natexlab{b}})},\ \Eprint
  {http://arxiv.org/abs/2401.08044} {arXiv:2401.08044 [quant-ph]} \BibitemShut
  {NoStop}%
\bibitem [{\citenamefont {Watson}\ \emph {et~al.}(2023)\citenamefont {Watson},
  \citenamefont {Bringewatt}, \citenamefont {Shaw}, \citenamefont {Childs},
  \citenamefont {Gorshkov},\ and\ \citenamefont {Davoudi}}]{Watson:2023oov}%
  \BibitemOpen
  \bibfield  {author} {\bibinfo {author} {\bibfnamefont {J.~D.}\ \bibnamefont
  {Watson}}, \bibinfo {author} {\bibfnamefont {J.}~\bibnamefont {Bringewatt}},
  \bibinfo {author} {\bibfnamefont {A.~F.}\ \bibnamefont {Shaw}}, \bibinfo
  {author} {\bibfnamefont {A.~M.}\ \bibnamefont {Childs}}, \bibinfo {author}
  {\bibfnamefont {A.~V.}\ \bibnamefont {Gorshkov}}, \ and\ \bibinfo {author}
  {\bibfnamefont {Z.}~\bibnamefont {Davoudi}},\ }\href@noop {} {\  (\bibinfo
  {year} {2023})},\ \Eprint {http://arxiv.org/abs/2312.05344} {arXiv:2312.05344
  [quant-ph]} \BibitemShut {NoStop}%
\bibitem [{\citenamefont {Kavaki}\ and\ \citenamefont
  {Lewis}(2024)}]{Kavaki:2024ijd}%
  \BibitemOpen
  \bibfield  {author} {\bibinfo {author} {\bibfnamefont {A.~H.~Z.}\
  \bibnamefont {Kavaki}}\ and\ \bibinfo {author} {\bibfnamefont
  {R.}~\bibnamefont {Lewis}},\ }\href@noop {} {\  (\bibinfo {year} {2024})},\
  \Eprint {http://arxiv.org/abs/2401.14570} {arXiv:2401.14570 [hep-lat]}
  \BibitemShut {NoStop}%
\bibitem [{\citenamefont {Ciavarella}\ and\ \citenamefont
  {Bauer}(2024)}]{Ciavarella:2024fzw}%
  \BibitemOpen
  \bibfield  {author} {\bibinfo {author} {\bibfnamefont {A.~N.}\ \bibnamefont
  {Ciavarella}}\ and\ \bibinfo {author} {\bibfnamefont {C.~W.}\ \bibnamefont
  {Bauer}},\ }\href {\doibase 10.1103/PhysRevLett.133.111901} {\bibfield
  {journal} {\bibinfo  {journal} {Phys. Rev. Lett.}\ }\textbf {\bibinfo
  {volume} {133}},\ \bibinfo {pages} {111901} (\bibinfo {year} {2024})},\
  \Eprint {http://arxiv.org/abs/2402.10265} {arXiv:2402.10265 [hep-ph]}
  \BibitemShut {NoStop}%
\bibitem [{\citenamefont {Illa}\ \emph {et~al.}(2024)\citenamefont {Illa},
  \citenamefont {Robin},\ and\ \citenamefont {Savage}}]{Illa:2024kmf}%
  \BibitemOpen
  \bibfield  {author} {\bibinfo {author} {\bibfnamefont {M.}~\bibnamefont
  {Illa}}, \bibinfo {author} {\bibfnamefont {C.~E.~P.}\ \bibnamefont {Robin}},
  \ and\ \bibinfo {author} {\bibfnamefont {M.~J.}\ \bibnamefont {Savage}},\
  }\href {\doibase 10.1103/PhysRevD.110.014507} {\bibfield  {journal} {\bibinfo
   {journal} {Phys. Rev. D}\ }\textbf {\bibinfo {volume} {110}},\ \bibinfo
  {pages} {014507} (\bibinfo {year} {2024})},\ \Eprint
  {http://arxiv.org/abs/2403.14537} {arXiv:2403.14537 [quant-ph]} \BibitemShut
  {NoStop}%
\bibitem [{\citenamefont {Farrell}\ \emph {et~al.}(2025)\citenamefont
  {Farrell}, \citenamefont {Illa},\ and\ \citenamefont
  {Savage}}]{Farrell:2024mgu}%
  \BibitemOpen
  \bibfield  {author} {\bibinfo {author} {\bibfnamefont {R.~C.}\ \bibnamefont
  {Farrell}}, \bibinfo {author} {\bibfnamefont {M.}~\bibnamefont {Illa}}, \
  and\ \bibinfo {author} {\bibfnamefont {M.~J.}\ \bibnamefont {Savage}},\
  }\href {\doibase 10.1103/PhysRevC.111.015202} {\bibfield  {journal} {\bibinfo
   {journal} {Phys. Rev. C}\ }\textbf {\bibinfo {volume} {111}},\ \bibinfo
  {pages} {015202} (\bibinfo {year} {2025})},\ \Eprint
  {http://arxiv.org/abs/2405.06620} {arXiv:2405.06620 [quant-ph]} \BibitemShut
  {NoStop}%
\bibitem [{\citenamefont {Lamm}\ \emph {et~al.}(2024)\citenamefont {Lamm},
  \citenamefont {Li}, \citenamefont {Shu}, \citenamefont {Wang},\ and\
  \citenamefont {Xu}}]{Lamm:2024jnl}%
  \BibitemOpen
  \bibfield  {author} {\bibinfo {author} {\bibfnamefont {H.}~\bibnamefont
  {Lamm}}, \bibinfo {author} {\bibfnamefont {Y.-Y.}\ \bibnamefont {Li}},
  \bibinfo {author} {\bibfnamefont {J.}~\bibnamefont {Shu}}, \bibinfo {author}
  {\bibfnamefont {Y.-L.}\ \bibnamefont {Wang}}, \ and\ \bibinfo {author}
  {\bibfnamefont {B.}~\bibnamefont {Xu}},\ }\href {\doibase
  10.1103/PhysRevD.110.054505} {\bibfield  {journal} {\bibinfo  {journal}
  {Phys. Rev. D}\ }\textbf {\bibinfo {volume} {110}},\ \bibinfo {pages}
  {054505} (\bibinfo {year} {2024})},\ \Eprint
  {http://arxiv.org/abs/2405.12890} {arXiv:2405.12890 [hep-lat]} \BibitemShut
  {NoStop}%
\bibitem [{\citenamefont {Li}\ \emph {et~al.}(2024)\citenamefont {Li},
  \citenamefont {Grabowska},\ and\ \citenamefont {Savage}}]{Li:2024lrl}%
  \BibitemOpen
  \bibfield  {author} {\bibinfo {author} {\bibfnamefont {Z.}~\bibnamefont
  {Li}}, \bibinfo {author} {\bibfnamefont {D.~M.}\ \bibnamefont {Grabowska}}, \
  and\ \bibinfo {author} {\bibfnamefont {M.~J.}\ \bibnamefont {Savage}},\
  }\href@noop {} {\  (\bibinfo {year} {2024})},\ \Eprint
  {http://arxiv.org/abs/2407.13835} {arXiv:2407.13835 [quant-ph]} \BibitemShut
  {NoStop}%
\bibitem [{\citenamefont {Florio}\ \emph {et~al.}(2024)\citenamefont {Florio},
  \citenamefont {Frenklakh}, \citenamefont {Ikeda}, \citenamefont {Kharzeev},
  \citenamefont {Korepin}, \citenamefont {Shi},\ and\ \citenamefont
  {Yu}}]{Florio:2024aix}%
  \BibitemOpen
  \bibfield  {author} {\bibinfo {author} {\bibfnamefont {A.}~\bibnamefont
  {Florio}}, \bibinfo {author} {\bibfnamefont {D.}~\bibnamefont {Frenklakh}},
  \bibinfo {author} {\bibfnamefont {K.}~\bibnamefont {Ikeda}}, \bibinfo
  {author} {\bibfnamefont {D.~E.}\ \bibnamefont {Kharzeev}}, \bibinfo {author}
  {\bibfnamefont {V.}~\bibnamefont {Korepin}}, \bibinfo {author} {\bibfnamefont
  {S.}~\bibnamefont {Shi}}, \ and\ \bibinfo {author} {\bibfnamefont
  {K.}~\bibnamefont {Yu}},\ }\href {\doibase 10.1103/PhysRevD.110.094029}
  {\bibfield  {journal} {\bibinfo  {journal} {Phys. Rev. D}\ }\textbf {\bibinfo
  {volume} {110}},\ \bibinfo {pages} {094029} (\bibinfo {year} {2024})},\
  \Eprint {http://arxiv.org/abs/2404.00087} {arXiv:2404.00087 [hep-ph]}
  \BibitemShut {NoStop}%
\bibitem [{\citenamefont {Kadam}\ \emph {et~al.}(2024)\citenamefont {Kadam},
  \citenamefont {Naskar}, \citenamefont {Raychowdhury},\ and\ \citenamefont
  {Stryker}}]{Kadam:2024zkj}%
  \BibitemOpen
  \bibfield  {author} {\bibinfo {author} {\bibfnamefont {S.~V.}\ \bibnamefont
  {Kadam}}, \bibinfo {author} {\bibfnamefont {A.}~\bibnamefont {Naskar}},
  \bibinfo {author} {\bibfnamefont {I.}~\bibnamefont {Raychowdhury}}, \ and\
  \bibinfo {author} {\bibfnamefont {J.~R.}\ \bibnamefont {Stryker}},\
  }\href@noop {} {\  (\bibinfo {year} {2024})},\ \Eprint
  {http://arxiv.org/abs/2407.19181} {arXiv:2407.19181 [hep-lat]} \BibitemShut
  {NoStop}%
\bibitem [{\citenamefont {Gustafson}\ \emph {et~al.}(2024)\citenamefont
  {Gustafson} \emph {et~al.}}]{Gustafson:2024bww}%
  \BibitemOpen
  \bibfield  {author} {\bibinfo {author} {\bibfnamefont {E.}~\bibnamefont
  {Gustafson}} \emph {et~al.},\ }\href@noop {} {\  (\bibinfo {year} {2024})},\
  \Eprint {http://arxiv.org/abs/2408.12641} {arXiv:2408.12641 [quant-ph]}
  \BibitemShut {NoStop}%
\bibitem [{\citenamefont {Fontana}\ \emph {et~al.}(2024)\citenamefont
  {Fontana}, \citenamefont {Riaza},\ and\ \citenamefont
  {Celi}}]{Fontana:2024rux}%
  \BibitemOpen
  \bibfield  {author} {\bibinfo {author} {\bibfnamefont {P.}~\bibnamefont
  {Fontana}}, \bibinfo {author} {\bibfnamefont {M.~M.}\ \bibnamefont {Riaza}},
  \ and\ \bibinfo {author} {\bibfnamefont {A.}~\bibnamefont {Celi}},\
  }\href@noop {} {\  (\bibinfo {year} {2024})},\ \Eprint
  {http://arxiv.org/abs/2409.04441} {arXiv:2409.04441 [quant-ph]} \BibitemShut
  {NoStop}%
\bibitem [{\citenamefont {Lee}\ \emph {et~al.}(2024)\citenamefont {Lee},
  \citenamefont {Turro},\ and\ \citenamefont {Yao}}]{Lee:2024jnt}%
  \BibitemOpen
  \bibfield  {author} {\bibinfo {author} {\bibfnamefont {K.}~\bibnamefont
  {Lee}}, \bibinfo {author} {\bibfnamefont {F.}~\bibnamefont {Turro}}, \ and\
  \bibinfo {author} {\bibfnamefont {X.}~\bibnamefont {Yao}},\ }\href@noop {} {\
   (\bibinfo {year} {2024})},\ \Eprint {http://arxiv.org/abs/2409.13830}
  {arXiv:2409.13830 [hep-ph]} \BibitemShut {NoStop}%
\bibitem [{\citenamefont {Burbano}\ and\ \citenamefont
  {Bauer}(2024)}]{Burbano:2024uvn}%
  \BibitemOpen
  \bibfield  {author} {\bibinfo {author} {\bibfnamefont {I.~M.}\ \bibnamefont
  {Burbano}}\ and\ \bibinfo {author} {\bibfnamefont {C.~W.}\ \bibnamefont
  {Bauer}},\ }\href@noop {} {\  (\bibinfo {year} {2024})},\ \Eprint
  {http://arxiv.org/abs/2409.13812} {arXiv:2409.13812 [hep-lat]} \BibitemShut
  {NoStop}%
\bibitem [{\citenamefont {Ciavarella}(2024)}]{Ciavarella:2024lsp}%
  \BibitemOpen
  \bibfield  {author} {\bibinfo {author} {\bibfnamefont {A.~N.}\ \bibnamefont
  {Ciavarella}},\ }\href@noop {} {\  (\bibinfo {year} {2024})},\ \Eprint
  {http://arxiv.org/abs/2411.05915} {arXiv:2411.05915 [quant-ph]} \BibitemShut
  {NoStop}%
\bibitem [{\citenamefont {Araz}\ \emph {et~al.}(2024)\citenamefont {Araz},
  \citenamefont {Grau}, \citenamefont {Montgomery},\ and\ \citenamefont
  {Ringer}}]{Araz:2024kkg}%
  \BibitemOpen
  \bibfield  {author} {\bibinfo {author} {\bibfnamefont {J.~Y.}\ \bibnamefont
  {Araz}}, \bibinfo {author} {\bibfnamefont {M.}~\bibnamefont {Grau}}, \bibinfo
  {author} {\bibfnamefont {J.}~\bibnamefont {Montgomery}}, \ and\ \bibinfo
  {author} {\bibfnamefont {F.}~\bibnamefont {Ringer}},\ }\href@noop {} {\
  (\bibinfo {year} {2024})},\ \Eprint {http://arxiv.org/abs/2410.07346}
  {arXiv:2410.07346 [quant-ph]} \BibitemShut {NoStop}%
\bibitem [{\citenamefont {Zhang}\ \emph {et~al.}(2024)\citenamefont {Zhang},
  \citenamefont {Guo}, \citenamefont {Wang},\ and\ \citenamefont
  {Xing}}]{Zhang:2024fgv}%
  \BibitemOpen
  \bibfield  {author} {\bibinfo {author} {\bibfnamefont {G.}~\bibnamefont
  {Zhang}}, \bibinfo {author} {\bibfnamefont {X.}~\bibnamefont {Guo}}, \bibinfo
  {author} {\bibfnamefont {E.}~\bibnamefont {Wang}}, \ and\ \bibinfo {author}
  {\bibfnamefont {H.}~\bibnamefont {Xing}} (\bibinfo {collaboration} {QuNu}),\
  }\href@noop {} {\  (\bibinfo {year} {2024})},\ \Eprint
  {http://arxiv.org/abs/2411.18869} {arXiv:2411.18869 [hep-ph]} \BibitemShut
  {NoStop}%
\bibitem [{\citenamefont {Dhaulakhandi}\ \emph {et~al.}(2024)\citenamefont
  {Dhaulakhandi}, \citenamefont {Das}, \citenamefont {Behera},\ and\
  \citenamefont {Seo}}]{Dhaulakhandi:2024tox}%
  \BibitemOpen
  \bibfield  {author} {\bibinfo {author} {\bibfnamefont {R.}~\bibnamefont
  {Dhaulakhandi}}, \bibinfo {author} {\bibfnamefont {R.}~\bibnamefont {Das}},
  \bibinfo {author} {\bibfnamefont {B.~K.}\ \bibnamefont {Behera}}, \ and\
  \bibinfo {author} {\bibfnamefont {F.~J.}\ \bibnamefont {Seo}},\ }\href
  {\doibase 10.1063/5.0231558} {\bibfield  {journal} {\bibinfo  {journal} {AIP
  Adv.}\ }\textbf {\bibinfo {volume} {14}},\ \bibinfo {pages} {125121}
  (\bibinfo {year} {2024})}\BibitemShut {NoStop}%
\bibitem [{\citenamefont {Zemlevskiy}(2024)}]{Zemlevskiy:2024vxt}%
  \BibitemOpen
  \bibfield  {author} {\bibinfo {author} {\bibfnamefont {N.~A.}\ \bibnamefont
  {Zemlevskiy}},\ }\href@noop {} {\  (\bibinfo {year} {2024})},\ \Eprint
  {http://arxiv.org/abs/2411.02486} {arXiv:2411.02486 [quant-ph]} \BibitemShut
  {NoStop}%
\bibitem [{\citenamefont {Mueller}\ \emph {et~al.}(2024)\citenamefont
  {Mueller}, \citenamefont {Wang}, \citenamefont {Katz}, \citenamefont
  {Davoudi},\ and\ \citenamefont {Cetina}}]{Mueller:2024mmk}%
  \BibitemOpen
  \bibfield  {author} {\bibinfo {author} {\bibfnamefont {N.}~\bibnamefont
  {Mueller}}, \bibinfo {author} {\bibfnamefont {T.}~\bibnamefont {Wang}},
  \bibinfo {author} {\bibfnamefont {O.}~\bibnamefont {Katz}}, \bibinfo {author}
  {\bibfnamefont {Z.}~\bibnamefont {Davoudi}}, \ and\ \bibinfo {author}
  {\bibfnamefont {M.}~\bibnamefont {Cetina}},\ }\href@noop {} {\  (\bibinfo
  {year} {2024})},\ \Eprint {http://arxiv.org/abs/2408.00069} {arXiv:2408.00069
  [quant-ph]} \BibitemShut {NoStop}%
\bibitem [{\citenamefont {Davoudi}\ \emph {et~al.}(2024)\citenamefont
  {Davoudi}, \citenamefont {Jarzynski}, \citenamefont {Mueller}, \citenamefont
  {Oruganti}, \citenamefont {Powers},\ and\ \citenamefont
  {Halpern}}]{Davoudi:2024osg}%
  \BibitemOpen
  \bibfield  {author} {\bibinfo {author} {\bibfnamefont {Z.}~\bibnamefont
  {Davoudi}}, \bibinfo {author} {\bibfnamefont {C.}~\bibnamefont {Jarzynski}},
  \bibinfo {author} {\bibfnamefont {N.}~\bibnamefont {Mueller}}, \bibinfo
  {author} {\bibfnamefont {G.}~\bibnamefont {Oruganti}}, \bibinfo {author}
  {\bibfnamefont {C.}~\bibnamefont {Powers}}, \ and\ \bibinfo {author}
  {\bibfnamefont {N.~Y.}\ \bibnamefont {Halpern}},\ }\href {\doibase
  10.1103/PhysRevLett.133.250402} {\bibfield  {journal} {\bibinfo  {journal}
  {Phys. Rev. Lett.}\ }\textbf {\bibinfo {volume} {133}},\ \bibinfo {pages}
  {250402} (\bibinfo {year} {2024})},\ \Eprint
  {http://arxiv.org/abs/2404.02965} {arXiv:2404.02965 [quant-ph]} \BibitemShut
  {NoStop}%
\bibitem [{\citenamefont {Jeyaretnam}\ \emph {et~al.}(2024)\citenamefont
  {Jeyaretnam}, \citenamefont {Bhore}, \citenamefont {Osborne}, \citenamefont
  {Halimeh},\ and\ \citenamefont {Papi\'c}}]{Jeyaretnam:2024tkj}%
  \BibitemOpen
  \bibfield  {author} {\bibinfo {author} {\bibfnamefont {J.}~\bibnamefont
  {Jeyaretnam}}, \bibinfo {author} {\bibfnamefont {T.}~\bibnamefont {Bhore}},
  \bibinfo {author} {\bibfnamefont {J.~J.}\ \bibnamefont {Osborne}}, \bibinfo
  {author} {\bibfnamefont {J.~C.}\ \bibnamefont {Halimeh}}, \ and\ \bibinfo
  {author} {\bibfnamefont {Z.}~\bibnamefont {Papi\'c}},\ }\href@noop {} {\
  (\bibinfo {year} {2024})},\ \Eprint {http://arxiv.org/abs/2409.08320}
  {arXiv:2409.08320 [quant-ph]} \BibitemShut {NoStop}%
\bibitem [{\citenamefont {Araz}\ \emph {et~al.}(2025)\citenamefont {Araz},
  \citenamefont {Bhowmick}, \citenamefont {Grau}, \citenamefont {McEntire},\
  and\ \citenamefont {Ringer}}]{Araz:2024bgg}%
  \BibitemOpen
  \bibfield  {author} {\bibinfo {author} {\bibfnamefont {J.~Y.}\ \bibnamefont
  {Araz}}, \bibinfo {author} {\bibfnamefont {S.}~\bibnamefont {Bhowmick}},
  \bibinfo {author} {\bibfnamefont {M.}~\bibnamefont {Grau}}, \bibinfo {author}
  {\bibfnamefont {T.~J.}\ \bibnamefont {McEntire}}, \ and\ \bibinfo {author}
  {\bibfnamefont {F.}~\bibnamefont {Ringer}},\ }\href {\doibase
  10.1103/PhysRevD.111.034506} {\bibfield  {journal} {\bibinfo  {journal}
  {Phys. Rev. D}\ }\textbf {\bibinfo {volume} {111}},\ \bibinfo {pages}
  {034506} (\bibinfo {year} {2025})},\ \Eprint
  {http://arxiv.org/abs/2407.17556} {arXiv:2407.17556 [quant-ph]} \BibitemShut
  {NoStop}%
\bibitem [{\citenamefont {Itou}\ \emph {et~al.}(2024)\citenamefont {Itou},
  \citenamefont {Matsumoto},\ and\ \citenamefont {Tanizaki}}]{Itou:2024psm}%
  \BibitemOpen
  \bibfield  {author} {\bibinfo {author} {\bibfnamefont {E.}~\bibnamefont
  {Itou}}, \bibinfo {author} {\bibfnamefont {A.}~\bibnamefont {Matsumoto}}, \
  and\ \bibinfo {author} {\bibfnamefont {Y.}~\bibnamefont {Tanizaki}},\ }\href
  {\doibase 10.1007/JHEP09(2024)155} {\bibfield  {journal} {\bibinfo  {journal}
  {JHEP}\ }\textbf {\bibinfo {volume} {09}},\ \bibinfo {pages} {155} (\bibinfo
  {year} {2024})},\ \Eprint {http://arxiv.org/abs/2407.11391} {arXiv:2407.11391
  [hep-lat]} \BibitemShut {NoStop}%
\bibitem [{\citenamefont {Lin}\ \emph {et~al.}(2024)\citenamefont {Lin},
  \citenamefont {Luo}, \citenamefont {Yao},\ and\ \citenamefont
  {Shanahan}}]{Lin:2024eiz}%
  \BibitemOpen
  \bibfield  {author} {\bibinfo {author} {\bibfnamefont {J.}~\bibnamefont
  {Lin}}, \bibinfo {author} {\bibfnamefont {D.}~\bibnamefont {Luo}}, \bibinfo
  {author} {\bibfnamefont {X.}~\bibnamefont {Yao}}, \ and\ \bibinfo {author}
  {\bibfnamefont {P.~E.}\ \bibnamefont {Shanahan}},\ }\href {\doibase
  10.1007/JHEP06(2024)211} {\bibfield  {journal} {\bibinfo  {journal} {JHEP}\
  }\textbf {\bibinfo {volume} {06}},\ \bibinfo {pages} {211} (\bibinfo {year}
  {2024})},\ \Eprint {http://arxiv.org/abs/2402.06607} {arXiv:2402.06607
  [hep-ph]} \BibitemShut {NoStop}%
\bibitem [{\citenamefont {Ballini}\ \emph {et~al.}(2024)\citenamefont
  {Ballini}, \citenamefont {Mildenberger}, \citenamefont {Wauters},\ and\
  \citenamefont {Hauke}}]{Ballini:2024qmr}%
  \BibitemOpen
  \bibfield  {author} {\bibinfo {author} {\bibfnamefont {E.}~\bibnamefont
  {Ballini}}, \bibinfo {author} {\bibfnamefont {J.}~\bibnamefont
  {Mildenberger}}, \bibinfo {author} {\bibfnamefont {M.~M.}\ \bibnamefont
  {Wauters}}, \ and\ \bibinfo {author} {\bibfnamefont {P.}~\bibnamefont
  {Hauke}},\ }\href@noop {} {\  (\bibinfo {year} {2024})},\ \Eprint
  {http://arxiv.org/abs/2412.07844} {arXiv:2412.07844 [quant-ph]} \BibitemShut
  {NoStop}%
\bibitem [{\citenamefont {Ciavarella}\ \emph {et~al.}(2025)\citenamefont
  {Ciavarella}, \citenamefont {Bauer},\ and\ \citenamefont
  {Halimeh}}]{Ciavarella:2025zqf}%
  \BibitemOpen
  \bibfield  {author} {\bibinfo {author} {\bibfnamefont {A.~N.}\ \bibnamefont
  {Ciavarella}}, \bibinfo {author} {\bibfnamefont {C.~W.}\ \bibnamefont
  {Bauer}}, \ and\ \bibinfo {author} {\bibfnamefont {J.~C.}\ \bibnamefont
  {Halimeh}},\ }\href@noop {} {\  (\bibinfo {year} {2025})},\ \Eprint
  {http://arxiv.org/abs/2502.03533} {arXiv:2502.03533 [quant-ph]} \BibitemShut
  {NoStop}%
\bibitem [{\citenamefont {Janik}\ \emph {et~al.}(2025)\citenamefont {Janik},
  \citenamefont {Nowak}, \citenamefont {Rams},\ and\ \citenamefont
  {Zahed}}]{Janik:2025bbz}%
  \BibitemOpen
  \bibfield  {author} {\bibinfo {author} {\bibfnamefont {R.~A.}\ \bibnamefont
  {Janik}}, \bibinfo {author} {\bibfnamefont {M.~A.}\ \bibnamefont {Nowak}},
  \bibinfo {author} {\bibfnamefont {M.~M.}\ \bibnamefont {Rams}}, \ and\
  \bibinfo {author} {\bibfnamefont {I.}~\bibnamefont {Zahed}},\ }\href@noop {}
  {\  (\bibinfo {year} {2025})},\ \Eprint {http://arxiv.org/abs/2502.12901}
  {arXiv:2502.12901 [hep-ph]} \BibitemShut {NoStop}%
\bibitem [{\citenamefont {Chandrasekharan}\ \emph {et~al.}(2025)\citenamefont
  {Chandrasekharan}, \citenamefont {Siew},\ and\ \citenamefont
  {Bhattacharya}}]{Chandrasekharan:2025smw}%
  \BibitemOpen
  \bibfield  {author} {\bibinfo {author} {\bibfnamefont {S.}~\bibnamefont
  {Chandrasekharan}}, \bibinfo {author} {\bibfnamefont {R.~X.}\ \bibnamefont
  {Siew}}, \ and\ \bibinfo {author} {\bibfnamefont {T.}~\bibnamefont
  {Bhattacharya}},\ }\href@noop {} {\  (\bibinfo {year} {2025})},\ \Eprint
  {http://arxiv.org/abs/2502.14175} {arXiv:2502.14175 [hep-lat]} \BibitemShut
  {NoStop}%
\bibitem [{\citenamefont {Turro}\ \emph {et~al.}(2022)\citenamefont {Turro},
  \citenamefont {Roggero}, \citenamefont {Amitrano}, \citenamefont {Luchi},
  \citenamefont {Wendt}, \citenamefont {Dubois}, \citenamefont {Quaglioni},\
  and\ \citenamefont {Pederiva}}]{turro_QITP_2022}%
  \BibitemOpen
  \bibfield  {author} {\bibinfo {author} {\bibfnamefont {F.}~\bibnamefont
  {Turro}}, \bibinfo {author} {\bibfnamefont {A.}~\bibnamefont {Roggero}},
  \bibinfo {author} {\bibfnamefont {V.}~\bibnamefont {Amitrano}}, \bibinfo
  {author} {\bibfnamefont {P.}~\bibnamefont {Luchi}}, \bibinfo {author}
  {\bibfnamefont {K.~A.}\ \bibnamefont {Wendt}}, \bibinfo {author}
  {\bibfnamefont {J.~L.}\ \bibnamefont {Dubois}}, \bibinfo {author}
  {\bibfnamefont {S.}~\bibnamefont {Quaglioni}}, \ and\ \bibinfo {author}
  {\bibfnamefont {F.}~\bibnamefont {Pederiva}},\ }\href {\doibase
  10.1103/PhysRevA.105.022440} {\bibfield  {journal} {\bibinfo  {journal}
  {Phys. Rev. A}\ }\textbf {\bibinfo {volume} {105}},\ \bibinfo {pages}
  {022440} (\bibinfo {year} {2022})}\BibitemShut {NoStop}%
\bibitem [{\citenamefont {Turro}(2023)}]{franceschino_thermal2023}%
  \BibitemOpen
  \bibfield  {author} {\bibinfo {author} {\bibfnamefont {F.}~\bibnamefont
  {Turro}},\ }\href {\doibase 10.48550/arXiv.2306.16580} {\enquote {\bibinfo
  {title} {Quantum imaginary time propagation algorithm for preparing thermal
  states},}\ } (\bibinfo {year} {2023}),\ \Eprint
  {http://arxiv.org/abs/2306.16580} {arXiv:2306.16580 [quant-ph]} \BibitemShut
  {NoStop}%
\bibitem [{\citenamefont {Motta}\ \emph {et~al.}(2019)\citenamefont {Motta},
  \citenamefont {Sun}, \citenamefont {Tan}, \citenamefont {Rourke},
  \citenamefont {Ye}, \citenamefont {Minnich}, \citenamefont {Brand\~ao},\ and\
  \citenamefont {Chan}}]{Motta:2019yya}%
  \BibitemOpen
  \bibfield  {author} {\bibinfo {author} {\bibfnamefont {M.}~\bibnamefont
  {Motta}}, \bibinfo {author} {\bibfnamefont {C.}~\bibnamefont {Sun}}, \bibinfo
  {author} {\bibfnamefont {A.~T.~K.}\ \bibnamefont {Tan}}, \bibinfo {author}
  {\bibfnamefont {M.~J.~O.}\ \bibnamefont {Rourke}}, \bibinfo {author}
  {\bibfnamefont {E.}~\bibnamefont {Ye}}, \bibinfo {author} {\bibfnamefont
  {A.~J.}\ \bibnamefont {Minnich}}, \bibinfo {author} {\bibfnamefont {F.~G.
  S.~L.}\ \bibnamefont {Brand\~ao}}, \ and\ \bibinfo {author} {\bibfnamefont
  {G.~K.-L.}\ \bibnamefont {Chan}},\ }\href {\doibase
  10.1038/s41567-019-0704-4} {\bibfield  {journal} {\bibinfo  {journal} {Nature
  Phys.}\ }\textbf {\bibinfo {volume} {16}},\ \bibinfo {pages} {205} (\bibinfo
  {year} {2019})},\ \Eprint {http://arxiv.org/abs/1901.07653} {arXiv:1901.07653
  [quant-ph]} \BibitemShut {NoStop}%
\bibitem [{\citenamefont {Low}\ and\ \citenamefont
  {Chuang}(2019)}]{Low2019hamiltonian}%
  \BibitemOpen
  \bibfield  {author} {\bibinfo {author} {\bibfnamefont {G.~H.}\ \bibnamefont
  {Low}}\ and\ \bibinfo {author} {\bibfnamefont {I.~L.}\ \bibnamefont
  {Chuang}},\ }\href {\doibase 10.22331/q-2019-07-12-163} {\bibfield  {journal}
  {\bibinfo  {journal} {{Quantum}}\ }\textbf {\bibinfo {volume} {3}},\ \bibinfo
  {pages} {163} (\bibinfo {year} {2019})}\BibitemShut {NoStop}%
\bibitem [{\citenamefont {Chen}\ \emph
  {et~al.}(2023{\natexlab{a}})\citenamefont {Chen}, \citenamefont {Kastoryano},
  \citenamefont {Brand\~ao},\ and\ \citenamefont {Gily\'en}}]{Chen:2023cuc}%
  \BibitemOpen
  \bibfield  {author} {\bibinfo {author} {\bibfnamefont {C.-F.}\ \bibnamefont
  {Chen}}, \bibinfo {author} {\bibfnamefont {M.~J.}\ \bibnamefont
  {Kastoryano}}, \bibinfo {author} {\bibfnamefont {F.~G. S.~L.}\ \bibnamefont
  {Brand\~ao}}, \ and\ \bibinfo {author} {\bibfnamefont {A.}~\bibnamefont
  {Gily\'en}},\ }\href@noop {} {\  (\bibinfo {year} {2023}{\natexlab{a}})},\
  \Eprint {http://arxiv.org/abs/2303.18224} {arXiv:2303.18224 [quant-ph]}
  \BibitemShut {NoStop}%
\bibitem [{\citenamefont {Chen}\ \emph
  {et~al.}(2023{\natexlab{b}})\citenamefont {Chen}, \citenamefont
  {Kastoryano},\ and\ \citenamefont {Gily\'en}}]{Chen:2023zpu}%
  \BibitemOpen
  \bibfield  {author} {\bibinfo {author} {\bibfnamefont {C.-F.}\ \bibnamefont
  {Chen}}, \bibinfo {author} {\bibfnamefont {M.~J.}\ \bibnamefont
  {Kastoryano}}, \ and\ \bibinfo {author} {\bibfnamefont {A.}~\bibnamefont
  {Gily\'en}},\ }\href@noop {} {\  (\bibinfo {year} {2023}{\natexlab{b}})},\
  \Eprint {http://arxiv.org/abs/2311.09207} {arXiv:2311.09207 [quant-ph]}
  \BibitemShut {NoStop}%
\bibitem [{\citenamefont {Ding}\ \emph {et~al.}(2024)\citenamefont {Ding},
  \citenamefont {Li},\ and\ \citenamefont {Lin}}]{Ding:2024mxo}%
  \BibitemOpen
  \bibfield  {author} {\bibinfo {author} {\bibfnamefont {Z.}~\bibnamefont
  {Ding}}, \bibinfo {author} {\bibfnamefont {B.}~\bibnamefont {Li}}, \ and\
  \bibinfo {author} {\bibfnamefont {L.}~\bibnamefont {Lin}},\ }\href@noop {} {\
   (\bibinfo {year} {2024})},\ \Eprint {http://arxiv.org/abs/2404.05998}
  {arXiv:2404.05998 [quant-ph]} \BibitemShut {NoStop}%
\bibitem [{\citenamefont {Chen}\ \emph {et~al.}(2024)\citenamefont {Chen},
  \citenamefont {Li}, \citenamefont {Lu},\ and\ \citenamefont
  {Ying}}]{Chen:2024btm}%
  \BibitemOpen
  \bibfield  {author} {\bibinfo {author} {\bibfnamefont {H.}~\bibnamefont
  {Chen}}, \bibinfo {author} {\bibfnamefont {B.}~\bibnamefont {Li}}, \bibinfo
  {author} {\bibfnamefont {J.}~\bibnamefont {Lu}}, \ and\ \bibinfo {author}
  {\bibfnamefont {L.}~\bibnamefont {Ying}},\ }\href@noop {} {\  (\bibinfo
  {year} {2024})},\ \Eprint {http://arxiv.org/abs/2407.06594} {arXiv:2407.06594
  [quant-ph]} \BibitemShut {NoStop}%
\bibitem [{\citenamefont {Brunner}\ \emph {et~al.}(2024)\citenamefont
  {Brunner}, \citenamefont {Coopmans}, \citenamefont {Matos}, \citenamefont
  {Rosenkranz}, \citenamefont {Sauvage},\ and\ \citenamefont
  {Kikuchi}}]{Brunner:2024ejl}%
  \BibitemOpen
  \bibfield  {author} {\bibinfo {author} {\bibfnamefont {E.}~\bibnamefont
  {Brunner}}, \bibinfo {author} {\bibfnamefont {L.}~\bibnamefont {Coopmans}},
  \bibinfo {author} {\bibfnamefont {G.}~\bibnamefont {Matos}}, \bibinfo
  {author} {\bibfnamefont {M.}~\bibnamefont {Rosenkranz}}, \bibinfo {author}
  {\bibfnamefont {F.}~\bibnamefont {Sauvage}}, \ and\ \bibinfo {author}
  {\bibfnamefont {Y.}~\bibnamefont {Kikuchi}},\ }\href@noop {} {\  (\bibinfo
  {year} {2024})},\ \Eprint {http://arxiv.org/abs/2412.17706} {arXiv:2412.17706
  [quant-ph]} \BibitemShut {NoStop}%
\bibitem [{\citenamefont {Lee}\ \emph {et~al.}(2023)\citenamefont {Lee},
  \citenamefont {Mulligan}, \citenamefont {Ringer},\ and\ \citenamefont
  {Yao}}]{Lee:2023urk}%
  \BibitemOpen
  \bibfield  {author} {\bibinfo {author} {\bibfnamefont {K.}~\bibnamefont
  {Lee}}, \bibinfo {author} {\bibfnamefont {J.}~\bibnamefont {Mulligan}},
  \bibinfo {author} {\bibfnamefont {F.}~\bibnamefont {Ringer}}, \ and\ \bibinfo
  {author} {\bibfnamefont {X.}~\bibnamefont {Yao}},\ }\href {\doibase
  10.1103/PhysRevD.108.094518} {\bibfield  {journal} {\bibinfo  {journal}
  {Phys. Rev. D}\ }\textbf {\bibinfo {volume} {108}},\ \bibinfo {pages}
  {094518} (\bibinfo {year} {2023})},\ \Eprint
  {http://arxiv.org/abs/2308.03878} {arXiv:2308.03878 [quant-ph]} \BibitemShut
  {NoStop}%
\bibitem [{\citenamefont {Romatschke}(2020)}]{Romatschke:2019nmo}%
  \BibitemOpen
  \bibfield  {author} {\bibinfo {author} {\bibfnamefont {P.}~\bibnamefont
  {Romatschke}},\ }\href {\doibase 10.1007/JHEP03(2020)174} {\bibfield
  {journal} {\bibinfo  {journal} {JHEP}\ ,\ \bibinfo {pages} {174}} (\bibinfo
  {year} {2020})},\ \Eprint {http://arxiv.org/abs/1910.09550} {arXiv:1910.09550
  [hep-lat]} \BibitemShut {NoStop}%
\bibitem [{\citenamefont {Ebner}\ \emph
  {et~al.}(2024{\natexlab{b}})\citenamefont {Ebner}, \citenamefont {Sch\"afer},
  \citenamefont {Seidl}, \citenamefont {M\"uller},\ and\ \citenamefont
  {Yao}}]{Ebner:2024mee}%
  \BibitemOpen
  \bibfield  {author} {\bibinfo {author} {\bibfnamefont {L.}~\bibnamefont
  {Ebner}}, \bibinfo {author} {\bibfnamefont {A.}~\bibnamefont {Sch\"afer}},
  \bibinfo {author} {\bibfnamefont {C.}~\bibnamefont {Seidl}}, \bibinfo
  {author} {\bibfnamefont {B.}~\bibnamefont {M\"uller}}, \ and\ \bibinfo
  {author} {\bibfnamefont {X.}~\bibnamefont {Yao}},\ }\href@noop {} {\
  (\bibinfo {year} {2024}{\natexlab{b}})},\ \Eprint
  {http://arxiv.org/abs/2401.15184} {arXiv:2401.15184 [hep-lat]} \BibitemShut
  {NoStop}%
\bibitem [{\citenamefont {Ebner}\ \emph
  {et~al.}(2024{\natexlab{c}})\citenamefont {Ebner}, \citenamefont {M\"uller},
  \citenamefont {Sch\"afer}, \citenamefont {Schmotzer}, \citenamefont {Seidl},\
  and\ \citenamefont {Yao}}]{Ebner:2024qtu}%
  \BibitemOpen
  \bibfield  {author} {\bibinfo {author} {\bibfnamefont {L.}~\bibnamefont
  {Ebner}}, \bibinfo {author} {\bibfnamefont {B.}~\bibnamefont {M\"uller}},
  \bibinfo {author} {\bibfnamefont {A.}~\bibnamefont {Sch\"afer}}, \bibinfo
  {author} {\bibfnamefont {L.}~\bibnamefont {Schmotzer}}, \bibinfo {author}
  {\bibfnamefont {C.}~\bibnamefont {Seidl}}, \ and\ \bibinfo {author}
  {\bibfnamefont {X.}~\bibnamefont {Yao}},\ }\href@noop {} {\  (\bibinfo {year}
  {2024}{\natexlab{c}})},\ \Eprint {http://arxiv.org/abs/2411.04550}
  {arXiv:2411.04550 [hep-lat]} \BibitemShut {NoStop}%
\end{thebibliography}%
\bibliographystyle{apsrev4-1}
\end{document}